\documentclass[twoside,12pt]{article}
\usepackage{epsfig}

\usepackage[margin=2cm]{geometry}
\usepackage{color}
\usepackage{graphicx}
\usepackage{amssymb}
\usepackage{amsmath} 
\usepackage{pifont} 
\usepackage{hyperref}
\usepackage[noadjust]{cite}
\usepackage[center,footnotesize,hang]{subfigure}
\usepackage{multirow}

\usepackage{braket}

\usepackage{mathtools}
\usepackage{amssymb}
\usepackage{amsfonts}
\usepackage[footnotesize]{caption}
\usepackage{color}
\usepackage{braket}
\usepackage{cite}
\usepackage{hyperref}
\usepackage{url}
\usepackage{multirow}
\usepackage{relsize}
\usepackage{fullpage}
\usepackage{makecell}
\usepackage{fullpage}

\newcommand{\be}{\begin{equation}}
\newcommand{\ee}{\end{equation}}
\newcommand{\bea}{\begin{eqnarray}}
\newcommand{\eea}{\end{eqnarray}}

\newcommand{\overbar}[1]{\mkern 1.5mu\overline{\mkern-1.5mu#1\mkern-1.5mu}\mkern 1.5mu}
\newcommand{\pmatr}[1]{\begin{pmatrix} #1 \end{pmatrix}}

\newcommand{\CP}{$\mathcal{CP}$\,}

\newcommand{\Tr}{\mathrm{Tr}\,}

\def\th#1#2{\ensuremath{\theta_{#1#2}}}

\def\Dm#1#2{\ensuremath{\Delta m^2_{#1#2}}}

\begin{document}

\title{ \vspace{1cm} \bf Unified Models of Neutrinos, Flavour \\ and CP Violation}
\author{S.F. King,$^{1}$ \\
\\
$^1$Physics and Astronomy, University of Southampton, \\ Southampton, SO17 1BJ, U.K.}
\maketitle
\begin{abstract} 
Recent data from neutrino experiments gives 
intriguing hints about the mass ordering, the CP 
violating phase and non-maximal atmospheric mixing.
There seems to be a (one sigma) preference for a normal ordered (NO) neutrino mass pattern, with a
CP phase $\delta = -100^{\circ}\pm 50 ^\circ$, and (more significantly) non-maximal atmospheric mixing. Global fits for the NO case yield lepton mixing angle one sigma ranges:
$\theta_{23}\approx 41.4^\circ \pm 1.6^\circ$, $\theta_{12}\approx 33.2^\circ \pm 1.2^\circ$, $\theta_{13}\approx 8.45^\circ \pm 0.15^\circ$.
Cosmology gives a limit on the total of the three masses to be below about 
$0.23$~eV, favouring hierarchical neutrino masses over quasi-degenerate masses.
Given such experimental advances, it seems an opportune moment to review the theoretical status of attempts
to explain such a pattern of neutrino masses and lepton mixing, focussing on
approaches based on the four pillars of:  {\em predictivity},  {\em minimality}, {\em robustness} and {\em unification}.
{\em Predictivity} can result from various mixing sum rules whose status is reviewed.
{\em Minimality} can follow from  
the type I seesaw mechanism, including constrained sequential dominance of right-handed (RH) neutrinos, and the littlest seesaw model. 
{\em Robustness} requires enforcing a discrete CP and non-Abelian family symmetry, 
spontaneously broken by flavons with the symmetry preserved in a semi-direct way.
{\em Unification} can account
for all lepton and quark masses, mixing angles
and CP phases, as in Supersymmetric Grand Unified Theories of Flavour,
with possible string theory origin.
\end{abstract}
\eject
\tableofcontents

\newpage

\section{Introduction}
\label{Intro}
Neutrino physics represents (at least so far) the first particle physics
beyond the Standard Model (BSM).
It gives tantalising new clues about the flavour puzzle which may lead to its eventual resolution.
The importance of neutrino mass and mixing was recently acknowledged by the Nobel Prize for Physics in 2015, awarded to 
Takaaki Kajita for the Super-Kamiokande (SK) Collaboration 
and to Arthur B. McDonald for the Sudbury Neutrino Observatory (SNO) Collaboration. The citation was 
``for the discovery of neutrino oscillations, which shows that neutrinos have mass'' and ``for their key contributions to the experiments which demonstrated that neutrinos change identities''.
Neutrino physics is 
the gift that keeps on giving, with new results and discoveries almost every year since 1998 \cite{nobel,expts,Neutrino2016}:
\begin{itemize}
\item 1998 Atmospheric $\nu_{\mu}$ disappear, implying large atmospheric mixing (SK) 
\item  2002 Solar $\nu_{e}$ disappear, implying large solar mixing (SK, after Homestake and Gallium) 
\item 2002 Solar $\nu_{e}$ appear as $\nu_{\mu}$ and $\nu_{\tau}$ (SNO)
\item 2004 Reactor $\overline{\nu_{e}}$ oscillations observed (KamLAND)
\item 2004 Accelerator $\nu_{\mu}$ disappear (K2K)
\item 2006 Accelerator $\nu_{\mu}$ disappearance confirmed and studied (MINOS)
\item 2010 Accelerator $\nu_{\mu}$ appear as $\nu_{\tau}$  (OPERA) 
\item 2011 Accelerator $\nu_{\mu}$ appear as $\nu_{e}$, hint for reactor mixing (T2K, MINOS)
\item 2012 Reactor $\overline{\nu_{e}}$ disappear, and reactor angle measured (Daya Bay, RENO)
\item 2014 Accelerator $\nu_{\mu}$ appear as $\nu_{e}$, hint for \CP violation (T2K)
\item 2015 Various $\nu_{\mu}$ disappearance hints for Normal Ordering (SK, T2K, NOvA)
\item 2016 Accelerator $\nu_{\mu}$ disappearance ``excludes maximal atmospheric mixing'' (NOvA)
\end{itemize}
What have we learned from this wealth of data?
We have learned that neutrinos have exceedingly small masses, all of them being much less than $m_e$. Not a strong neutrino mass hierarchy, at least as compared to the charged lepton or quark masses.
Neutrino masses break individual lepton numbers $L_e , L_{\mu} , L_{\tau}$, however the jury is still
out on whether they break total 
lepton number $L=L_e + L_{\mu} + L_{\tau}$, which would be a signal that 
neutrinos are Majorana, rather than Dirac.
Furthermore, neutrinos are observed to mix a lot, much more than the quarks; indeed the smallest lepton mixing angle is comparable to the 
largest quark mixing angle. In fact we have learned quite a lot about the lepton mixing angles and neutrino masses (or rather, their mass squared differences),
as we shall discuss later.
But before getting too carried away, it is worth summarising what we still don't know:
\begin{itemize}
\item Is leptonic \CP symmetry violated?
\item Does $\theta_{23}$ belong to the first octant or the second octant?
\item Are the neutrino mass squareds normal ordered (NO)? 
\item What is the lightest neutrino mass value?
\item Are the neutrino masses of the Dirac or Majorana type?
\end{itemize}

Before entering into such details about the neutrino mass and mixing, and the emerging hints arising from the latest data for the what this pattern looks like,
it is important the emphasise that we are dealing with BSM physics.
To understand why this 
is evidence for BSM physics, we recall that, in the SM,
neutrinos are massless for three reasons:
\begin{itemize}
\item There are no RH (sterile) neutrinos $\nu_R$ in the SM;
\item In the SM there are no Higgs in $SU(2)_L$ triplet representations;
\item The SM Lagrangian is renormalisable.
\end{itemize}
In the SM, there are three neutrinos $\nu_e$, $\nu_{\mu}$, $\nu_{\tau}$ which 
all massless and are
distinguished by separate lepton numbers $L_e$, $L_{\mu}$,
$L_{\tau}$. The neutrinos and antineutrinos are distinguished by total lepton number 
$L=\pm 1$. 
Clearly we must go beyond the SM to understand the origin 
of the tiny neutrino masses, so that at least 
one of the above should not apply. 

For instance, if RH (sterile) neutrinos $\nu_R$
are included, then the usual Higgs mechanism of the SM
yields Dirac neutrino mass in the same way as for the electron mass $m_e$. 
This would break
$L_e$, $L_{\mu}$, $L_{\tau}$, but preserve $L$. According to 
this simplest possibility, the Yukawa term would be
$Y_{\nu}\overline{L}H\nu_R$, in the standard notation,
and $Y_{\nu}\sim 10^{-12}$. 
By comparison, the electron mass has a 
Yukawa coupling eigenvalue $Y_e$ of about $10^{-6}$.

Alternatively, neutrinos may have a Majorana mass
which would break $L$, leading to neutrinoless double beta decay.
Indeed, having introduced RH neutrinos (also called sterile neutrinos,
since they are SM singlets), something must prevent 
(large) Majorana mass terms $M_{R} \nu_R \nu_R$ where $M_{R}$ could take any value up to 
the Planck scale. 
A conserved symmetry such as $U(1)_{B-L}$ would forbid RH neutrino masses, 
but if gauged (in order to be a robust symmetry) it would have to be broken at the TeV scale or higher,
allowing Majorana masses at the $U(1)_{B-L}$ breaking scale. 

According to the above argument, SM singlet RH neutrino Majorana masses seem difficult to avoid.
However, it is also possible to generate left-handed (LH) Majorana neutrino masses,
which may arise 
even without RH neutrinos. 
Such masses are allowed below the electroweak (EW) scale since 
neutrinos do not carry electric charge.
For instance, introducing a Higgs triplet $\Delta$ (written as a $2\times 2$ matrix), 
LH Majorana neutrino masses arise from the term $y_M  L^T (\Delta )  L $,
where $y_M$ is a dimensionless coupling. Majorana masses occur 
once the lepton doublets $L$ are contracted with the neutral component of the
Higgs triplet which develops a vacuum expectation value (VEV).

Alternatively, Majorana mass can arise 
from dimension five operators first proposed by Weinberg~\cite{Weinberg:1980bf},
\begin{equation}
-\frac{1}{2}\left( \frac{\lambda}{\Lambda} \right) L^T (HH) L ,
\label{dim5}
\end{equation}
where $\lambda $ is a dimensionless coupling constant, $\Lambda$ is a mass scale and $(HH)$ is an $SU(2)_L$ triplet combination of two Higgs doublets (written as a $2\times 2$ matrix).
This is a non-renormalisable operator, which is the lowest dimension operator which 
may be added to the renormalisable SM
Lagrangian. We require 
$\left( \frac{\lambda}{\Lambda} \right) \sim 1/(10^{14} {\rm GeV}) $ for $m^{\nu}\sim 0.1$ eV.
The elegant type I seesaw mechanism \cite{seesaw} identifies 
the mass scale $\Lambda$ with the RH neutrino Majorana mass $\Lambda = M_R$,
and $\lambda $ with the product of Dirac Yukawa couplings $\lambda = Y_{\nu}^2$.
Of course the situation is rather more complicated in practice since there may be three RH neutrinos
and both $M_R$ and $Y_{\nu}$ may be a $3\times 3 $ matrices.
In this case, after integrating out the RH neutrinos \cite{seesaw},
we arrive at a more complicated version of Eq.\ref{dim5}:
\begin{equation}
-\frac{1}{2}\left(Y_{\nu}M_R^{-1}Y_{\nu}^T\right) L^T (HH) L .
\label{dim5s}
\end{equation}

In general there are three classes of proposals in the literature 
for the new physics at the scale $\Lambda$:
\begin{itemize}
\item Three types of seesaw mechanisms~\cite{seesaw,type2,Foot:1989type3}; 
also in addition low (TeV) scale seesaw mechanisms \cite{King:2004cx}
(with the Weinberg operator resulting from the mass $M$ of a heavy particle exchanged at tree-level with $\Lambda=M$); 
\item $R$-parity violating supersymmetry~\cite{Drees:1997id}
( $\Lambda=$TeV Majorana mass neutralinos $\chi$); 
\item Loop mechanisms involving scalars with masses of order the TeV-scale ~\cite{Zee:1980ai}
(in which the Weinberg operator arises from loop diagrams involving 
additional Higgs doublets/singlets);
\end{itemize}
In addition there are two classes of early 
\footnote{We shall discuss some recent developments later.}
string-inspired explanations for neutrino mass:
\begin{itemize}
\item Extra dimensions~\cite{Arkani-Hamed:1998vp} with RH neutrinos in the bulk
leading to suppressed Dirac Yukawa $Y_{\nu}$;
\item Stringy mechanisms~\cite{Mohapatra:bd}. 
\end{itemize}

In this review we shall focus on the type I seesaw mechanism
(for a full discussion of other neutrino mass mechanisms see e.g.~\cite{Bandyopadhyay:2007kx}).
Whatever its origin, 
the observation of neutrino mass and mixing implies around seven new parameters beyond those in the SM, namely:
3 neutrino masses (or maybe 2 if one neutrino is massless), 3 lepton mixing angles, plus at least 1 phase which is \CP violating.
If there are 3 Majorana neutrino masses, then there will be 2 further \CP violating phases.
The existence of these extra seven (more or less) parameters, adding to the already twenty or so parameters of the minimal SM, 
means that we now have approaching thirty parameters describing our supposedly fundamental theory of quarks and leptons.
In the words of Feynman~\cite{Feynman}: {\it ``Nature gives us such wonderful puzzles! Why does She repeat the electron at 206 times and 3,640 times its mass?"} 
Feynman goes on to say that there are many such numbers that are not understood,
but although we use these numbers all the time we have no understanding of where they come from.
He is of course referring to the flavour puzzle which is not addressed by the SM.

We define the flavour puzzle as a collection of related questions:
\begin{itemize}
\item What is the origin of the three lepton and quark families?
\item Why are $d,s,b$ quark and $e,\mu , \tau$ lepton masses hierarchical?
\item Why are $u,c,t$ quark masses the most hierarchical?
\item Why are the two heavier neutrino masses less hierarchical?
\item What is the theory behind the neutrino masses?
\item Are neutrinos mainly Dirac or Majorana particles?
\item Why are neutrino masses so small?
\item What is the reason for large lepton mixing?
\item What is the physics behind \CP violation?
\end{itemize}

Neutrinos with mass and mixing exacerbates the flavour puzzle,
but also provides fresh opportunities to resolve it. 
Indeed, as we shall see, the key observations are {\em small} neutrino masses
and {\em large} lepton mixing. 
In this review article we shall be concerned with the impact of neutrino physics on models which address the flavour problem. 
We shall also consider how these theories fit into 
the quest for the unification of all particle forces, begun by Maxwell in his c.1865 unification of electricity and magnetism, and continued with
the c.1965 electroweak unification of the SM. 

To set the scene for the present review, let us briefly review 
neutrino model building, starting from 1998, as traced by a selection of earlier review articles
\cite{Altarelli:1999gu,King:2003jb,Mohapatra:2006gs,Altarelli:2010gt, Ishimori:2010au,  King:2013eh,King:2014nza,King:2015aea}. 
The earliest review \cite{Altarelli:1999gu} which considered models with both small and large solar mixing,
with mass matrix textures enforced by a $U(1)$ family symmetry, already considered how such models could be extended into Grand Unified Theories (GUTs).
Another review \cite{King:2003jb}, written shortly after large solar mixing was established in 2002, 
focussed on the idea of a seesaw mechanism in which there is a sequential dominance
(SD) of the RH neutrinos \cite{King:1998jw,King:1999mb,King:2002nf}, where 
a family symmetry such as $SU(3)$ \cite{King:2001uz}
(continuous and non-Abelian) 
is required to simultaneously explain
large solar and atmospheric mixing. The seesaw mechanism was also 
emphasised in \cite{Mohapatra:2006gs}.
It is worth emphasising that, back in the day, SD predicted a NO neutrino masses with $m_1\ll m_2 < m_3$ (i.e. hierarchical as well as NO)
and allowed a sizeable reactor angle 
$\theta_{13}\lesssim m_2/m_3$ \cite{King:1998jw,King:1999mb,King:2002nf},
consistent with current data.

The next period in model building witnessed the rise of 
tri-bimaximal lepton mixing with a large number of such  
models being enforced by discrete non-Abelian family symmetries
\cite{Ma:2001dn}, enforced by vacuum alignment \cite{Altarelli:2005yp}, 
as reviewed in \cite{Altarelli:2010gt,Ishimori:2010au}.
The discrete symmetry was linked to simple mixing patterns such as tri-bimaximal mixing.
However, Nature turned out to be not so simple, and the discovery of a sizeable reactor angle in 2012 ruled out tri-bimaximal mixing, along with many of these models. Yet the idea of a simple discrete non-Abelian family symmetry 
survived, as exemplified by the post-2012 literature on models based on $S_3$ \cite{survey-SM-S3},
$A_4$ \cite{survey-SM-A4} or $S_4$ \cite{survey-SM-S4}.

Indeed, as the subsequent review articles
\cite{King:2013eh,King:2014nza} showed, although tri-bimaximal lepton mixing is excluded, 
tri-bimaximal {\em neutrino} mixing is still possible in conjunction with 
charged lepton corrections. Another idea is to preserve either column 1 or 2 of the TB mixing matrix,
called trimaximal ($\text{TM}_1$ or $\text{TM}_2$) lepton mixing.
In such cases the structures may still be 
enforced by discrete non-Abelian family symmetry.
For example, 
several model building approaches were discussed in \cite{King:2013eh,King:2014nza} classified as:
direct (involving a large discrete symmetry with a fixed reactor angle); semi-direct (with a small discrete symmetry but an undetermined reactor angle);
indirect (again with a small discrete symmetry but novel vacuum alignments with a fixed reactor angle); or anarchy (no symmetry at all).
Predictions of the \CP phase resulting from the interplay of the discrete \CP symmetry and the discrete family symmetry of semi-direct models were reviewed in \cite{King:2015aea}. The present situation in neutrino theory is a bit like an orchestra tuning up, with everyone playing a different tune.
Hopefully this is just a prelude to a new movement in neutrino theory, as we will discuss in this review.

The present review will focus on classes of models 
which are based on the four pillars of:
\begin{itemize}
\item
{\em Predictivity} (it must be possible to exclude such models by experiment);
\item
 {\em Minimality} (models must be simple/elegant enough to have a chance of being correct);
 \item
 {\em Robustness} (models must be firmly based on some theoretical symmetry and/or dynamics);
 \item
 {\em Unification} (models must be capable of being embedded into a unified theory).
 \end{itemize}
The first requirement of {\em predictivity} immediately excludes, for example, the idea of anarchy 
\cite{Hall:1999sn} which, along with many other flavour models,
are not {\em sufficiently} predictive to enable them to be definitively tested.
Similarly we shall regard models with very large discrete symmetry as failing the second test of {\em minimality}. 
Finally we consider {\it ad hoc} texture models based on mass matrices not enforced by symmetry,
or models with unsubstantiated assumptions, as failing the third test of  {\em robustness}.
We also reject models which do not allow the gauge group to be unified.
Examples of models which pass all four tests are the semi-direct models,
mentioned above, based on smaller discrete 
symmetries, including those combined with spontaneous \CP violation. 
Such models generally lead to mixing sum rules which can be subject to definitive
experimental tests. There are a large number of such scenarios, based purely on symmetry arguments.
We also go beyond symmetry arguments, and consider the dynamics of flavons (the Higgs which break the family symmetry)
and the simplest type I seesaw mechanism, based on tree-level RH neutrino exchange.
We show how such models may be extended to the quark sector, as well as the lepton
sector, by embedding the Standard Model into a Supersymmetric Grand Unified Theory (SUSY GUT), 
augmented by a discrete non-Abelian family symmetry. 
Such models offer the promise of describing both quark
and lepton masses, as well as their mixing angles
and \CP phases, in a single unified framework. 
Finally we speculate on the possible string theory origins of such theories, including gravity.

The layout of the remainder of this review is as follows.
Following the pedagogical Introduction,
in section \ref{numassmixing} we give an overview of neutrino mixing and mass, including the latest global fits and the emerging hints
from the latest neutrino data.
The next four sections review the four pillars of:  {\em predictivity},  {\em minimality}, {\em robustness} and {\em unification}.
In section \ref{patterns} on {\em predictivity} we describe some of the
simpler ideas for lepton mixing, including the bimaximal, golden ratio and tri-bimaximal
schemes. Although they are 
not viable by themselves, they may be corrected by 
charged lepton mixing, resulting in solar mixing sum rules.
Alternatively, simple sub-structures may be partly preserved as in the case of 
trimaximal lepton mixing, resulting in atmospheric mixing rules.
In section~\ref{seesawsection} on {\em minimality} we review the elegant type I seesaw mechanism, including the
one RH neutrino (RHN) and two RHN models, 
as well as the idea of sequential dominance of three RH neutrinos,
constrained sequential dominance and the littlest seesaw (LS) model.
Section~\ref{sec:GUTxFam} on {\em robustness}
is devoted to a brief review of discrete CP and non-Abelian family symmetry,
spontaneously broken by flavons, in a
semi-direct way.
In section \ref{ToF} on {\em unification}
we briefly review GUTs and we then give examples of SUSY GUTs of flavour,
which incorporate many of the preceding ideas, then 
speculate about the possible string theory origin of such theories.
Section \ref{conclusion} concludes this review.

\section{Neutrino Mass and Mixing}
\label{numassmixing}

\subsection{The Neutrino Parameters}
\begin{figure}[t]
\centering
\includegraphics[width=0.6\textwidth]{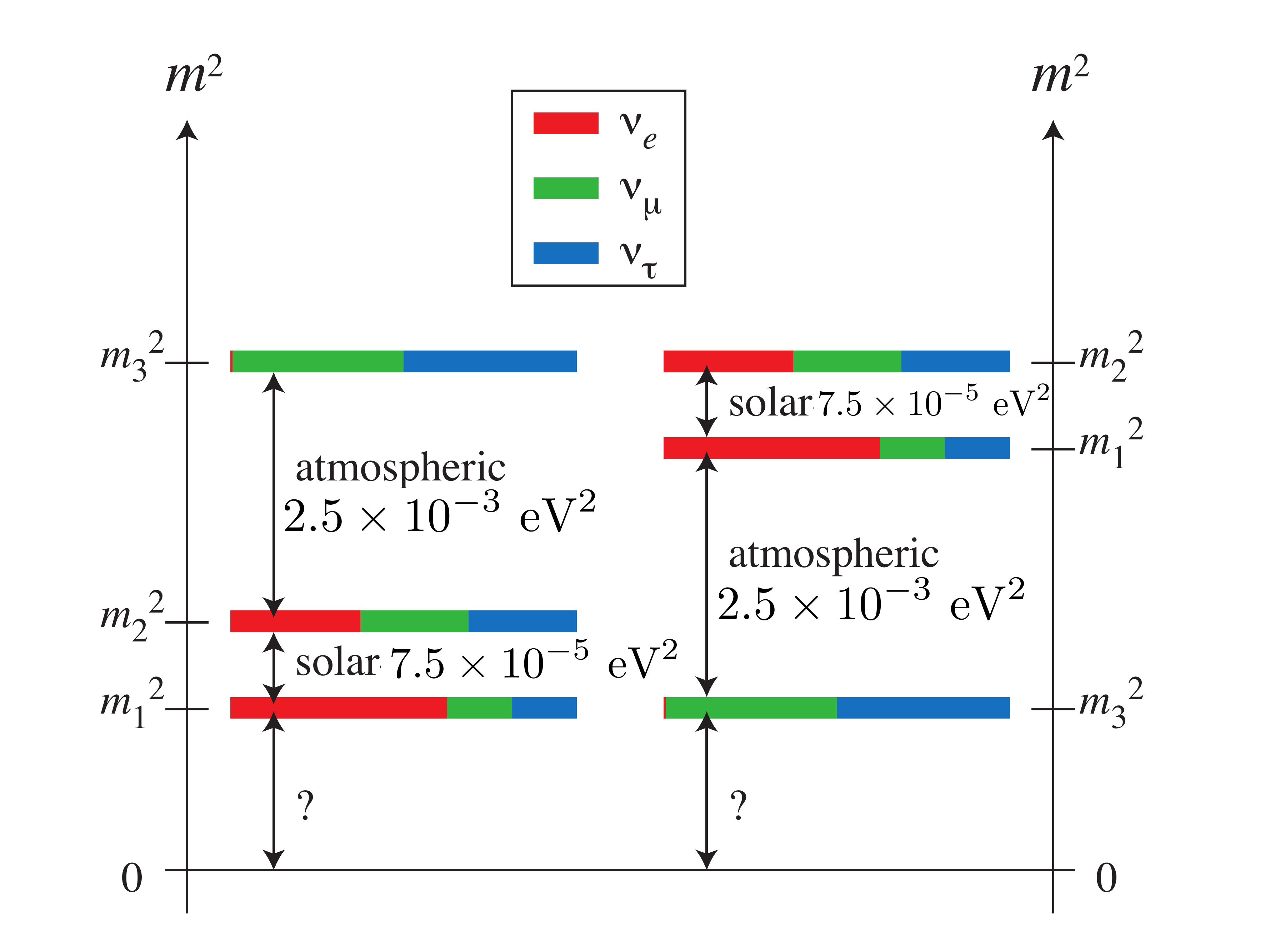}
\caption{\label{mass}\small{On the left is the normal ordering (NO),
while on the right is the inverted ordering (IO).
The probability that a neutrino state of mass (squared) $m_i^2$ contains each of 
$(\nu_e, \nu_{\mu}, \nu_{\tau})$ is proportional to the length
of the respective coloured band. Oscillation experiments only determine $m_i^2-m_j^2$.
}}
\end{figure}

Neutrino oscillation experiments are not sensitive absolute neutrino masses, only the 
neutrino mass squared differences: 
\be
\Delta m_{ij}^2=m_i^2-m_j^2.
\ee
There are two possible orderings, as shown in Fig.\ref{mass},
where the coloured bands indicate the probabilty that a particular neutrino mass eigenstate
is composed of the various flavour or weak eigenstates $(\nu_e, \nu_{\mu}, \nu_{\tau})$,
which are defined as the 
upper components of $SU(2)_L$ doublets in the diagonal
charged lepton mass basis.
One of the eigenstates is seen to contain roughly equal
amounts of $(\nu_{\mu}, \nu_{\tau})$, which, if accurately realised,
is known as bimaximal mixing. This (approximately) bimaximally mixed state
may be either identifed as
the heaviest mass eigenstate of mass $m_3$
(as shown in the left-half, called normal ordering (NO))
or such a state may be identified
as the lightest one of mass $m_1$
(as shown in the right-half, called inverted ordering (IO)).
One of the mass eigenstates is seen to contain roughly equal
amounts of all three of the weak eigenstates $(\nu_e, \nu_{\mu}, \nu_{\tau})$, which, if accurately realised,
is known as trimaximal mixing. Thus the neutrino mixing pattern is at least approximately 
characterised as being a tri-bimaximal mixing pattern, although reactor neutrino oscillation
experiments in 2012 indicated
a small but non-zero admixture of 
$\nu_e$ in the approximately bimaximally mixed $(\nu_{\mu}, \nu_{\tau})$ state, so really the pattern
of neutrino mixing should be referred to as a tri-bimaximal-reactor mixing pattern
\cite{King:2009qt,King:2011ab}.

The normal ordered (NO) pattern (positive $\Delta m_{31}^2$) seems to be slightly preferred by current data
\cite{Neutrino2016}.
The best fit mass squared differences are: 
$\Delta m_{21}^2= (7.45^{+0.25\phantom{0}}_{-0.25})10^{-5}$ eV$^2$ and 
$\Delta m_{31}^2= (2.55^{+0.05}_{-0.05})10^{-3}$ eV$^2$,
according to the global fits
\cite{Gonzalez-Garcia:2014bfa,Capozzi:2013csa,Forero:2014bxa}, 
updated after Neutrino 2016 (and ICHEP 2016).
These values and ranges are extracted from two of the updated global fits 
for the NO case as shown in Table~\ref{tab:bfp}. 
There is a 
cosmological limit on the sum total of the three
neutrino masses: $m_1+m_2+m_3 <0.23$ eV \cite{Ade:2015xua}.
Prospects for future cosmological limits approaching this value are discussed 
in \cite{Hannestad:2016mvv}.
However, there is some cosmological model dependence in these determinations,
as discussed in \cite{Hannestad:2016mvv}.
For a recent discussion of mass varying neutrinos 
which would evade these cosmological limits see \cite{Ghalsasi:2014mja}.
In this review we shall sometimes focus on models with zero lightest neutrino mass.
We stress that this is motivated purely by {\em minimality} rather than any 
definitive experimental indication.
In this spirit, we note that, if $m_1=0$, then NO would give $m_2=0.0086$ eV
and $m_3=0.050$ eV, hence $m_1+m_2+m_3\approx 0.06$ eV.
While for IO with $m_3=0$, we would find $m_2\approx m_1=0.050$ eV,
hence $m_1+m_2+m_3\approx 0.10$ eV.

 \begin{table}[ht]
\centering\begin{tabular}{lrrc}
		\hline
		 &  \multicolumn{1}{c}{NuFIT 3.0} &  \multicolumn{1}{c}{Capozzi et al} &  \multicolumn{1}{c}{Range} \\
				\hline
		$\th12$ [$^\circ$] & $33.56^{+0.77\phantom{0}}_{-0.75}$ & $33.02^{+1.06\phantom{0}}_{-1.01}$ &$33.2^{+1.2\phantom{0}}_{-1.2}$\\
		$\th13$ [$^\circ$] & $8.46^{+0.15\phantom{0}}_{-0.15}$  & $8.43^{+0.14\phantom{0}}_{-0.14}$ & $8.45^{+0.15\phantom{0}}_{-0.15}$ \\
		$\th23$ [$^\circ$] & $41.6^{+1.5\phantom{00}}_{-1.2}$  & $40.5^{+1.4\phantom{00}}_{-0.7}$& $41.4^{+1.6\phantom{00}}_{-1.6}$\\
		$\delta$ [$^\circ$] & $-99^{+51\phantom{.00}}_{-59}$ & $-108^{+38\phantom{.00}}_{-40}$ & $-100^{+50\phantom{.00}}_{-50}$  \\
		$\Dm21$ [$10^{-5}\text{eV}^2$] & $7.50^{+0.19\phantom{0}}_{-0.17}$ & $7.37^{+0.17\phantom{0}}_{-0.16}$ & $7.45^{+0.25\phantom{0}}_{-0.25}$\\
		$\Dm31$ [$10^{-3}\text{eV}^2$] & $2.524^{+0.039}_{-0.040}$ & $2.56^{+0.05}_{-0.03}$ &$2.55^{+0.05}_{-0.05}$ \\
		\hline
		\hline
\end{tabular}
\caption{ The results of the global fits for the normal ordered (NO) case.
The Gonzalez-Garcia et al NuFIT 3.0 (November 2016) give values of the above angles directly, whiile we deduced
the angles for the Capozzi et al fit from the one sigma ranges of the squared sines of the angles.
Capozzi et al give results for $\Delta m^2 = \Dm31 - (\Dm21 /2)$ from which we deduce the above values for $\Dm31$.
We also extract the combined 1$\sigma$ ranges which we derive from the two fits.}
\label{tab:bfp}
\end{table}

\begin{figure}[t]
\centering
\includegraphics[width=0.80\textwidth]{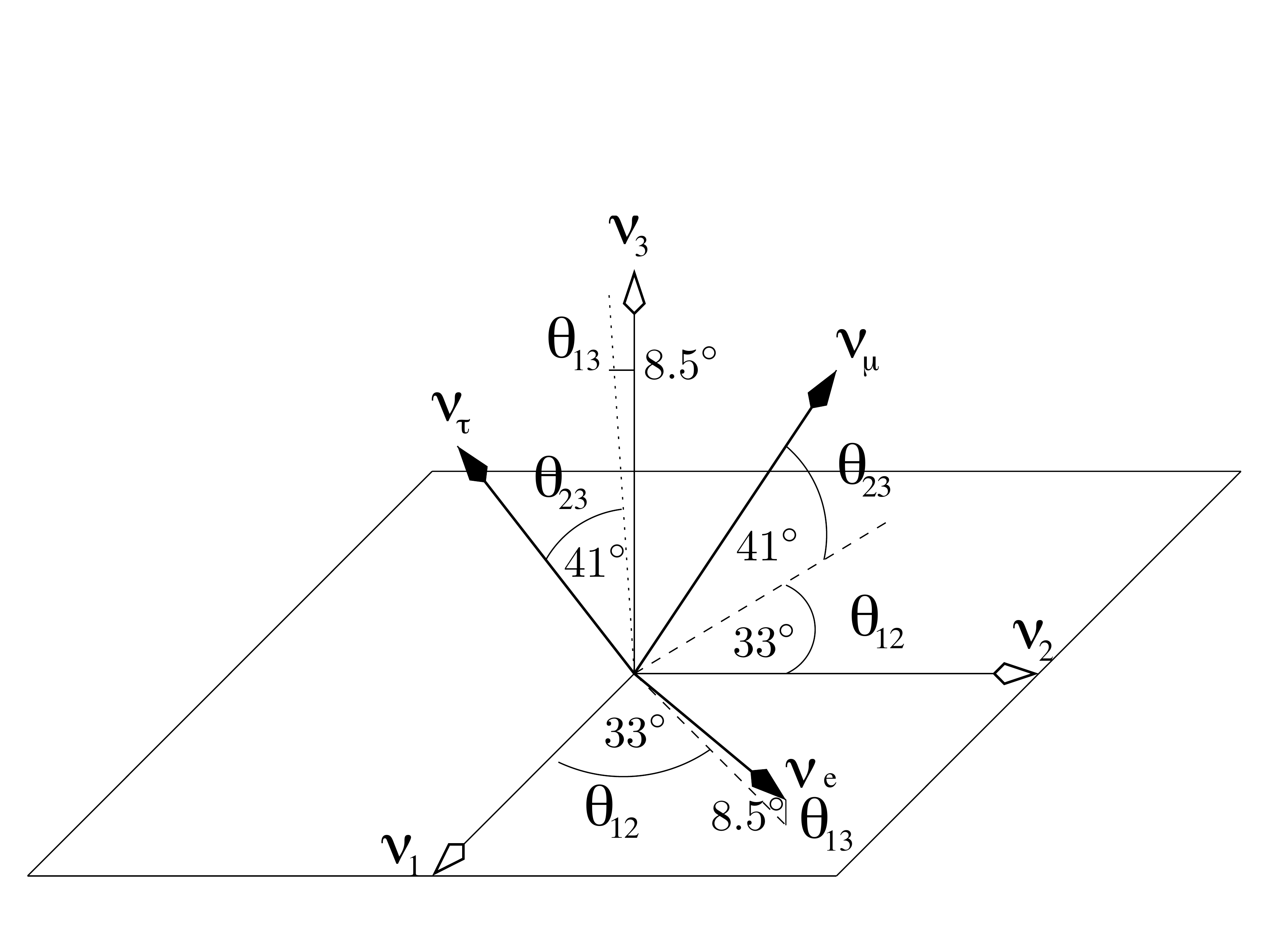}
    \caption{Neutrino mixing angles represented as a product of Euler
    rotations: $(\nu_e, \nu_{\mu}, \nu_{\tau})^T=R_{23} R_{13} R_{12}(\nu_1, \nu_2, \nu_3)^T$. Some representative values of the angles are shown for the NO case.} \label{angles}
\end{figure}

Lepton mixing (analagous to similar mixing in the
quark sector), may be parametrised by three lepton mixing angles.
The lepton mixing matrix (assuming zero \CP violation) relates the
neutrino flavour or weak eigenstates $(\nu_e, \nu_{\mu}, \nu_{\tau})$ (defined above)
to the neutrino mass eigenstate basis states
    $(\nu_1, \nu_2, \nu_3)$, according to:
    $(\nu_e, \nu_{\mu}, \nu_{\tau})^T=R_{23} R_{13} R_{12}(\nu_1, \nu_2, \nu_3)^T$ 
    where $R_{ij}$ is a real orthogonal rotation matrix in the $ij$ plane, as shown in 
    Eq.\ref{eq:matrix} (with the phase set to zero) and depicted in Fig.\ref{angles}.
    
\begin{figure}[t]
\centering
\includegraphics[width=0.30\textwidth]{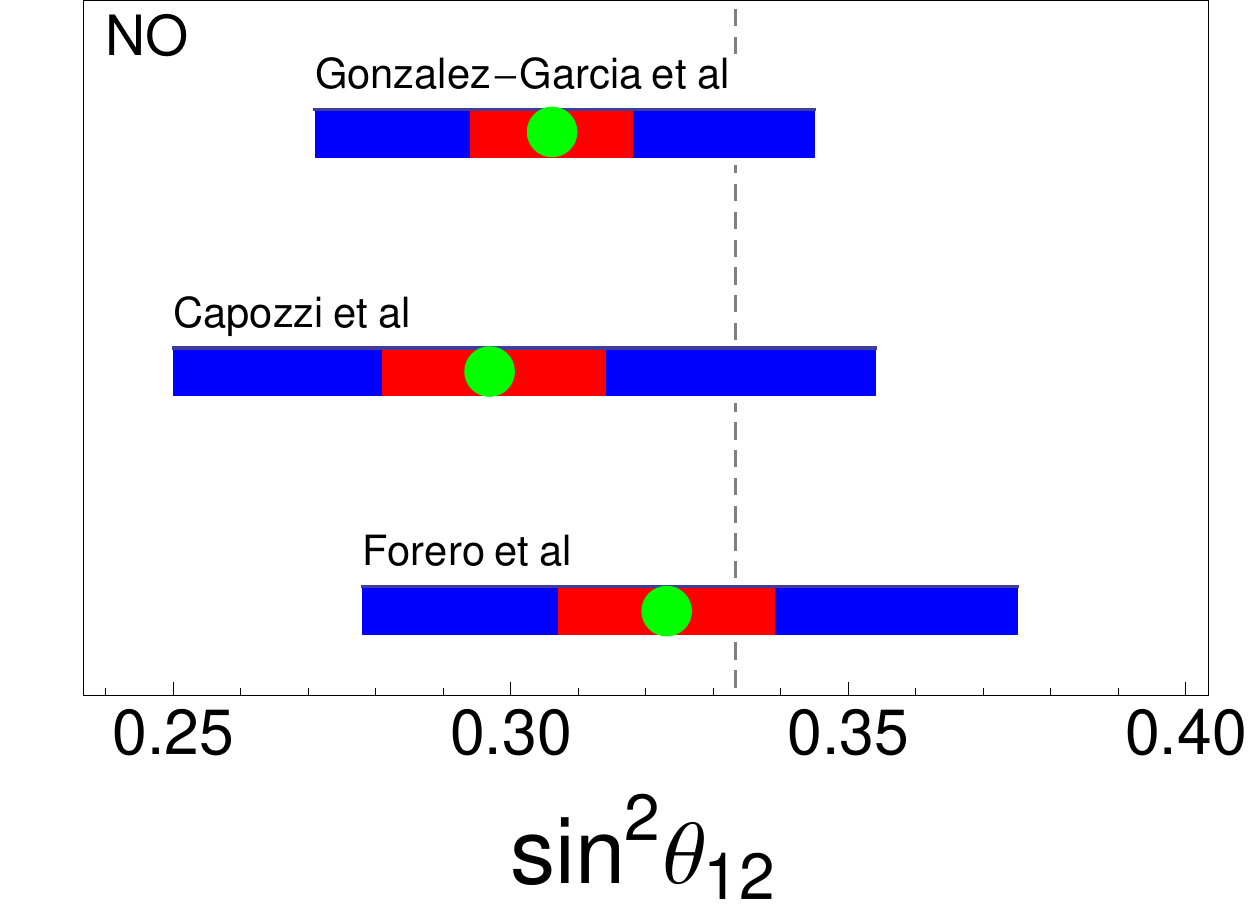}
\includegraphics[width=0.30\textwidth]{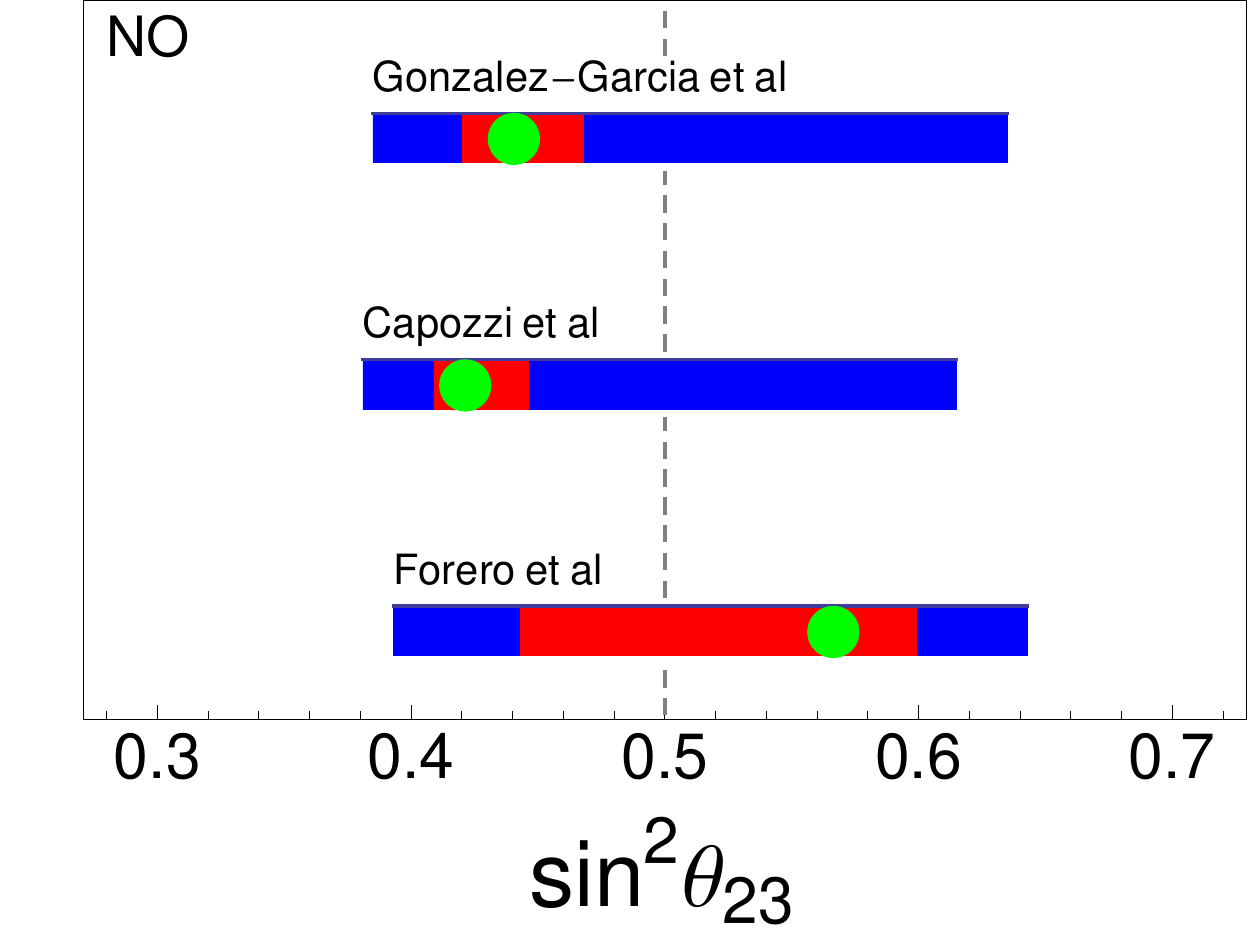}
\includegraphics[width=0.30\textwidth]{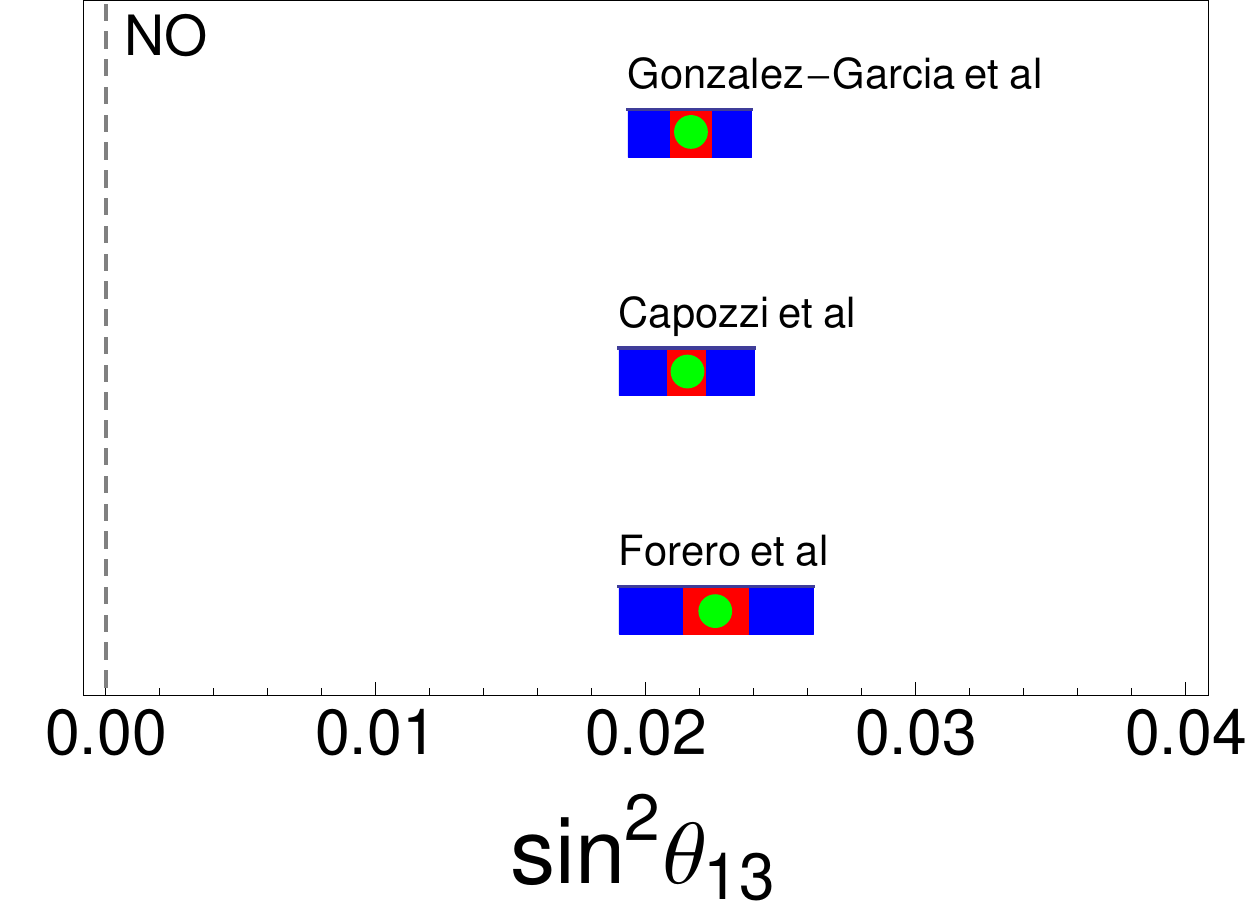}
\includegraphics[width=0.30\textwidth]{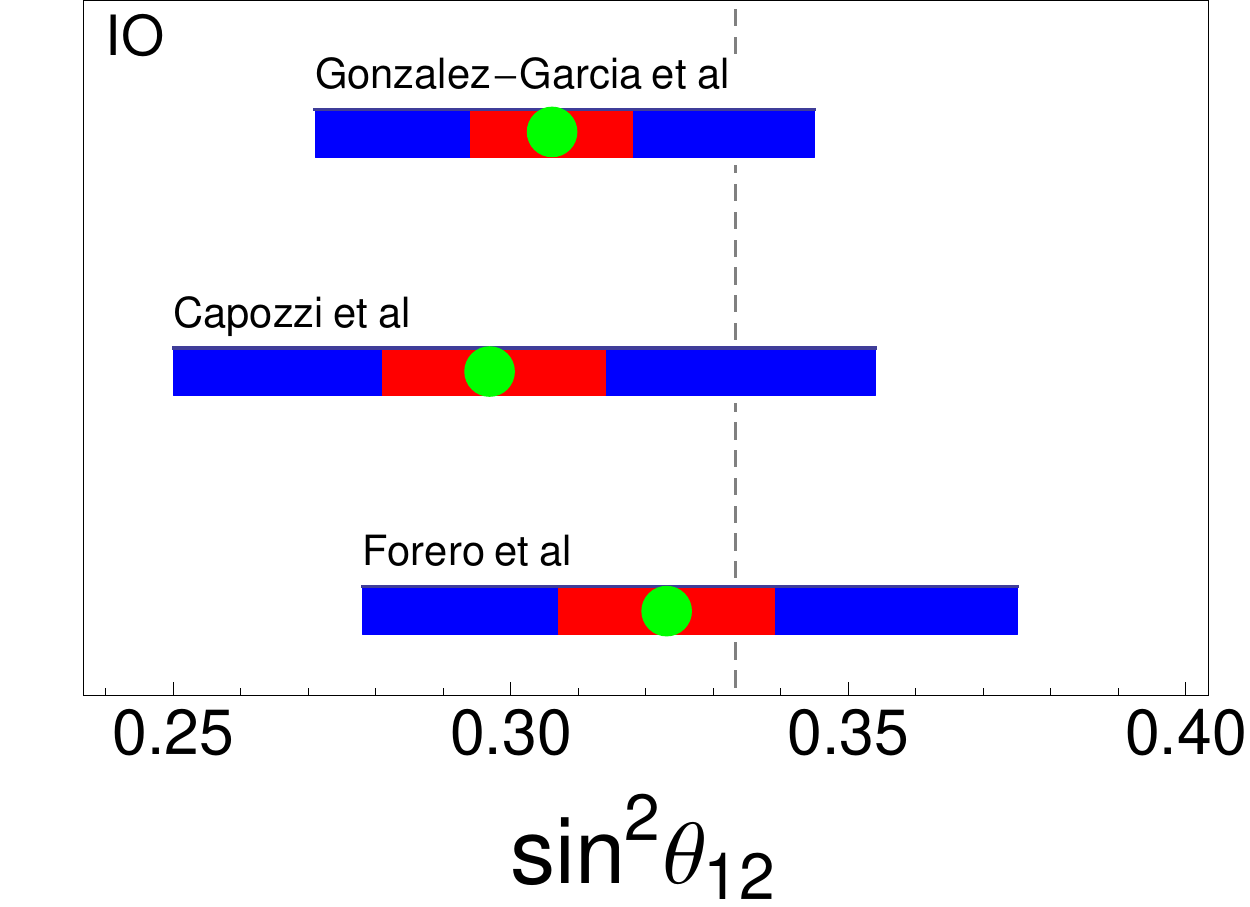}
\includegraphics[width=0.30\textwidth]{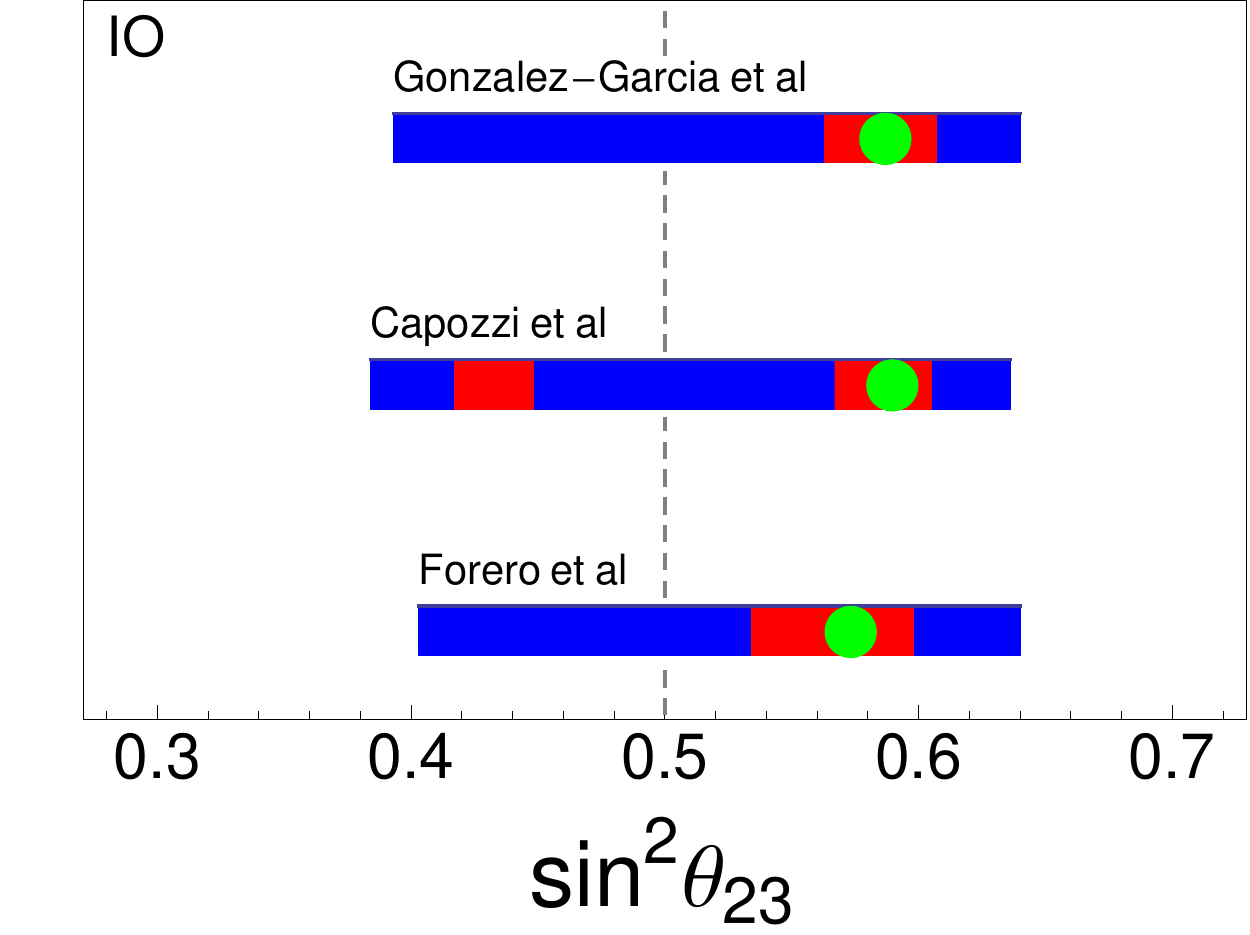}
\includegraphics[width=0.30\textwidth]{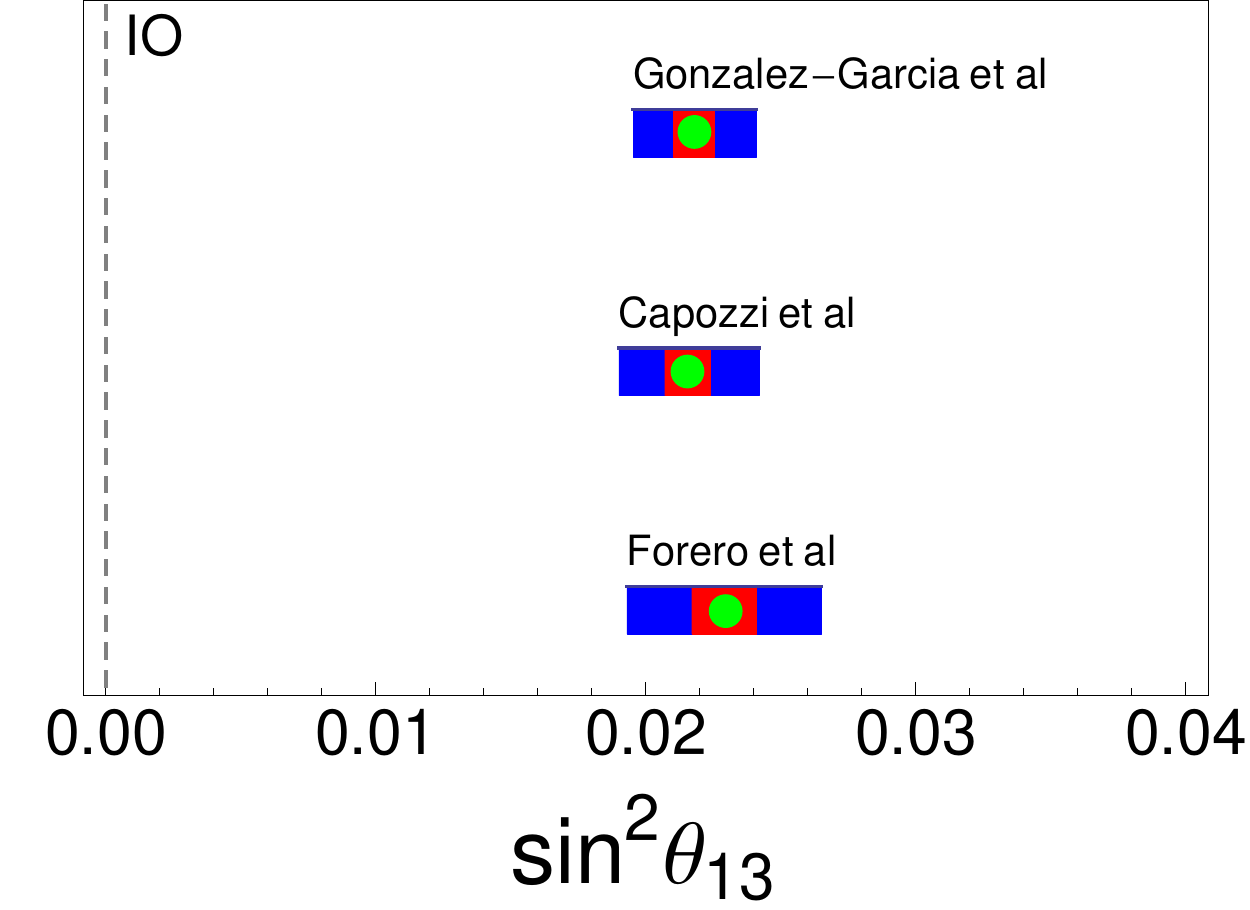}
    \caption{The current global fits of neutrino oscillation data
    (\cite{Gonzalez-Garcia:2014bfa,Capozzi:2013csa,Forero:2014bxa}) for NO (upper panels) and IO (lower panels).
    The green circles represent the best fit points, while the red/blue areas are the one/three sigma ranges.
        The results from Gonzalez-Garcia et al and Capozzi et al have been updated following the new results from the Neutrino 2016 and ICHEP 2016 conferences. The Forero et al results are from 2014.
        The dashed lines correspond to tri-bimaximal lepton mixing: 
    $s^2_{12}=1/3$, $s^2_{23}=1/2$
    $s^2_{13}=0$. 
} \label{global}
\vspace*{-2mm}
\end{figure}

   The measured mixing angles depend on whether the neutrino masses are in the NO
   or the IO pattern as shown in Fig.\ref{global}.
   Tri-bimaximal mixing would correspond to $\sin^2\theta_{23}=1/2$ and $\sin^2\theta_{13}=1/3$,
   and indicated by the dashed lines in Fig.\ref{global}, which translates into 
   $\theta_{23}= 45^\circ$, $\theta_{12}= 35.26^\circ $.
The current best 
   lepton mixing angle one sigma ranges 
      are displayed in Table~\ref{tab:bfp}
for the NO case:
$\theta_{23}\approx 41.4^\circ \pm 1.6^\circ$, $\theta_{12}\approx 33.2^\circ \pm 1.2^\circ$, $\theta_{13}\approx 8.45^\circ \pm 0.15^\circ$. These values are extracted from the two recently updated global fits of 
\cite{Gonzalez-Garcia:2014bfa,Capozzi:2013csa}
\footnote{At the time of writing \cite{Forero:2014bxa} has not yet been updated.}
The non-zero reactor angle excludes tri-bimaximal mixing.
The alternative tri-bimaximal-reactor mixing is evidently excluded by about two sigma.
In addition, there is weak evidence for a non-zero \CP violating phase.
Present data (slightly) prefers a normal ordered (NO) neutrino mass pattern, with a
CP phase $\delta = -100^{\circ}\pm 50 ^\circ$, and (more significantly) non-maximal atmospheric mixing. 
The meaning of the \CP phase $\delta$ is discussed below.

\subsection{Comparing the CKM and PMNS mixing matrices}
 
The PDG \cite{pdg} standardizes the parameterisation of the 
CKM and the PMNS mixing matrices in terms of 
unitary matrices consisting of a product of matrices $R_{23} U_{13} R_{12}$:
\begin{eqnarray}
 \label{eq:matrix}
\left(\begin{array}{ccc}
1 & 0 & 0 \\
0 & c_{23} & s_{23} \\
0 & -s_{23} & c_{23} \\
\end{array}\right)
\left(\begin{array}{ccc}
c_{13} & 0 & s_{13}e^{-i\delta} \\
0 & 1 & 0 \\
-s_{13}e^{i\delta} & 0 & c_{13} \\
\end{array}\right)
\left(\begin{array}{ccc}
c_{12} & s_{12} & 0 \\
-s_{12} & c_{12} & 0\\
0 & 0 & 1 \\
\end{array}\right) \nonumber
=
\\
 \left(\begin{array}{ccc}
    c_{12} c_{13}
    & s_{12} c_{13}
    & s_{13} e^{-i\delta}
    \\
    - s_{12} c_{23} - c_{12} s_{13} s_{23} e^{i\delta}
    & \hphantom{+} c_{12} c_{23} - s_{12} s_{13} s_{23}
    e^{i\delta}
    & c_{13} s_{23} \hspace*{5.5mm}
    \\
    \hphantom{+} s_{12} s_{23} - c_{12} s_{13} c_{23} e^{i\delta}
    & - c_{12} s_{23} - s_{12} s_{13} c_{23} e^{i\delta}
    & c_{13} c_{23} \hspace*{5.5mm}
    \end{array}\right)
   \end{eqnarray}
where $\delta \equiv \delta_{CP}$ is the \CP violating phase,
relevant to the particular sector (either quark or lepton).
We follow the short-hand notation
$s_{13}=\sin \theta_{13}$, etc, with small quark mixing, 
\begin{equation}
s^q_{12}= \lambda , \ \  s^q_{23}\sim \lambda^2, \ \  s^q_{13}\sim \lambda^3
\end{equation}
where the Wolfenstein parameter is $\lambda = 0.226\pm 0.001$ \cite{pdg}.
From Table~\ref{tab:bfp} we have large lepton mixing,
\begin{equation}
s^l_{13}\sim \lambda /\sqrt{2} , \ \  s^l_{23}\sim 1/\sqrt{2}, \ \  s^l_{12}\sim 1/\sqrt{3}.
\end{equation}
The smallest lepton mixing angle 
\footnote{When distinguishing leptons from quarks we use the superscripts $l$ and $q$,
but when it is obvious we are referring to leptons we often simply drop the superscript $l$.}
$\theta^l_{13}$ (the reactor angle), is of order the largest quark mixing angle 
$\theta_C=\theta^q_{12}$ (the Cabibbo angle).
There have been attempts to relate quark and lepton mixing angles such as
postulating $\theta^l_{13}=\theta_C/\sqrt{2}$
\cite{Minakata:2004xt}, however this relation is now experimentally excluded,
along with all models which predicted this relation. 
This is a good example of how predictive models can be excluded by accurate 
experiments. Later we shall discuss other predictive models, some of which are excluded and some which are not yet excluded.

The \CP violating quark phase $\delta^q \sim (\pi/2)/ \sqrt{2}$,
which is close to maximal
\footnote{Interestingly, in
Kobayashi-Maskawa's parametrisation,
$\delta^q \sim \pi/2$ is identified as the angle $\alpha \sim \pi/2$,
where $\alpha$ is one of the angles in the unitarity triangle
corresponding to the $1st$ and $3rd$ columns of the CKM matrix being orthogonal \cite{pdg}.}, is
reminiscent of the hint for the \CP violating lepton phase $\delta^l \sim -\pi/2$.

\subsection{Constructing the PMNS Lepton Mixing Matrix}
In this subsection we discuss lepton mixing from first principles.
For definiteness we consider Majorana masses, since Dirac neutrinos 
are completely analogous to 
the SM description of quarks. 
Consider the effective Lagrangian,
\begin{equation}
	{\cal L}^{\rm mass}_{\rm lepton} = 
	-v_dY^e_{ij}\overline e^i_{\mathrm{L}} e^j_{\mathrm{R}}  
	-\frac{1}{2} m^{\nu_e}_{ij} \overline{{\nu}_e^i}_{L} {\nu}_{eL}^{cj} 
	+ \mathrm{H.c.}
\label{lepton}
\end{equation}
which is valid below the EW symmetry breaking scale, 
where $i,j$ are flavour indices.
We do not yet specify the mechanism responsible for the above 
Majorana neutrino masses.
The mass matrices may be diagonalised by unitary matrices,
\begin{eqnarray}
U_{e_L}Y^eU_{e_R}^{\dagger}=
\left(\begin{array}{ccc}
y_e&0&0\\
0&y_{\mu}&0\\
0&0&y_{\tau}
\end{array}\right), \ \ \ \ 
U_{{\nu_e}_L}m^{\nu_e}U_{{\nu_e}_L}^{T}=
\left(\begin{array}{ccc}
m_1&0&0\\
0&m_2&0\\
0&0&m_3
\end{array}\right).
\end{eqnarray}
The charged current (CC) couplings to $W^-$ in the flavour basis are given by 
$-\frac{g}{\sqrt{2}}\overline{e}^i_L\gamma^{\mu}W_{\mu}^-{\nu}^i_{eL}$,
which becomes in the mass basis,
\begin{eqnarray}
{\cal L}^{CC}_{\rm lepton}= -\frac{g}{\sqrt{2}}
\left(\begin{array}{ccc}
\overline{e}_L  & \overline{\mu}_L &  \overline{\tau}_L
\end{array}\right)
U_{\rm PMNS}
\gamma^{\mu}W_{\mu}^-
\left(\begin{array}{c}
{\nu}_{1L} \\ 
{\nu}_{2L} \\ 
{\nu}_{3L}
\end{array}\right)+H.c.
\end{eqnarray}
where we the lepton mixing matrix is identified as,
\footnote{Different physically equivalent conventions appear in the literature, we follow the conventions in
\cite{King:2002nf}.},
\begin{equation}
\label{enu}
U_{\rm PMNS}=U_{e_L}U_{{\nu}_{eL}}^{\dagger}.
\end{equation}
It is possible to remove three of the lepton phases, using the phase invariance of $m_e,m_{\mu}, m_{\tau}$.
For example,
$m_e\overline e_{\mathrm{L}}e_{\mathrm{R}}$,
is unchanged by $e_{\mathrm{L}}\rightarrow e^{i\phi_e}e_{\mathrm{L}}$
and $e_{\mathrm{R}}\rightarrow e^{i\phi_e}e_{\mathrm{R}}$. The three such phases 
$\phi_e, \phi_{\mu}, \phi_{\tau}$
may be chosen in various ways to yield an assortment of possible PMNS parametrisations
one of which is the PDG standard choice discussed below).
This does not apply to the Majorana mass terms 
$-\frac{1}{2} {m_i} \overline{{\nu}_{iL}} {\nu}_{iL}^{c}$ where $m_i$ are real and positive, and thus the PMNS matrix
may be parametrised as in Eq.\ref{eq:matrix} but with an extra Majorana phase matrix \cite{pdg}:
\begin{eqnarray}
 \label{eq:matrix_pmns}
\!\!\!\!\!\!\!\!\!\!\!\!\!\!\!\!\!
U_{\rm PMNS}=\left(\begin{array}{ccc}
    c_{12} c_{13}
    & s_{12} c_{13}
    & s_{13} e^{-i\delta}
    \\
    - s_{12} c_{23} - c_{12} s_{13} s_{23} e^{i\delta}
    & \hphantom{+} c_{12} c_{23} - s_{12} s_{13} s_{23}
    e^{i\delta}
    & c_{13} s_{23} \hspace*{5.5mm}
    \\
    \hphantom{+} s_{12} s_{23} - c_{12} s_{13} c_{23} e^{i\delta}
    & - c_{12} s_{23} - s_{12} s_{13} c_{23} e^{i\delta}
    & c_{13} c_{23} \hspace*{5.5mm}
    \end{array}\right)
    \left(\begin{array}{ccc}
   1
    & 0     & 0
    \\
   0
    &    e^{i\frac{\alpha_{21}}{2}}
    &  0 
    \\
   0    & 0    & e^{i\frac{\alpha_{31}}{2}}
    \end{array}\right),
   \end{eqnarray}
where $\alpha_{21}$ and $\alpha_{31}$ are irremovable Majorana phases.
The mixing angles $\theta_{13}$ and $\theta_{23}$ must lie between $0$ and ${\pi}/{2}$,
while (after reordering the masses) $\theta_{12}$ lies between $0$ and ${\pi}/{4}$.
The phases all lie between $0$ and $2 \pi$, however we shall equivalently express $\delta$ 
in the range $-\pi$ to $\pi$.
There is no current constraint on the Majorana phases, $\alpha_{21}$ and $\alpha_{31}$,
nor is there likely to be in the forseeable future. The first step will be to experimentally show that neutrinos
are Majorana particles, which will most likely require neutrinoless double beta decay to be discovered.
Then, only after precision studies of neutrinoless double beta decay rates, will there be any hope of 
determining the Majorana phases $\alpha_{21}$ and $\alpha_{31}$ \cite{King:2013psa}.

\section{{\em Predictivity}: Lepton Mixing Patterns and Sum Rules}
\label{patterns}
In this section we discuss some simple structures for $U_{\rm PMNS}$.
Although some are excluded by $\theta_{13}$,
they will motivate approaches which involve a non-zero $\theta_{13}$.
Others such as trimaximal mixing allow an undetermined reactor angle $\theta_{13}$.
An important point to emphasise is that all such ansatze may be enforced by a some small
discrete non-Abelian family symmetry, making these predictions 
robust, as discussed in the following section.

\subsection{Bimaximal and Golden Ratio Mixing}

Bimaximal (BM) mixing is defined as
$s^2_{13}=0$ and $s^2_{12}=s^2_{23}=1/2$. It is sometimes enforced by 
$S_4$. It corresponds to a matrix of
the form \cite{Barger:1998ta},
\begin{equation}\label{BM}
U_{\rm BM} =
\left(
\begin{array}{ccc}
\frac{1}{\sqrt{2}} & \frac{1}{\sqrt{2}} & 0\\
-\frac{1}{2} & \frac{1}{2} & \frac{1}{\sqrt{2}}\\
\frac{1}{2} & -\frac{1}{2} & \frac{1}{\sqrt{2}}
\end{array}
\right).
\end{equation}%

Another pattern of lepton mixing associates the golden ratio
$\varphi=\frac{1+\sqrt{5}}{2}$ with $\theta_{12}$.
It is sometimes enforced by $A_5$~\cite{Datta:2003qg}. It also predicts
$s^2_{13}=0$ and $s^2_{23}=1/2$, but differs by having $\theta_{12}$ given by
$t_{12}=1/\varphi$, i.e. $\theta_{12}\approx 31.7^\circ$.
It corresponds to the mixing matrix,
\footnote{An alternative GR scheme has also been proposed with $c_{12}=\varphi/2$
\cite{Rodejohann:2008ir}.}
\begin{equation}\label{GR}
U_{\rm GR} =
\left(
\begin{array}{ccc}
\frac{\varphi}{\sqrt{2+\varphi}} &
\frac{1}{\sqrt{2+\varphi}} &  0 \\ -\frac{1}{\sqrt{4+2\varphi}}&
\frac{\varphi}{\sqrt{4+2\varphi}} & \frac{1}{\sqrt{2}} \\
\frac{1}{\sqrt{4+2\varphi}} & -\frac{\varphi}{\sqrt{4+2\varphi}} &
\frac{1}{\sqrt{2}}
\end{array}
\right).
\end{equation}%

\subsection{Tri-bimaximal lepton mixing and deviation parameters }

The Tribimaximal (TB) mixing matrix \cite{Harrison:2002er} is 
predicts $s^2_{13}=0$ and $s^2_{23}=1/2$ but differs since it 
predicts $s_{12}=1/\sqrt{3}$,
i.e. $\theta_{12}\approx 35.3^\circ$. 
It may be enforced by $S_4$, or sometimes $A_4$ with suitable field content. 
It corresponds to a mixing matrix,
\begin{equation}\label{TB}
U_{\rm TB} =
\left(
\begin{array}{ccc}
\sqrt{\frac{2}{3}} &  \frac{1}{\sqrt{3}}
&  0 \\ - \frac{1}{\sqrt{6}}  & \frac{1}{\sqrt{3}} &  \frac{1}{\sqrt{2}} \\
\frac{1}{\sqrt{6}} & -\frac{1}{\sqrt{3}} &  \frac{1}{\sqrt{2}}    
\end{array}
\right).
\end{equation}%
It is excluded by $\theta_{13}$, with $\theta_{12}$ and $\theta_{23}$ being more or less consistent.
The deviation of the mixing angles from TB mixing may be parametrised as
\cite{King:2007pr,Pakvasa:2008zz}:
\begin{eqnarray}
\sin\theta_{12} &=& \frac{1}{\sqrt{3}} (1+s),\\
\sin\theta_{23} &=& \frac{1}{\sqrt{2}} (1+a),\\
\sin\theta_{13} &=& \frac{r}{\sqrt{2}},
\end{eqnarray}
in terms of the ($s$)olar, ($a$)tmospheric and ($r$)eactor deviation parameters.
Current global fits for the NO case
yield $s\approx -0.057$, $a\approx -0.063$, $r\approx 0.21$, which shows that the reactor angle represents the most significant deviation from TB mixing.

\subsection{Trimaximal lepton mixing and sum rules}
\label{atmospheric}
Trimaximal $\text{TM}_1$ or $\text{TM}_2$ lepton mixing preserves the first or the second column of Eq.\ref{TB} \cite{Albright:2008rp},
{\small
\begin{equation}\label{TMM}
\!\!\!\!\!\!\!\!
|U_{\rm  TM_1}| =
\left(
\begin{array}{ccc}
\frac{2}{\sqrt{6}} &  - &  - \\ 
\frac{1}{\sqrt{6}} &  - &  - \\
\frac{1}{\sqrt{6}} &  - &  -  
\end{array}
\right),\ \ \ \ 
|U_{\rm TM_2}| =
\left(
\begin{array}{ccc}
- &  \frac{1}{\sqrt{3}} &  - \\ 
- & \frac{1}{\sqrt{3}} &  - \\
- & \frac{1}{\sqrt{3}} &  -  
\end{array}
\right).
\end{equation}%
}
The reactor angle is a free parameter.
The unfilled entries are fixed when the reactor angle
is specified. It is important to emphasise that these forms are more than simple ansatze, since they may be enforced by discrete non-Abelian family symmetry.
For example, $\text{TM}_2$ mixing can be realised by $A_4$ or $S_4$ symmetry \cite{King:2011zj},
while $\text{TM}_1$ mixing can be realised by $S_4$ symmetry
\cite{Luhn:2013vna}. A general group theory analysis of semi-direct symmetries 
was given in \cite{Hernandez:2012ra}.

Eq.\ref{TMM} evidently implies the relations
\begin{eqnarray} 
\!\!\!\!\!\!\!\!{\rm TM_1}:\ \ \ \ \left | U_{e1} \right | = 
\sqrt{\frac{2}{3}} \ \  {\rm and}\ \   \left|U_{\mu1}\right|
=\left|U_{\tau1}\right| =\frac{1}{\sqrt{6}}\ ;  \label{TM1}\\
\!\!\!\!\!\!\!\!\!\!\!\!\!\!\!\!{\rm TM_2}:\ \ \ \ \left|U_{e2}\right| =  \left|U_{\mu2}\right| = \left|U_{\tau2}\right| =
\frac{1}{\sqrt{3}}\ .  \label{TM2} 
\end{eqnarray}
The above $\text{TM}_1$ relations above imply three equivalent relations:
\be
\tan \theta_{12} = \frac{1}{\sqrt{2}}\sqrt{1-3s^2_{13}}\ \ \ \ {\rm or} \ \ \ \ 
\sin \theta_{12}= \frac{1}{\sqrt{3}}\frac{\sqrt{1-3s^2_{13}}}{c_{13}} \ \ \ \ {\rm or} \ \ \ \ 
\cos \theta_{12}= \sqrt{\frac{2}{3}}\frac{1}{c_{13}}
\label{t12p}
\ee
leading to a prediction for $\theta_{12}\approx 34^{\circ}$,
in agreement with current global fits, assuming $\theta_{13}\approx 8.5^{\circ}$.
By contrast, the corresponding $\text{TM}_2$ relations imply $\theta_{12}\approx 36^{\circ}$ \cite{Albright:2008rp}, which is in tension with current global best fit value $\theta_{12}\approx 33.2^\circ \pm 1.2^\circ$. 
$\text{TM}_1$ mixing also leads to an exact sum rule relation relation for $\cos \delta$ in terms of the other lepton mixing angles
\cite{Albright:2008rp},
\be
\cos \delta = - \frac{\cot 2\theta_{23}(1-5s^2_{13})}{2\sqrt{2}s_{13}\sqrt{1-3s^2_{13}}},
\label{TM1sum}
\ee
which, for approximately maximal atmospheric mixing, predicts $\cos \delta \approx 0$,
or $\delta \approx \pm 90^{\circ}$, in accord with the recent hints.
Incidentally the reason why $\cos \delta$ (not $\sin \delta$) is predicted is because such predictions follow from 
$|U_{ij}|$ being predicted, where $U_{ij}=a+be^{i\delta}$, where $a,b$ are real functions of angles in Eq.\ref{eq:matrix}.

Eqs.\ref{TM1},\ref{TM2}, can be expanded to leading order in the TB deviation parameters as~\cite{King:2007pr}, 
\begin{equation} 
a= \lambda r\cos\delta ,
\end{equation} 
where
$\lambda=(1,-\frac{1}{2})$ for ($\text{TM}_1$,$\text{TM}_2$), respectively. 
Such sum rules may be tested
in future experiments~\cite{Ballett:2013wya}. However, as noted above, 
even the current determination of $\theta_{12}$ strongly favours $\text{TM}_1$ mixing over $\text{TM}_2$ mixing.

\subsection{Charged lepton mixing corrections and sum rules}
\label{solar}

Recall that the physical PMNS matrix in Eq.\ref{enu} is given
by $U_{\rm PMNS}= U^e U^{\nu}_{\rm TB}$.
Now suppose that 
$U^{\nu}_{\rm TB}$ is the TB matrix in Eq.\ref{TB} while
$U^e$ corresponds to small but unknown charged lepton corrections.
This was first discussed in 
\cite{King:2005bj,Masina:2005hf,Antusch:2005kw,Antusch:2007rk}
where the following sum rule involving the lepton mixing parameters, including crucially the \CP phase $\delta$,
was first derived:
\begin{equation} 
\!\!\!\!\!\!\!\!\!\!
\theta_{12}\approx  35.26^o + \theta_{13}\cos\delta ,
\label{eq:linearSSR} 
\end{equation}
where $35.26^o=\sin ^{-1}\frac{1}{\sqrt{3}}$.
Eq.\ref{eq:linearSSR} may be recast in terms of TB deviation parameters as \cite{King:2007pr},
\begin{equation} 
s = r \cos \delta .
\label{eq:linearSSR2} 
\end{equation}

To derive this sum rule, let us consider the case of 
the charged lepton mixing corrections involving only (1,2) mixing,
so that the PMNS matrix is given by \cite{Antusch:2007rk},
\begin{equation}
U_{\mathrm{PMNS}} = \left(\begin{array}{ccc}
\!c^e_{12}& s^e_{12}e^{-i\delta^e_{12}}&0\!\\
\!-s^e_{12}e^{i\delta^e_{12}}&c^e_{12} &0\!\\
\!0&0&1\!
\end{array}
\right)
\left( \begin{array}{ccc}
\sqrt{\frac{2}{3}} & \frac{1}{\sqrt{3}} & 0 \\
-\frac{1}{\sqrt{6}}  & \frac{1}{\sqrt{3}} & \frac{1}{\sqrt{2}} \\
\frac{1}{\sqrt{6}}  & -\frac{1}{\sqrt{3}} & \frac{1}{\sqrt{2}}
\end{array}
\right)
= \left(\begin{array}{ccc}
\! \cdots & \ \ 
\! \cdots&
\! \frac{s^e_{12}}{\sqrt{2}}e^{-i\delta^e_{12}} \\
\! \cdots
& \ \
\! \cdots
&
\! \frac{c^e_{12}}{\sqrt{2}}
\!\\
\frac{1}{\sqrt{6}}  & -\frac{1}{\sqrt{3}} & \frac{1}{\sqrt{2}}
\end{array}
\right)
\label{Ucorr}
\end{equation}
Comparing to the PMNS parametrisation in Eq.\ref{eq:matrix} we identify
the exact sum rule relations \cite{Antusch:2007rk},
\begin{eqnarray}
\label{Eq:Sumrule4}  |U_{e3}|= s_{13} &=&  \frac{s^e_{12}}{\sqrt{2}}  \; , \\
\label{Eq:Sumrule1} |U_{\tau 1}|= |s_{23}s_{12}-s_{13}c_{23}c_{12}e^{i\delta} |    &=&     \frac{1}{\sqrt{6}}\; , \\
\label{Eq:Sumrule2} |U_{\tau 2}|=  |- c_{12} s_{23} - s_{12} s_{13} c_{23} e^{i\delta}| &=&     \frac{1}{\sqrt{3}}\; , \\
|U_{\tau 3}|=c_{13}c_{23} &=& \frac{1}{\sqrt{2}}.   
\end{eqnarray}
The first equation implies a reactor angle $\theta_{13}\approx 8.45^{\circ}$
if $\theta_e\approx 12^\circ$, just a little smaller than the Cabibbo angle.
The second and third equations, after eliminating $\theta_{23}$, yield
a new relation between the PMNS parameters, $\theta_{12}$, $\theta_{13}$ and $\delta$.
Expanding to first order gives the approximate solar sum rule relations in
Eq.\ref{eq:linearSSR} \cite{King:2005bj}.

The above derivation assumes only (1,2) charged lepton corrections.
However it is possible to derive an accurate sum rule which is valid for both (1,2) and (2,3) charged lepton corrections (while keeping $\theta^e_{13}=0$). 
Indeed, using a similar matrix multiplication method to that employed above
leads to 
the exact result  \cite{Ballett:2014dua}:
\begin{equation}
\frac{\left | U_{\tau 1} \right |}{\left | U_{\tau 2} \right |
}=\frac{|s_{12} s_{23} - c_{12} s_{13} c_{23} e^{i\delta}|}
    {|- c_{12} s_{23} - s_{12} s_{13} c_{23} e^{i\delta}|}
= \frac{1}{\sqrt{2}}\;. \label{sol3}
\end{equation}
This may also be obtained by taking the ratio
of Eqs.~\ref{Eq:Sumrule1} and \ref{Eq:Sumrule2}.
Therefore it applies to the previous case
with $\theta^e_{23}=0$.
However, since $\theta^e_{23}$ cancels in the ratio, it also applies for $\theta^e_{23}\neq 0$.
It is not fully general however since we are always assuming 
$\theta^e_{13}=0$.

After some algebra, Eq.\ref{sol3} leads to \cite{Ballett:2014dua},
\begin{equation}
\cos \delta =\frac
{t_{23}s^2_{12}+s^2_{13}c^2_{12}/t_{23}-\frac{1}{3}(t_{23}+s^2_{13}/t_{23})}
{\sin 2\theta_{12}s_{13}}. \label{sol4}
\end{equation}
To leading order in $\theta_{13}$, Eq.\ref{sol4} returns the sum rule in Eq.\ref{eq:linearSSR},
from which we find 
$\cos \delta \approx 0$ if $\theta_{12}\approx 35^o$,
consistent with $\delta \sim -\pi /2$. 
This can also be understood directly from 
Eq.\ref{sol4} where we see that for $s_{12}^2=1/3$
the leading terms $t_{23}s^2_{12}$ and $\frac{1}{3}t_{23}$ cancel in the numerator, 
giving $\cos \delta = s_{13}/(2\sqrt{2}t_{23})\approx 0.05$ to be compared to  $\cos \delta \approx 0$ in the linear approximation.
In general the error induced by using the linear sum rule instead of the exact one has been shown to be
$\Delta(\cos\delta) \lesssim 0.1$  \cite{Ballett:2014dua} for the TB sum rule.

Recently there has been much activity in exploring the phenomenology of various such 
{\em solar mixing sum rules},
arising from charged lepton corrections to simple neutrino mixing,
not just TB neutrino mixing, but other simple neutrino mixing,
including BM and GR mixing, allowing more general charged lepton corrections, renormalisation
group running and so on \cite{Marzocca:2013cr}.

It is important to distinguish {\em solar mixing sum rules} discussed here 
from {\em atmospheric mixing sum rules} discussed previously.
The physics is different: here we consider charged lepton corrections to TB neutrino mixing,
while previously we considered two forms of the physical
trimaximal lepton mixing matrix.

\section{{\em Minimality}: The Type I Seesaw Mechanism }
\label{seesawsection}

\subsection{The type I seesaw mechanism with one RH neutrino}
The LH Majorana masses are given by,
\begin{equation}
\mathcal{L}^{LL}_\nu = 
-\frac{1}{2}m^{\nu} \overline{{\nu}_{L}} {\nu}_{L}^{c} + \mathrm{H.c.}
\label{mLL}
\end{equation}
where $\nu_L^c$ is a RH antineutrino field, which is
the CP conjugate of the LH neutrino field $\nu_L$.
Majorana masses are possible below the electroweak symmetry (EW) breaking scale 
since the neutrino has zero electric charge.
Majorana neutrino masses
violate lepton number conservation, and are
forbidden above the EW
breaking scale.
The type I seesaw mechanism assumes
that Majorana neutrino mass terms are zero to begin with, but are generated effectively
by RH neutrinos \cite{seesaw}.

If we introduce one RH neutrino field $\nu_R$,
\footnote{A single RH neutrino is sufficient to account for atmospheric neutrino oscillations
if it couples approximately equally to $\nu_{\mu}$ and $\nu_{\tau}$ as 
discussed in \cite{King:1998jw}.}
then there are two possible additional neutrino mass terms. First there are
Majorana masses,
\begin{equation}
\mathcal{L}^{R}_\nu = -\frac{1}{2} M_{R} \overline{\nu_R^c} \nu_{R} + \mathrm{H.c.} 
\label{MRR}
\end{equation}
Secondly, there are
Dirac masses,
\begin{equation}
\mathcal{L}^{D}_\nu = -m_D\overline{\nu_L}\nu_R + \mathrm{H.c.} .
\label{mLR}
\end{equation}
Dirac mass terms arise from Yukawa couplings to a Higgs doublet,
$H_u$, 
\begin{equation}
\mathcal{L}^{\rm Yuk} =- H_u Y^{\nu}\overline L \nu_{R} + \mathrm{H.c.}
\label{yuk}
\end{equation}
where we write $H_u$ rather than $H$ in anticipation of a two Higgs doublet
extension of the SM, 
with $m_D=v_uY^{\nu}$ where $v_u= \langle H_u \rangle$.

\begin{figure}[t]
\centering
\includegraphics[width=0.60\textwidth]{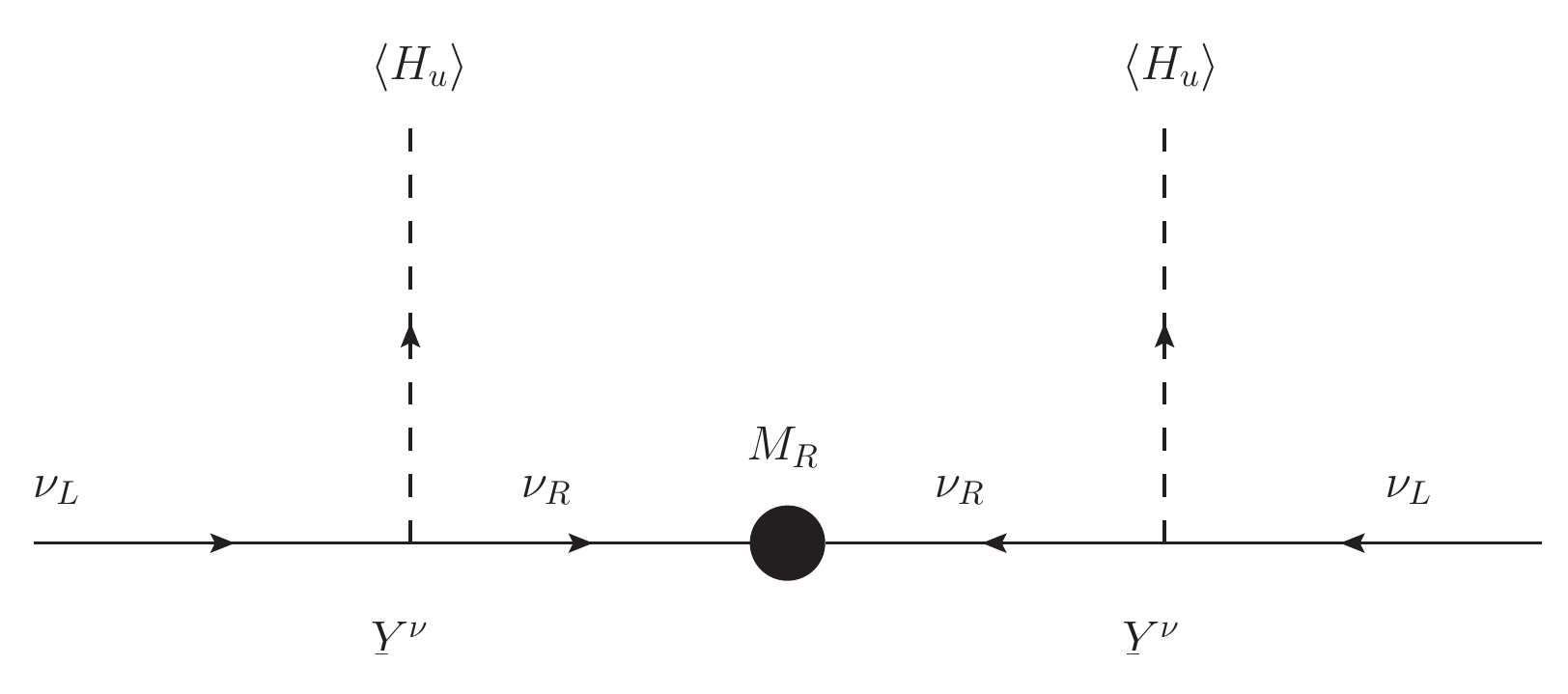}
    \caption{The seesaw mass insertion diagram responsible for the light effective LH Majorana neutrino mass 
    $m^{\nu}=-m_DM_{R}^{-1}(m_D)^T$ where the Dirac neutrino mass is $m_D=Y^{\nu}\langle H_u \rangle 
    =Y^{\nu}v_u$.    } \label{seesawfig}
\end{figure}

Collecting together 
Eqs.\ref{MRR},\ref{mLR} (assuming Eq.\ref{mLL} terms to be absent) we have the seesaw mass matrix,
\begin{equation}
\left(\begin{array}{cc} \overline{\nu_L} & \overline{\nu^c_R}
\end{array} \\ \right)
\left(\begin{array}{cc}
0 & m_D\\
(m_D)^T & M_{R} \\
\end{array}\right)
\left(\begin{array}{c} \nu_L^c \\ \nu_R \end{array} \\ \right).
\label{matrix}
\end{equation}
Since the RH neutrinos are electroweak singlets
the Majorana masses of the RH neutrinos $M_{R}$
may be orders of magnitude larger than the electroweak
scale. In the approximation that $M_{R}\gg m_D$
the matrix in Eq.\ref{matrix} may be diagonalised to
yield effective Majorana masses of the type in Eq.\ref{mLL},
\begin{equation}
m^{\nu}=-m_DM_{R}^{-1}(m_D)^T.
\label{seesaw}
\end{equation}
The seesaw mechanism formula is represented by the mass insertion diagram in Fig.\ref{seesawfig}.
This formula is valid below the EW scale. Above the EW scale, but below the scale $M_R$,
the seesaw mechanism is represented by the Weinberg operator in Eq.\ref{dim5s},
whose coefficient has the same structure as the seesaw formula in Eq.\ref{seesaw}.

\begin{figure}[t]
\centering
\includegraphics[width=0.45\textwidth]{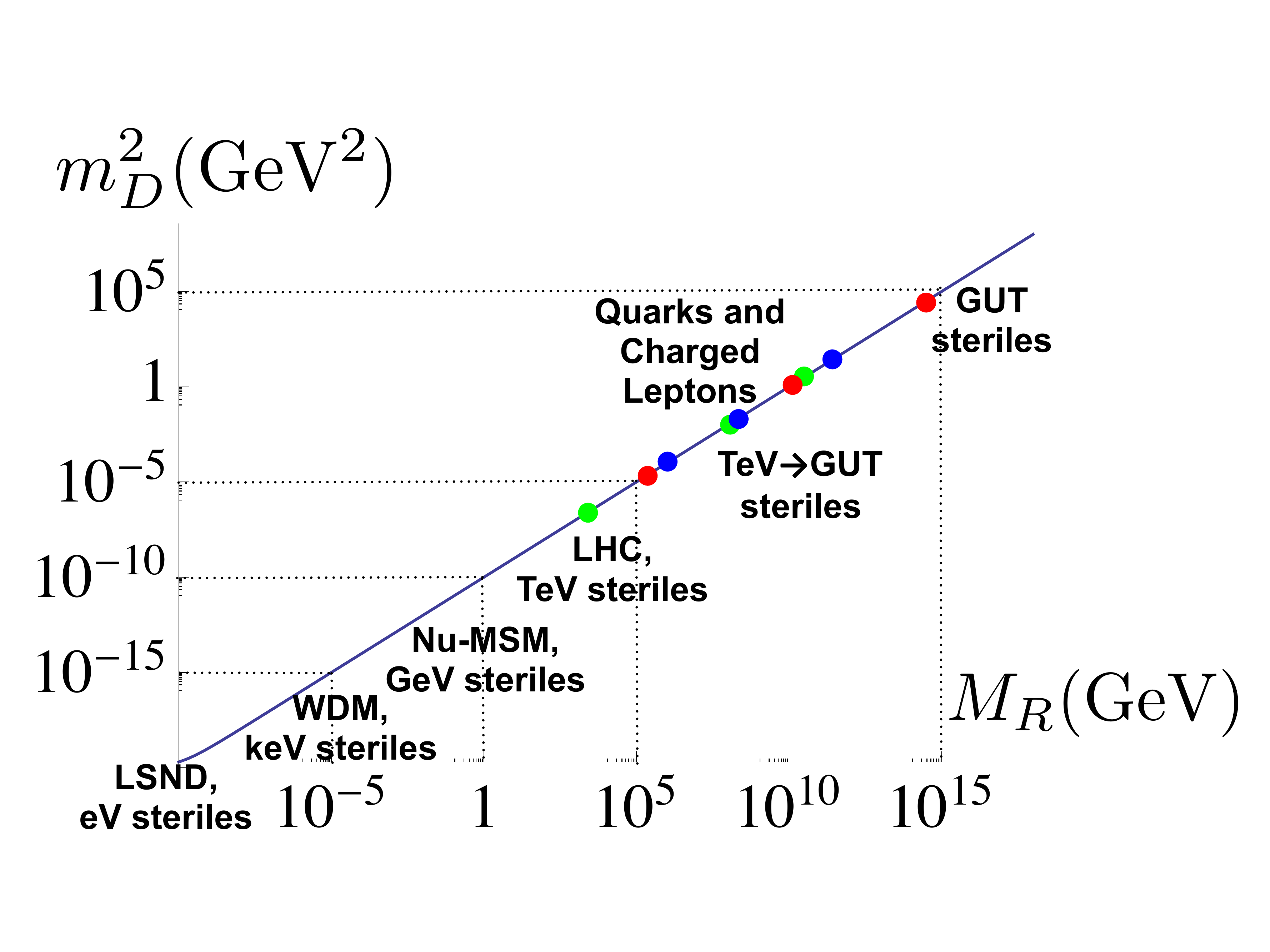}
\includegraphics[width=0.45\textwidth]{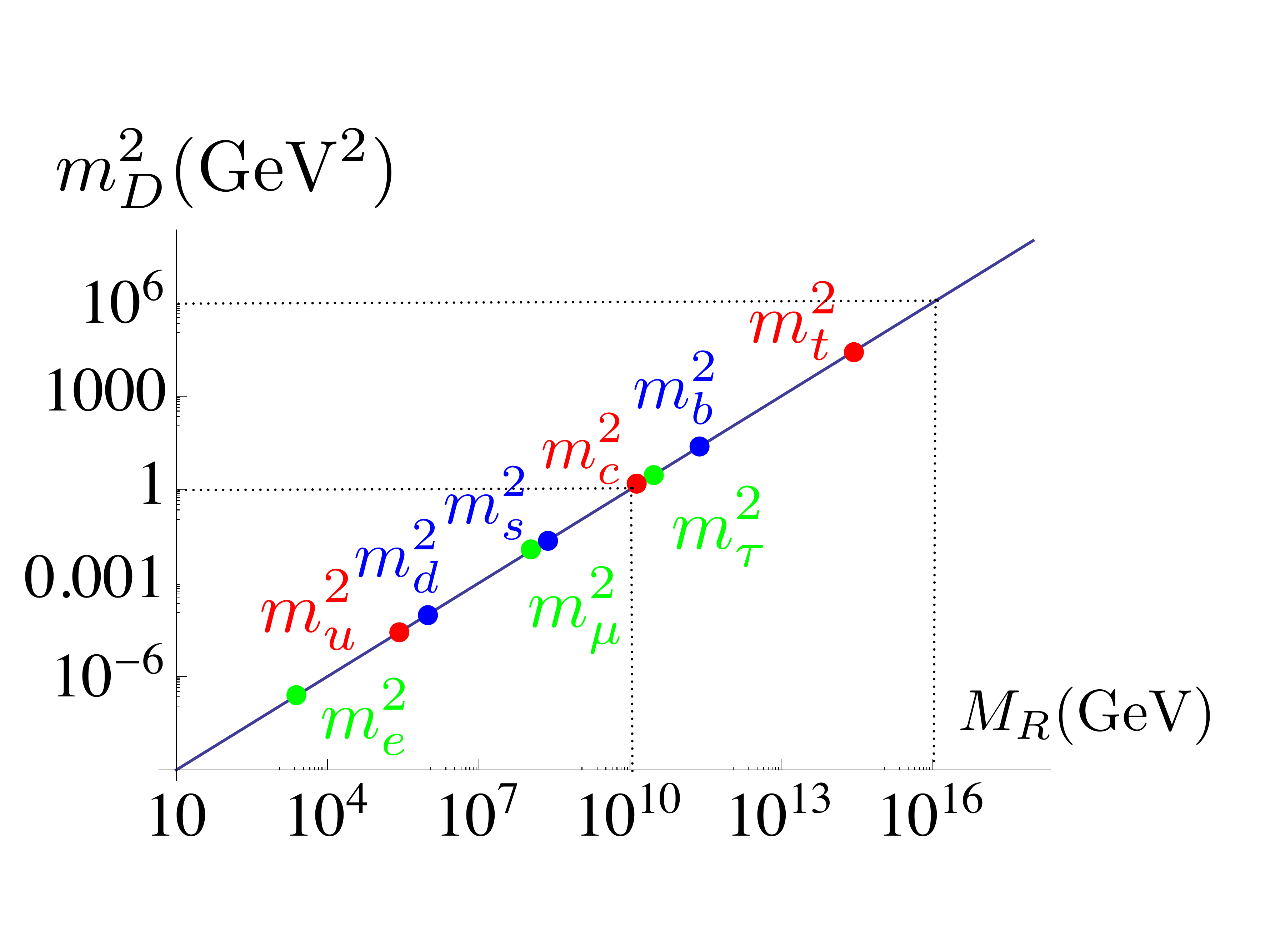}
    \caption{ The required values of $m_D^2$ in order to obtain a physical
    light neutrino mass of $m^{\nu}=10^{-10}$ GeV as a function of the RH neutrino mass $M_R$. 
    The left panel shows a large range of masses $M_R$, ranging from 1 eV up to the GUT scale.
    The right panel zooms in on the values of $m_D^2$ identified with 
    the quark and charged lepton mass squared values, leading to TeV-GUT scale sterile neutrino masses
    which is what we focus on in this review.
    In constructing the plots we used the more accurate 
eigenvalue formula of the seesaw matrix (ignoring phases): $m_D^2=m_{\nu}M_R+m_{\nu}^2$,
with $m^{\nu}=10^{-10}$ GeV,
but the usual seesaw formula $m_D^2\approx m_{\nu}M_R$ 
(fixing $m^{\nu}=0.1$ eV) is a good approximation over the ranges shown.
     } \label{seesawplot}
\end{figure}

The light effective LH neutrino Majorana mass $m^{\nu}$ is naturally
suppressed by the heavy scale $M_{R}$, but its precise value depends on the Dirac neutrino mass $m_D$.
Suppose we fix the desired physical neutrino mass to be $m^{\nu}=0.1$ eV, then the seesaw formula in 
Eq.\ref{seesaw}
relates the possible values of $m_D$ to $M_R$ as shown in Fig.\ref{seesawplot}.
This illustrates the huge range of allowed values of $m_D$ and $M_R$ consistent with an
observed neutrino mass of 0.1 eV, with $M_R$ ranging from 1 eV up to the GUT scale,
leading to many different types of seesaw models and phenomenology,
including eV mass LSND sterile neutrinos, keV mass sterile neutrinos suitable for warm dark matter (WDM),
GeV mass sterile neutrinos suitable for resonant leptogenesis and TeV mass sterile neutrinos possibly 
observable at the LHC (for a review see e.g. \cite{Drewes:2015jna} and references therein).
In this review we shall focus on the case of Dirac neutrino masses identified with charged quark and lepton masses,
leading to a wide range of RH neutrino (or sterile neutrino) masses from the TeV scale
to the GUT scale, which we refer to as the classic seesaw model. 
For example, if we take $m_D$ to be 1 GeV (roughly equal to the charm
quark mass) then a neutrino mass of 0.1 eV requires a RH (sterile) neutrino mass of 
$10^{10}$ GeV.

\subsection{The type I seesaw mechanism with two RH neutrinos }
The type I see-saw neutrino model involving just 
{\em two} RH neutrinos was introduced in \cite{King:1999mb}.
This is the minimal case sufficient to account for all neutrino oscillation data, and makes the prediction that the lightest neutrino mass is zero since the resulting light neutrino mass matrix $m^{\nu}$ is rank two \cite{King:1999mb}.
In this case the neutrino masses are hierarchical (since the lightest mass is zero) 
and we can alternatively refer to normal ordered (NO) mass squareds
as a normal hierarchy (NH) and inverted ordered (IO) mass squareds as an inverted hierarchy (IH).

Assuming the charged lepton mass matrix is diagonal,
the two RH neutrinos
$\nu^{\rm sol}_R$ and $\nu^{\rm atm}_R$ have Yukawa couplings \cite{King:1999mb},
\begin{equation}
\mathcal{L}^{Yuk} =(H_u/v_u)(a\overline L_e+ b\overline L_{\mu} +c\overline L_{\tau}){\nu^{\rm sol}_R}
+ (H_u/v_u)(d\overline L_e+ e\overline L_{\mu} +f\overline L_{\tau}){\nu^{\rm atm}_R}+H.c.,
\end{equation}
where $L_{e,\mu , \tau}$ are the lepton doublets containing the $e_L,\mu_L , \tau_L$ mass eigenstates,
and $v_u$ the VEV of the $H_u$ Higgs doublet.
The Majorana Lagrangian is,
\begin{equation}
\mathcal{L}^{R}_\nu = M_{\rm sol}\overline{\nu^{\rm sol }_R}({\nu^{\rm sol}_R})^c
+M_{\rm atm}\overline{\nu^{\rm atm }_R}({\nu^{\rm atm}_R})^c +H.c..
\end{equation}
In the convention that the rows are 
$(\overline \nu_{eL}, \overline \nu_{\mu L}, \overline \nu_{\tau L})$ and 
the columns are ${\nu^{\rm atm}_R}, {\nu^{\rm sol}_R}$, we find the Dirac mass matrix,
\begin{equation}
m_D=
\left( \begin{array}{cc}
d & a \\
e & b \\
f & c
\end{array}
\right),\ \ \ \ 
(m_D)^T=
\left( \begin{array}{ccc}
d & e & f\\
a & b& c
\end{array}
\right)
\label{mD}
\end{equation}

The RH neutrino Majorana mass matrix $M_{R}$
with rows $(\overline{\nu^{\rm atm}_R}, \overline{\nu^{\rm sol}_R})^T$ and columns $(\nu^{\rm atm}_R, \nu^{\rm sol}_R)$
is,
\begin{equation}
M_{R}=
\left( \begin{array}{cc}
M_{\rm atm} & 0 \\
0 & M_{\rm sol}
\end{array}
\right),\ \ \ \ 
M^{-1}_{R}=
\left( \begin{array}{cc}
M^{-1}_{\rm atm} & 0 \\
0 & M^{-1}_{\rm sol}
\end{array}
\right)
\label{mR}
\end{equation}
The see-saw formula in Eq.\ref{seesaw} \cite{seesaw} gives a light neutrino mass matrix,
\begin{equation}
m^{\nu}=-m^DM^{-1}_{R}(m^D)^T.
\label{seesaw2}
\end{equation}
This is the effective Majorana mass matrix for LH neutrinos, and is the relevant mass matrix
for the light neutrino states which appear dominantly in neutrino oscillations.
The overall minus sign is not physical and can be safely dropped.
In left-right (LR) convention,
$m_D$ is the Dirac mass matrix.
$M_R$ is the Majorana
mass matrix, which typically involves masses higher than the EW scale.
The physical neutrino mass matrix is obtained using the matrices in Eqs.\ref{mD},\ref{mR},
\begin{equation}
m^{\nu}=
\frac{1}{M_{\rm atm}}
\left( \begin{array}{ccc}
d^2& de &df  \\
de & e^2 &ef \\
df & ef & f^2
\end{array}
\right)
+
\frac{1}{M_{\rm sol}}
\left( \begin{array}{ccc}
a^2& ab &ac  \\
ab & b^2 &bc \\
ac & bc & c^2
\end{array}
\right).
\label{2rhn}
\end{equation}
The main prediction of the two RH neutrino (2RHN) model is that the lightest neutrino mass is zero,
but it does not distinguish between NO and IO, or make any further predictions.
In order to make the 2RHN model more predictive one must make further assumptions.
For example, the 2RHN model
with two texture zeros \cite{Frampton:2002qc} turns out to be only consistent with current data for
the case of IO \cite{Harigaya:2012bw}.
However the 2RHN model with one texture zero \cite{King:1999mb} is viable for the NO case.

\begin{figure}[t]
\centering
\includegraphics[width=0.4\textwidth]{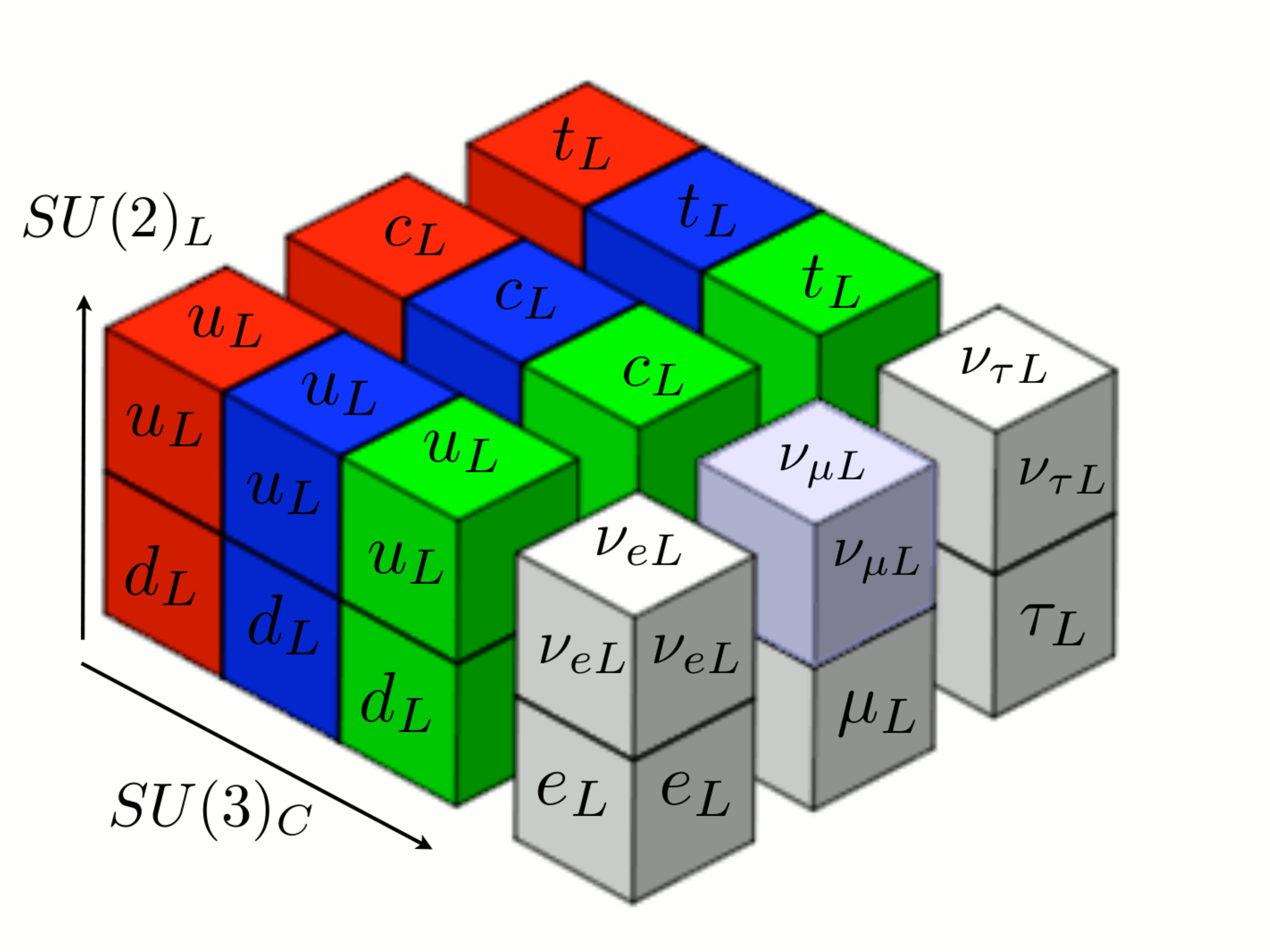}
\includegraphics[width=0.4\textwidth]{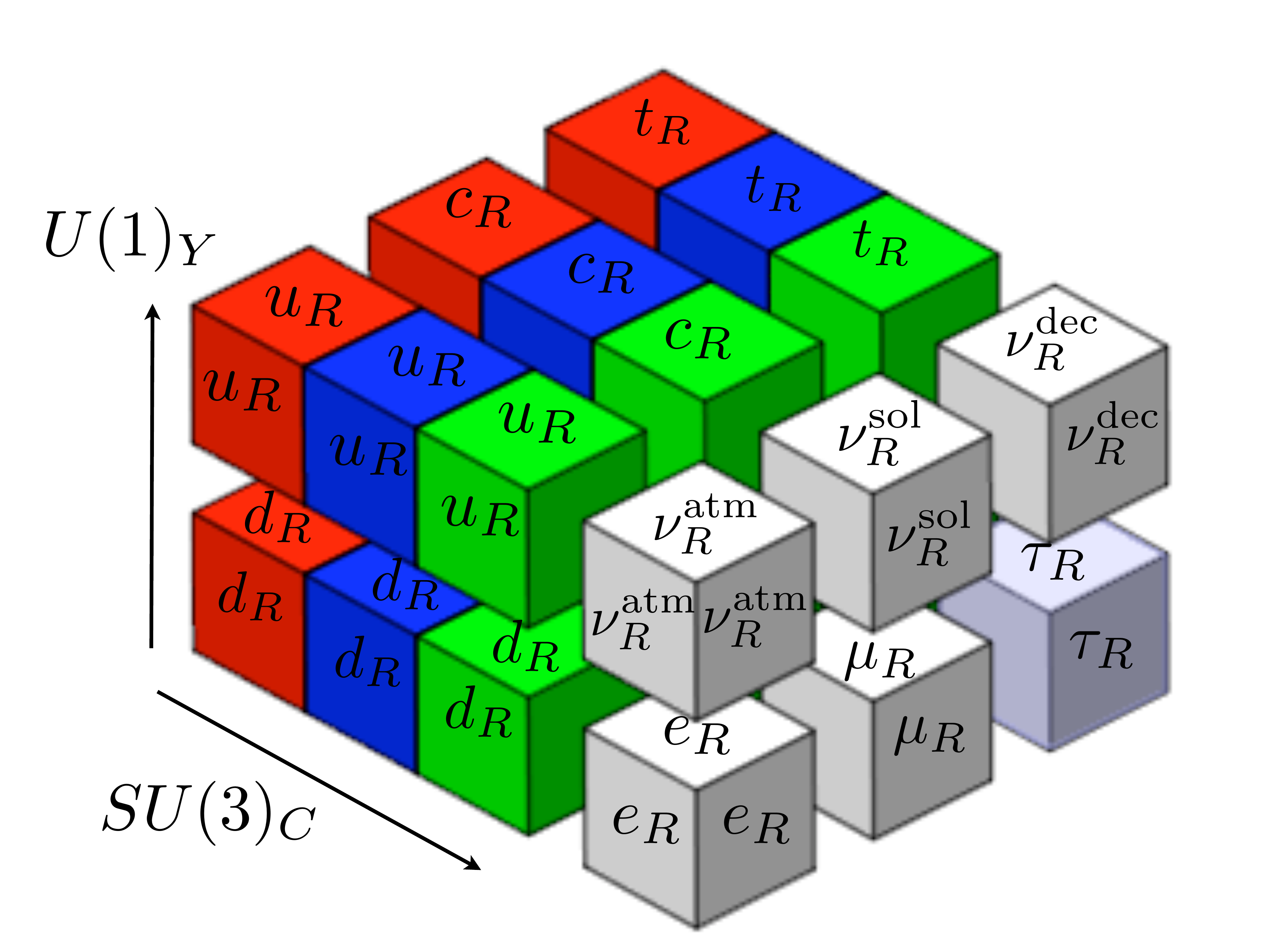}
\vspace*{-4mm}
    \caption{The LH and RH quarks and leptons are represented 
    by stacked cubes which transform under the SM gauge group as indicated.
    Three RH neutrinos have been added to the SM, namely 
    $(\nu_R^{\rm atm},\nu_R^{\rm sol},\nu_R^{\rm dec})$ which in sequential dominance are mainly responsible for the $m_3,m_2,m_1$ physical neutrino masses, respectively.} \label{SM}
\vspace*{-2mm}
\end{figure}

\subsection{The type I seesaw mechanism with three RH neutrinos and sequential dominance}
\begin{figure}[t]
\centering
\includegraphics[width=0.4\textwidth]{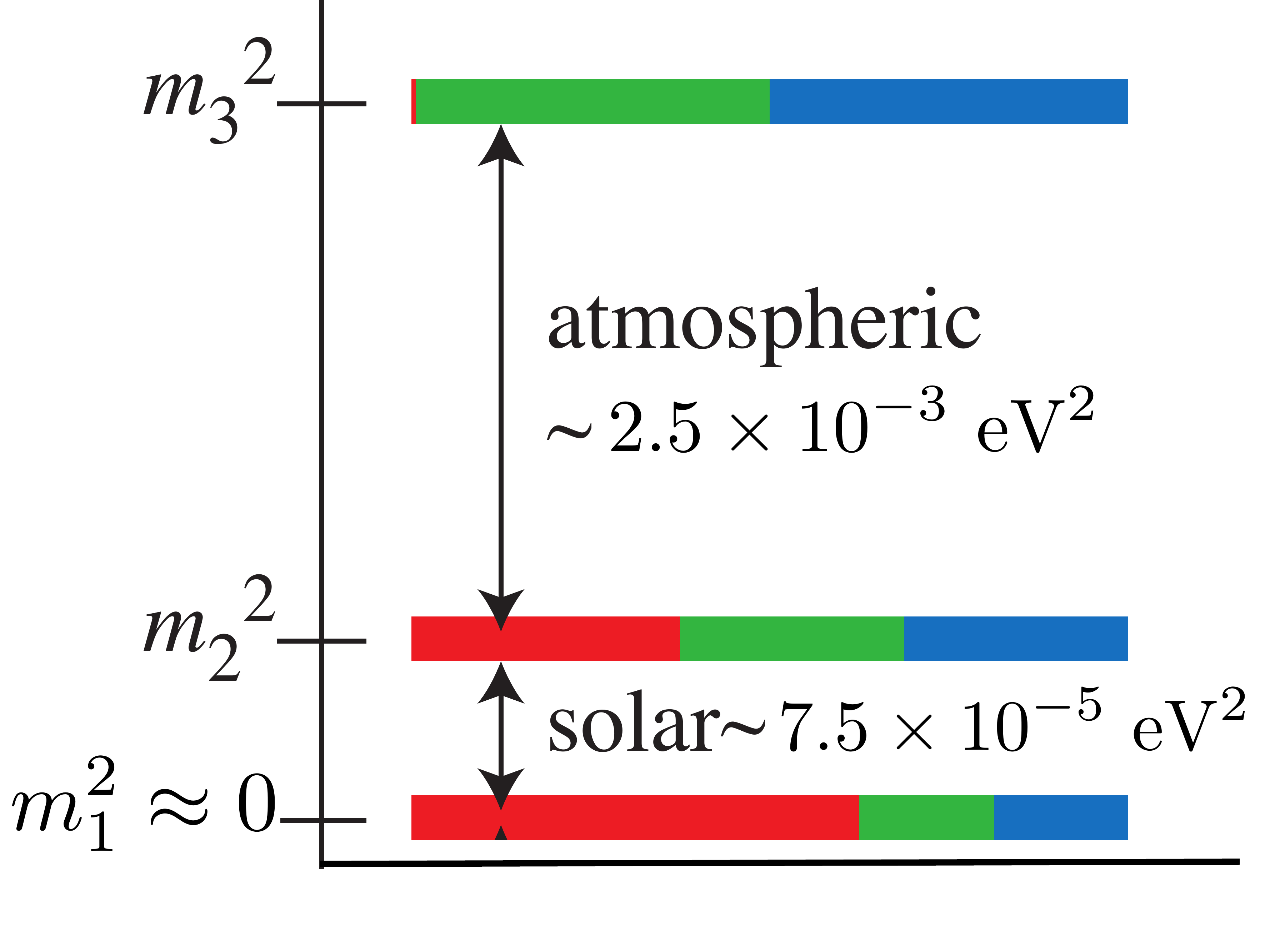}
\caption{\label{mass2}\small{A normal hierarchy as predicted by SD.
In this case the neutrino masses are approxiately just the square roots of the mass squared differences,
$m_2=\sqrt{\Dm21}$ and $m_3=\sqrt{\Dm31}$.
According to the global fit values in Table~\ref{tab:bfp}, with $m_1=0$, we find $m_2=8.6\pm 0.2$ meV
and $m_3=50\pm 1$ meV.}}
\end{figure}

More generally there may be three RH neutrinos,
$\nu^{\rm atm}_R$, $\nu^{\rm sol}_R$ and $\nu^{\rm dec}_R$,
with large Majorana masses 
$M_{\rm atm},M_{\rm sol}$ and $M_{\rm dec}$, respectively.
They are sometimes called sterile neutrinos since they transform as 
singlets under the SM gauge group (see Fig.\ref{SM}). 
If the third RH neutrino 
$\nu^{\rm dec}_R$ makes a negligible contribution to the seesaw mechanism,
either due to its high mass or its small Yukawa couplings $a',b',c'$, or both, 
then it will approximately decouple from the seesaw mechanism, 
and we return back to the 
two RH neutrino mass matrix in Eq.\ref{2rhn}.
This decoupling may be part of a sequential dominance (SD) of the three RH neutrinos
to the seesaw mechanism \cite{King:1998jw,King:1999mb},
\begin{equation}
 \frac{(d,e,f)^2}{M_{\rm atm}} \gg \frac{(a,b,c)^2}{M_{\rm sol}}  \gg \frac{(a',b',c')^2}{M_{\rm dec}}.
\label{SD0}
\end{equation}
At this stage, the SD condition in Eq.\ref{SD0} is just an assumption.
However it may emerge in a robust way in the context of particular flavour models where the 
Yukawa couplings and RH neutrino masses are predicted from symmetry and dynamics,
as is the case in the models discussed later.
Eq.\ref{SD0} implies a strong and normal mass hierarchy:
\be
m_3 \gg m_2 \gg m_1
\label{NH}
\ee
where $m_3$ arises mainly from Fig.\ref{seesawfig} with $\nu^{\rm atm}_R$
exchange, while $m_2$ is dominated by $\nu^{\rm sol}_R$ exchange
and the lightest neutrino mass $m_1$ arises from $\nu^{\rm dec}_R$.
The smallest physical neutrino mass $m_1$ vanishes in the limit that 
the primed couplings vanish, since then the model reduces to the
two RH neutrino model in Eq.\ref{2rhn} for which $\det m^{\nu} = 0$.
Furthermore in the single RH neutrino approximation \cite{King:1998jw},
we have $m_3\gg m_2$.
Hence SD in Eq.\ref{SD0} {\it implies a normal neutrino mass hierarchy} as in Eq.\ref{NH}
and Fig.~\ref{mass2}.
We emphasise that the prediction of a NH is two predictions: both NO and a {\em hierarchy},
specifically a very small value of lightest neutrino mass $m_1$.

If in addition it is assumed that $d=0$ in the diagonal charged lepton mass basis,
then Eq.\ref{SD0} implies
\cite{King:1998jw,King:1999mb},
\begin{equation}
\tan \theta_{23}\sim \frac{e}{f}, \ \ \ \ \tan \theta_{12} \sim \frac{\sqrt{2}a}{b-c},
\label{t12}
\end{equation}
\begin{equation}
\theta_{13} \lesssim m_2/m_3.
\label{13}
\end{equation}
Eq.\ref{13}, which shows that the reactor angle may be quite sizeable,
was written down a decade before the angle was measured,
and may be counted as a success of the SD approach.
However to understand the reason why this bound is approximately saturated,
we need to consider constrained forms of SD.

\subsection{Constrained sequential dominance}
\label{CSDn}
In the previous subsection, we saw that a simple constraint on SD, namely $d=0$ in the diagonal charged lepton mass basis, led to some remarkably simple results for lepton mixing. 
One may go further and impose other constraints on the couplings (later enforced by symmetry) 
in order to enhance predictivity still further.
The first example of 
such constrained sequential dominance (CSD) \cite{King:2005bj} was to impose the constraints
$d=0$, $e=f$ and $a=b = -c$ leading to precise tri-bimaximal mixing.
This can readily be seen by inserting these constrained Yukawa couplings into Eq.\ref{2rhn}, then showing that the resulting mass matrix is 
exactly diagonalised by the TB mixing matrix in Eq.\ref{TB}, as follows,
\begin{equation}
U_{\mathrm{TB}}^T m^{\nu} U_{\mathrm{TB}}=
\left( \begin{array}{ccc}
0 & 0 & 0  \\
0 & \frac{3a^2}{M_{\rm sol}} & 0 \\
0 & 0 & \frac{2e^2}{M_{\rm atm}}  
\end{array}
\right).
\end{equation}

Although the above choice of Yukawa couplings leads to the undesirable result $\theta_{13}=0$,
other choices lead to values of reactor angle in agreement with experiment,
while preserving the good values of atmospheric and solar mixing.
The above CSD can be generalised to CSD($n$)
\cite{Antusch:2011ic,King:2013iva,King:2013xba,King:2013hoa,King:2014iia,
Bjorkeroth:2014vha,Bjorkeroth:2015tsa,King:2015dvf,King:2016yvg,King:2016yef} in which 
$b=na$ and  $c=(n-2)a$ (case A), or alternatively $b=(n-2)a$ and $c=na$ (case B),
for any postive integer $n$. The other couplings are as before, $d=0$ and $e=f$.
Applying these constraints to Eq.\ref{mD} gives, 
\begin{equation}
m^A_D=
\left( \begin{array}{cc}
0 & a \\
e & na \\
e & (n-2)a
\end{array}
\right),\ \ \ \ {\rm or} \ \ \ \ 
m^B_D=
\left( \begin{array}{cc}
0 & a \\
e & (n-2)a \\
e & na
\end{array}
\right).
\label{mDCSD}
\end{equation}
The Dirac mass matrices above are is in the diagonal RH neutrino and charged lepton mass basis. According to Eq.\ref{t12}, we expect approximate TB mixing with these constraints applied.
Exact analytic results for lepton mixing angles and phases based on CSD($n$) have been derived in
\cite{King:2015dvf,King:2016yvg}.
The reactor angle has an approximate $n$ dependence of,
\be
\theta_{13}\sim (n-1)\frac{\sqrt{2}}{3}\frac{m_2}{m_3}.
\ee
The predictions are sensitive to 
the relative phase between the complex masses $e$ and $a$. 
\footnote{This is the only physical phase in the lepton sector of 2RHN CSD($n$) models and as such may be identified
with the leptogenesis phase, which requires a lightest RHN mass of $M_1\sim 4.10^{10}$ GeV \cite{Bjorkeroth:2015tsa}. }
The choice $n=1$ returns us to the original CSD discussed at the start of this subsection, with 
a zero reactor angle. Choosing $n=2$ gives 
CSD(2) which fails to give a large enough reactor angle
for all choices of phase \cite{Antusch:2011ic}.
The simplest viable case is CSD(3) \cite{King:2013iva}, while CSD(4) is also possible
\cite{King:2013xba,King:2013hoa}, and CSD($n\geq 5$) \cite{Bjorkeroth:2014vha}
is disfavoured due to the reactor angle being too large.
In the next subsection we focus on the simplest viable case of CSD(3), 
with a fixed relative phase, which has been called the Littlest 
Seesaw (LS) \cite{King:2015dvf,King:2016yvg}, since the resulting neutrino mass matrix involves the smallest number of 
free parameters.
The theoretical justification for the constrained choice of Yukawa couplings relies of vacuum
alignment and in particular $S_4$ symmetry as shown in \cite{King:2015dvf,King:2016yvg} and discussed later.

\subsection{The Littlest Seesaw }

The Littlest Seesaw (LS) model is the minimal viable seesaw model corresponding to a 2RHN model with CSD($3$). In the basis where the charged leptons have a diagonal mass matrix, and the RH neutrino mass
matrix is also diagonal,
CSD($3$) corresponds to the Dirac mass matrix in Eq.\ref{mDCSD} with $n=3$ \cite{King:2013iva}:
\begin{equation}
m^A_D=
\left( \begin{array}{cc}
0 & a \\
e & 3a \\
e & a
\end{array}
\right),\ \ \ \ {\rm or} \ \ \ \ 
m^B_D=
\left( \begin{array}{cc}
0 & a \\
e & a \\
e & 3a
\end{array}
\right).
\label{mDCSD3}
\end{equation}
After the seesaw mechanism has been implemented, the low energy effective LH Majorana neutrino mass matrix in the two RH neutrino case may be written as (see in Eq.\ref{2rhn}),

\begin{align}
m^\nu_\text{LSA}&=m_a
\begin{pmatrix}
0 & 0 & 0 \\
0 & 1 & 1 \\
0 & 1 & 1
\end{pmatrix}
+m_be^{i\eta}
\begin{pmatrix}
1 & 3 & 1\\
3 & 9 & 3\\
1 & 3 & 1
\end{pmatrix},\label{eq:matrix_LSA}\\
m^\nu_\text{LSB}&=m_a
\begin{pmatrix}
0 & 0 & 0 \\
0 & 1 & 1 \\
0 & 1 & 1
\end{pmatrix}
+m_be^{i\eta}
\begin{pmatrix}
1 & 1 & 3\\
1 & 1 & 3\\
3 & 3 & 9
\end{pmatrix}.\label{eq:matrix_LSB}
\end{align}
where we have written $m_a=\frac{|e|^2}{M_{\rm atm}}$ and $m_b=\frac{|a|^2}{M_{\rm sol}}$.
The phase $\eta$ is physical: it is given by $\arg (a/e)$.
The LS model is {\em the minimal currently viable seesaw model} since it involves 
only three real parameters $m_a, m_b, \eta$. These parameters fix the three neutrino masses
(one of which is zero) and fully determine the PMNS matrix (three angles and three phases, one of which is unphysical due to the zero neutrino mass).

\begin{figure}[t]
\centering
\includegraphics[width=0.80\textwidth]{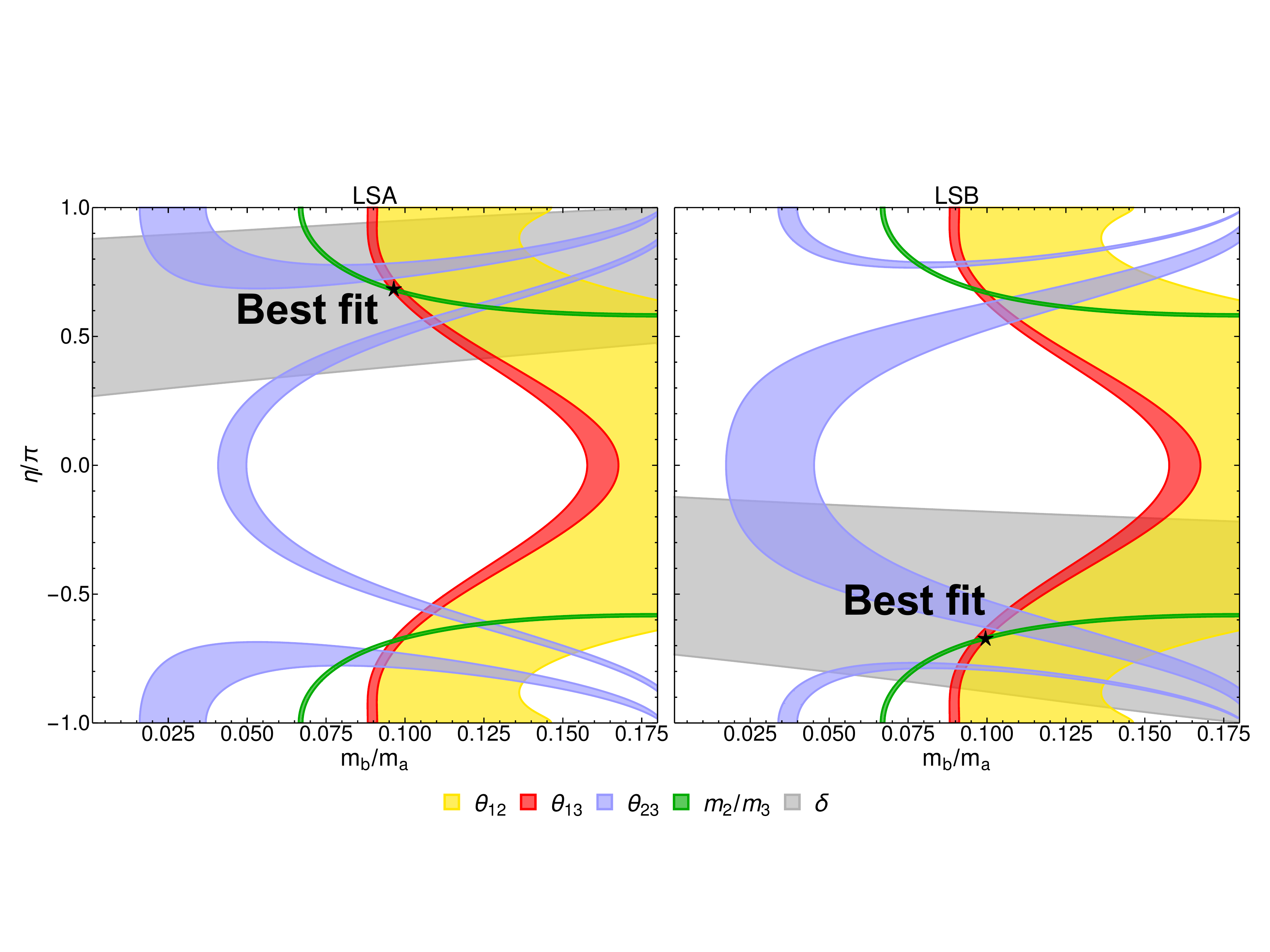}
\vspace*{-12mm}
    \caption{The parameter space in the plane of $m_b/m_a$ vs $\eta$ of the LSA (left) and LSB (right).
    The coloured bands show the present one sigma ranges of the parameters shown, using NuFIT 3.0.
    Note that, since $m_1=0$, the value of $m_2/m_3$ is the square root of $\Delta m_{21}^2/\Delta m_{31}^2$.
    The best fit point shown
    with a star is close to $m_b/m_a = 0.1$ and $\eta = \pm 2\pi/3$.
    There is a mild tension at the one sigma level with the atmospheric and solar angles.
    This figure is adapted from \cite{Ballett:2016yod}.} \label{review1}
\end{figure}

Both LSA and LSB allow good fits of current data \cite{Ballett:2016yod} as seen in 
Figure~\ref{review1}. The intersection of the accurately measured reactor angle one sigma band (in red)
with the accurarely measured mass ratio one sigma band (in green) is instrumental in determining the
model parameters $\eta$ and $m_b/m_a$.
There is a mild tension at the one sigma level with the less accurarely measured atmospheric and solar angles,
so a better future determation of these angles could exclude the model.
The best fit points in Figure~\ref{review1} indicated by stars are 
close to $m_b/m_a = 0.1$ and $\eta = \pm 2\pi/3$.

In Table~\ref{tab:bfp2}, we show the best fit values of $m_a$ and $m_b$ 
with $\eta$ either free or held fixed at $\eta = \pm 2\pi/3$,
together with the corresponding values of mixing parameters.
The results are taken from \cite{Ballett:2016yod} where more details may be found.

\begin{table}[t]
\centering\begin{tabular}{lrrrrr}
		\hline
		 & \multicolumn{2}{c}{LSA} & \multicolumn{2}{c}{LSB} & \multicolumn{1}{c}{NuFIT 3.0} \\
		 & \multicolumn{1}{c}{$\eta$ free} & \multicolumn{1}{c}{$\eta$ fixed} & \multicolumn{1}{c}{$\eta$ free} & \multicolumn{1}{c}{$\eta$ fixed} & \multicolumn{1}{c}{global fit} \\
		\hline
		$m_a$ [meV] & 27.22 & 26.78 & 27.14 & 26.77 & \\
		$m_b$ [meV] & 2.653 & 2.678 & 2.658 & 2.681 & \multicolumn{1}{c}{---} \\
		$\eta$ [rad] & $0.680\pi$ & $2\pi/3$ & $-0.678\pi$ & $-2\pi/3$ &  \\
		\hline
		$\th12$ [$^\circ$] & 34.37 & 34.34 & 34.36 & 34.33 & $33.72^{+0.79\phantom{0}}_{-0.76}$ \\
		$\th13$ [$^\circ$] & 8.45 & 8.58 & 8.48 & 8.59 & $8.46^{+0.14\phantom{0}}_{-0.15}$ \\
		$\th23$ [$^\circ$] & 45.01 & 45.69 & 44.87 & 44.30 & $41.5^{+1.3\phantom{00}}_{-1.1}$ \\
		$\delta$ [$^\circ$] & -89.9 & -87.0 & -90.6 & -93.1 & $-71^{+38\phantom{.00}}_{-51}$ \\
		$\Dm21$ [$10^{-5}\text{eV}^2$] & 7.499 & 7.362 & 7.482 & 7.379 & $7.49^{+0.19\phantom{0}}_{-0.17}$ \\
		$\Dm31$ [$10^{-3}\text{eV}^2$] & 2.505 & 2.515 & 2.505 & 2.515 & $2.526^{+0.039}_{-0.037}$ \\
		\hline
		$\Delta\chi^2$\,/\,d.o.f & 4.7\,/\,3 & 6.4\,/\,4 & 4.5\,/\,3 & 5.1\,/\,4 & \multicolumn{1}{c}{---} \\
\hline
\end{tabular}
\caption{Results of a fit of existing data to LSA and LSB with $\eta$ left free and for $\eta=\frac{2\pi}{3}$ for LSA and $\eta=-\frac{2\pi}{3}$ for LSB \cite{Ballett:2016yod}. The results of the NuFIT 3.0 (2016) global fit to standard neutrino mixing are shown for the normal ordering case for comparison. Notice that there are two ($m_a$, $m_b$) or three 
(if $\eta$ is left free) input parameters describing six observables,
so that the number of degrees of freedom (d.o.f.) is either three (six minus three) or four (six minus two).}
\label{tab:bfp2}
\end{table}

Both LSA and LSB both predict a NO neutrino mass pattern with a zero neutrino mass, $m_1=0$,
and $\text{TM}_1$ lepton mixing as in Eq.\ref{TMM}. To understand this, first observe that,
\begin{equation}
m^\nu_\text{LSA}
\left(
\begin{array}{c}
2 \\
-1\\
1
\end{array}
\right)
=
\left(
\begin{array}{c}
0 \\
0\\
0
\end{array}
\right), \ \ \ \ {\rm or} \ \ \ \ 
m^\nu_\text{LSB}
\left(
\begin{array}{c}
2 \\
1\\
-1
\end{array}
\right)
=
\left(
\begin{array}{c}
0 \\
0\\
0
\end{array}
\right).
\label{CSD(n)a}
\end{equation}
These equations show that the column vectors on the left are eigenvectors with zero eigenvalues,
and in each case they may be identified as the first column 
of the PMNS mixing matrix associated with the zero neutrino mass $m_1=0$,
yielding the $\text{TM}_1$ mixing form as in Eq.\ref{TMM}.
\footnote{In fact a NH neutrino mass pattern with a zero neutrino mass, $m_1=0$,
and $\text{TM}_1$ lepton mixing is a prediction of all CSD($n$) 2RHN seesaw models,
by a similar argument.}
It can readily be verified that the best fit predictions in Table~\ref{tab:bfp} respect the $\text{TM}_1$ mixing sum rules
in Eqs.\ref{t12p}, \ref{TM1sum}.

The baryon asymmetry of the Universe
(BAU) resulting from leptogenesis has the right sign, consistent with an excess of matter over antimatter,
only if the lightest RH neutrino is
$N_{\rm atm}$, so that $M_{\rm atm} <  M_{\rm sol}$. It was estimated for LSA \cite{Bjorkeroth:2015tsa}:
\begin{equation}
	Y^{\rm LSA}_B \approx 2.5 \times 10^{-11}\sin \eta \left[\frac{M_{\rm atm}}{10^{10} ~\mathrm{GeV}} \right].
\label{BAUA}
\end{equation}
Using $\eta = 2\pi/3$ (preferred by the fit with $\delta \approx -90^{\circ} $),
the correct baryon asymmetry requires,
\begin{equation}
	M_{\rm atm}  \approx 3.9 \times 10^{10} ~\mathrm{GeV}.
	\label{Matm}
\end{equation}
Note that $\eta$, which is the phase in the neutrino mass matrix in Eq.\ref{eq:matrix_LSA},
is also the leptogenesis phase in Eq.\ref{BAUA}.
There is only one phase in LSA which controls everything: \CP violation
in the laboratory and in the Universe. This is a very attractive feature of the LS model.

For LSB, using $\eta = -2\pi/3$ (preferred by the fit with $\delta \approx -90^{\circ} $)
in order to achieve
a positive matter-antimatter asymmetry we would require the lightest RH neutrino to be $N_{\rm sol}$,
so that $M_{\rm sol} <  M_{\rm atm}$
\cite{King:2016yvg}.

Renormalisation group (RG) corrections have been studied for LSA and LSB with both $M_{\rm atm} <  M_{\rm sol}$
and $M_{\rm sol} <  M_{\rm atm}$  \cite{King:2016yef}.
\footnote{For a recent review of RG corrections in general neutrino mass models, with original references
see e.g. \cite{Ohlsson:2013xva}.}
 It has been shown that,
if the predictions in Table~\ref{tab:bfp2} are valid at high energies, such as the GUT scale, then the low energy angles 
are rather stable under radiative corrections. For example, the atmospheric angle receives corrections of 
$\Delta \theta_{23} \lesssim 1^{\circ}$ \cite{King:2016yef}, with the effect of radiative corrections
tending to increase the low energy atmospheric angle compared to its GUT scale prediction.

\subsection{Precision neutrino experiments vs the Littlest Seesaw}
In the previous subsection we saw that both versions of the Littlest Seesaw, LSA and LSB,
are consistent with current neutrino oscillation data. 
We noted that the well measured neutrino mass squared differences,
when combined with the accurately determined reactor angle $\theta_{13}$,
were sufficient to precisely fix the parameters of the model $\eta$, $m_a$ and $m_b$.
However we also saw that there 
is a mild tension at the one sigma level with the less accurarely measured atmospheric and solar angles,
so a better future determation of these angles by future experiments could exclude the model.

This raises the general question of how precise experimental measurements need to be before qualitative progress can be made for flavour models. 
This type of input is very important to the experimental community and it can be fully addressed within a particular model
such as the Littlest Seesaw.
Indeed the prospects for excluding LSA and LSB in future neutrino oscillation experiments have
recently been analysed \cite{Ballett:2016yod}, and we shall briefly review the results
of that study. 
We should say at the outset that one way to exclude the LS models is via its prediction
of a NO mass spectrum with $m_1=0$. 
For example a determination of an IO would exclude these models, as would any signal from
neutrinoless double beta decay experiments, or cosmology, which are not currently capable of 
seeing a signal for $m_1=0$.
In the following we assume that data continues to be consistent
with such a neutrino mass spectrum and consider the prospects for excluding the models by precision
measurements of the two mass squared differences, the three angles and the \CP violating oscillation phase $\delta$.

\begin{figure}[t]
\centering
\includegraphics[width=0.60\textwidth]{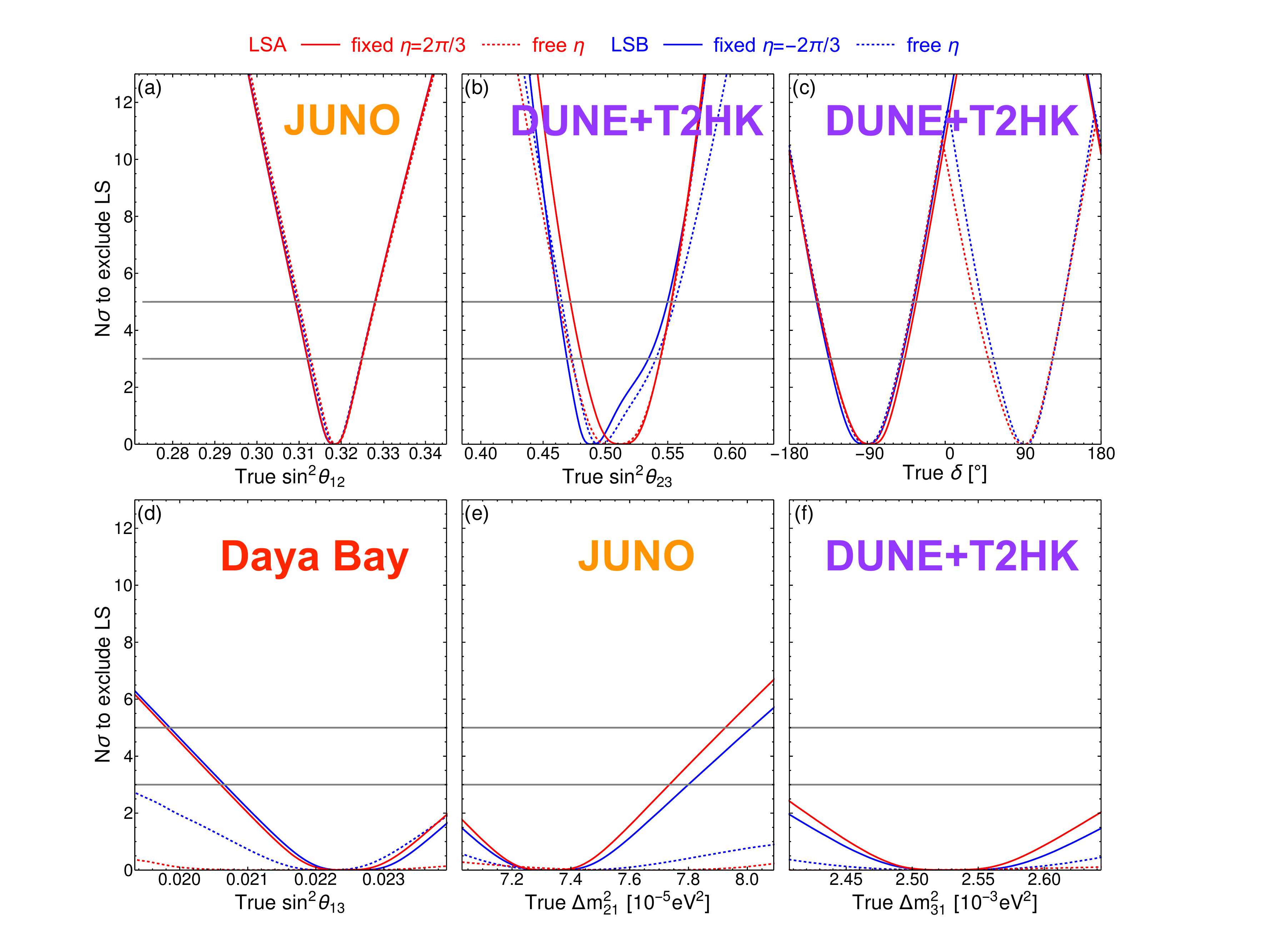}
    \caption{The prospects for excluding the Littlest Seesaw models using data from future precision
    neutrino oscillation experiments Daya Bay, JUNO, DUNE and T2HK. 
    The horizontal axes shows the possible true values of the 
    six oscillation parameters, and the vertical axes show the exclusions possible in each case
    for LSA and LSB, both with $\eta$ fixed at $\pm 2\pi /3$ and free.
    The range of the horizontal axes is chosen to show the currently allowed 3$\sigma$
    ranges of the six oscillation parameters.  
    The lower horizontal grey line shows the possible 3$\sigma$ exclusions for each parameter.
    This figure is adapted from \cite{Ballett:2016yod}.} \label{review2}
\vspace*{-8mm}
\end{figure}

We shall focus on the future precision neutrino oscillation experiments 
Daya Bay, JUNO, DUNE and T2HK, as recently discussed in 
\cite{Ballett:2016daj}. First we briefly summarise the plans for these experiments.
Daya Bay is a short baseline neutrino oscillation experiment which detects anti-electron neutrinos from various nuclear reactors in China at distances between 1.5 km and 1.9 km, near the first atmospheric oscillation maxiimum.
Daya Bay has currently the best precision on $\sin^2 \theta_{13}$ and in the future aims to acheive an accuracy of about 3\% .
JUNO is a medium baseline reactor experiment planned to have a basline of 53 km from two planned nuclear
reactors in China, corresponding to the first solar oscillation maximum. The longer baseline would allow
sensitive measurements of $\sin^2 \theta_{12}$ and $\Delta m_{21}^2$ accurate to about 0.5\% .
DUNE is a long baseline experiment which would use an accelerator at Fermilab to direct a wide band 
beam of muon (anti-)neutrinos with energies between 0.5 GeV and 5 GeV which are observed  
using a liquid Argon detector at Sanford located at a distance of 1300 km,
near the first atmospheric oscillation maximum. T2HK is also a long baseline experiment which would use an accelerator at Tokai to direct a narrow 2.5 degree off-axis 
beam of muon (anti-)neutrinos with energies around 0.6 GeV which are observed  
using large water Cerenkov detectors in Kamioka located at a distance of 295 km,
near the first atmospheric oscillation maximum. 
The muon disappearance and electron appearance channels of both DUNE and T2HK
allow precise measurements of $\sin^2 \theta_{23}$, the sign and magnitude of $\Delta m_{31}^2$ and the \CP phase $\delta$.
In Figure~\ref{review2} we show the prospects for excluding the Littlest Seesaw models using data from these future precision
    neutrino oscillation, where 
        the top three panels show that JUNO, DUNE and T2HK are capable of excluding
    LSA and LSB over much of the currently allowed range of the solar and atmospheric angles and the \CP phase.

\section{{\em Robustness}: Discrete non-Abelian family symmetry models}
\label{sec:GUTxFam}

\begin{figure}[t]
\centering
\includegraphics[width=0.60\textwidth]{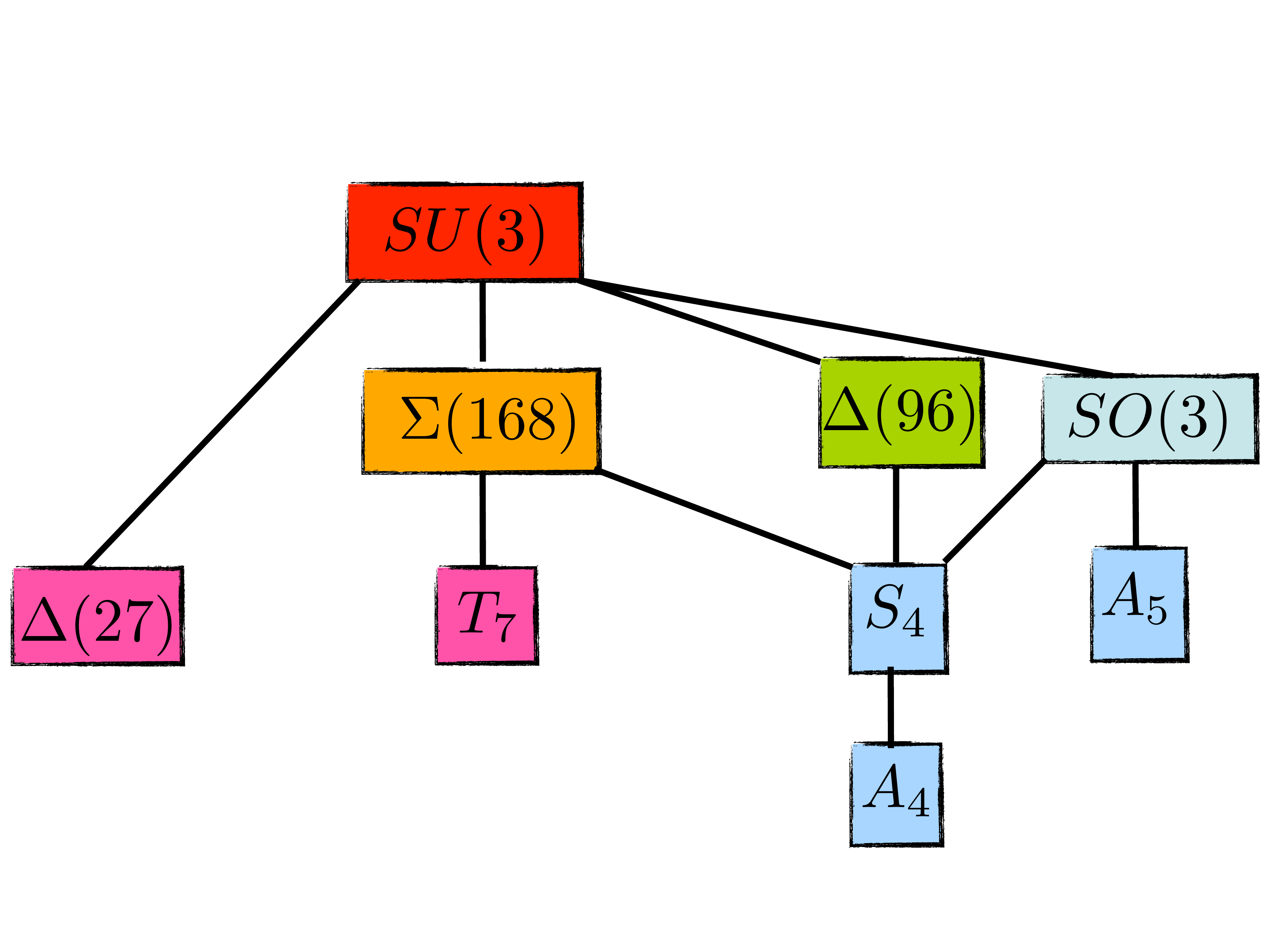}
\vspace*{-4mm}
    \caption{Some subgroups of $SU(3)$ which involve triplet representations. 
    The simplest groups $S_4$, $A_4$~\cite{Ma:2001dn}, $A_5$ (in pale blue) are related to BM, TB and GR mixing. 
    $\Delta (96)$ is an example of the $\Delta (6n^2)$ series~\cite{Escobar:2008vc}, while 
    $\Delta (27)$~\cite{deMedeirosVarzielas:2006fc} is an example of the $\Delta (3n^2)$ series~\cite{Luhn:2007uq}.
    $\Sigma (168)$, also called $PSL_2(7)$~\cite{Luhn:2007yr}, is a simple group, with a subgroup 
    $T_7$~\cite{Luhn:2007sy}.} \label{discrete}
\vspace*{-2mm}
\end{figure}

\subsection{Finite group theory}

\begin{figure}[t]
\centering
\includegraphics[width=0.60\textwidth]{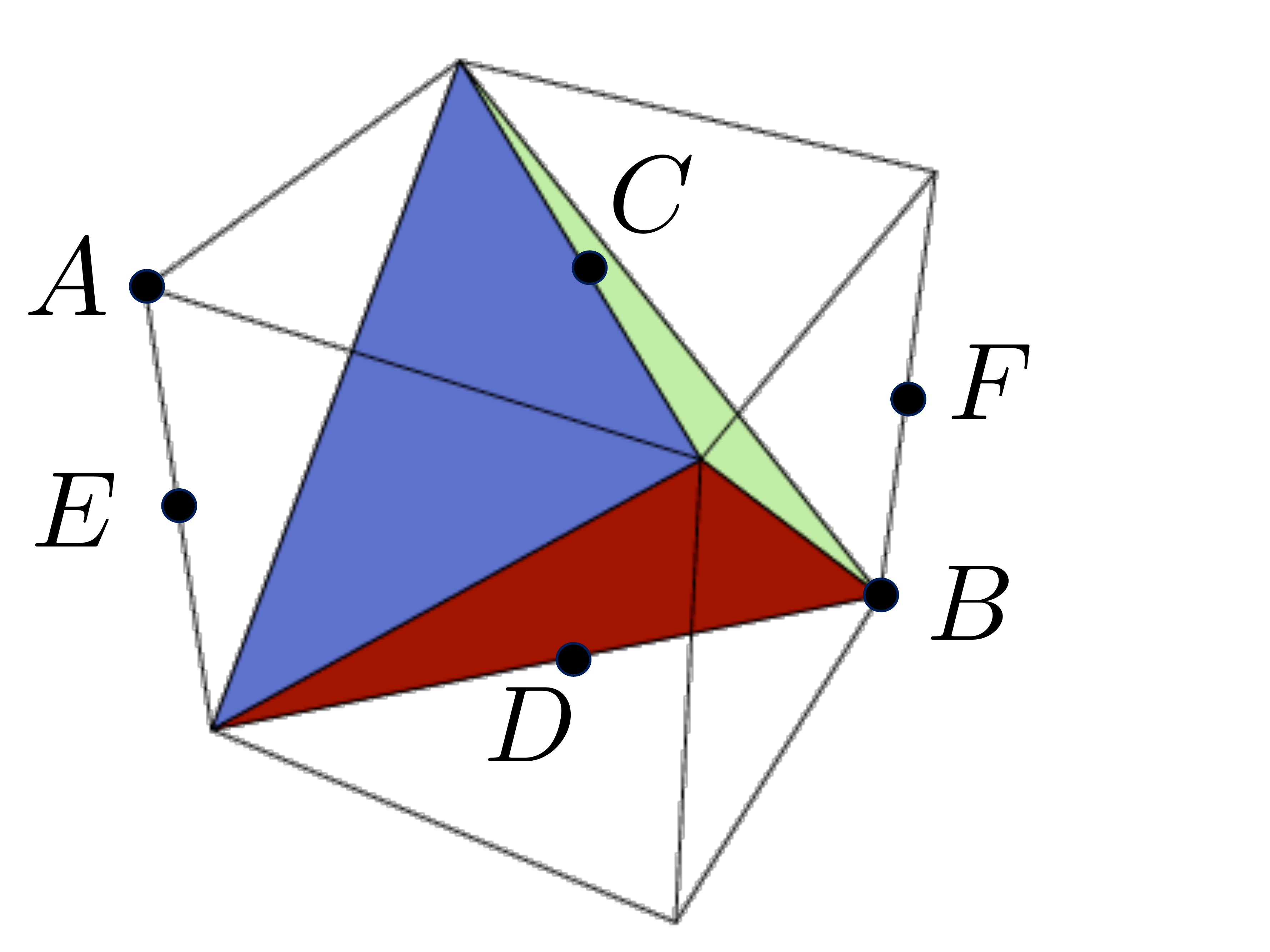}
\vspace*{-4mm}
    \caption{The groups $S_4$ and $A_4$ correspond to the rigid rotational symmetries of a 
    cube and tetrahedron, respectively, with $A_4$ being a subgroup of $S_4$ as seen geometrically by inscribing the tetrahedron inside the cube as shown.
    The rotation by 180 degrees about the axis EF is an example of a $U$-type symmetry of the cube but not
    the tetrahedron.} \label{s4a4}
\vspace*{-2mm}
\end{figure}

For a comprehensive
introduction to (finite) group theory we refer the reader to 
\cite{Ishimori:2010au}. Here we shall only recall a few basic features.
\begin{itemize}
\item A finite group $G$ contains of a finite number of elements $g$ together with a multiplication
law between any two of the elements so that it yields another element of $G$. 
\item
A group must
include the identity element $e$.
\item
For every element
$g$ there must be an inverse $g^{-1}$. 
\item
The product
of three elements satisfies $(g_1 g_2)g_3=g_1(g_2g_3)$ (associative). 
\item
Groups are called
Abelian if all the elements commute, $g_1g_2=g_2g_1$.
Non-Abelian groups have elements which do not commute. 
\end{itemize}

Abelian groups such as $Z_n$ have elements which commute and may be represented by complex numbers, $e^{2\pi i/n}$ of unit modulus.
Some non-Abelian groups subgroups 
of $SU(3)$ which contain triplet representations
are depicted in Fig.~\ref{discrete}. 
The simplest groups $S_4$, $A_4$~\cite{Ma:2001dn}, $A_5$ (in pale blue) are related to BM, TB and GR mixing. 
    $\Delta (96)$ is an example of the $\Delta (6n^2)$ series~\cite{Escobar:2008vc}, while 
    $\Delta (27)$~\cite{deMedeirosVarzielas:2006fc} is an example of the $\Delta (3n^2)$ series~\cite{Luhn:2007uq}.
    $\Sigma (168)$, also called $PSL_2(7)$~\cite{Luhn:2007yr}, is a simple group, with a subgroup 
    $T_7$~\cite{Luhn:2007sy}.
For example, $S_4$ is the rigid rotation group of a cube,
while $A_4$ is that of the tetrahedron, where $A_4$ is a subgroup of $S_4$,
as seen geometrically by inscribing the tetrahedron inside the cube as shown in Fig.\ref{s4a4}.

For a tetrahedron, there are twelve independent transformations (group elements of $A_4$)
as follows (see Fig.\ref{s4a4}):
\begin{itemize}
\item 4 rotations by 120$^\circ$ clockwise (about axes like AB) which are $T$-type 
\item 4 rotations by 120$^\circ$  anti-clockwise (about axes like AB) which are $T$-type  
\item 3 rotations by 180$^\circ$  (about axes like CD) which are $S$-type 
\item 1 unit operatator $\mathcal{I}$
\end{itemize}

For  a cube there are 24 independent transformations (group elements of $S_4$)
of which 12 are symmetries of $A_4$ (as above) and the remaining 
12 are not symmetries of $A_4$ and are as follows (see Fig.\ref{s4a4}):
\begin{itemize}
\item 3 rotations by 90$^\circ$  clockwise (about axes like CD) 
\item 3 rotations by 90$^\circ$  anti-clockwise (about axes like CD) 
\item 6 rotations by  180$^\circ$  (about axes like EF) which are $U$-type
\end{itemize}

Although a group is specified by its multiplication
table, the definition of a finite group this way 
becomes unweildy with increasing order (number of group elements) of $G$.
Another way is to use the 
``presentation'' of the group,
where the generators (subsets of elements 
from which all elements of the group can be obtained by multiplication) have to respect certain rules.
For example, 
the permutation group of four objects $S_4$, which is equivalent to the rigid symmetry group of the cube,
 can be defined by the presentation rules \cite{Hagedorn:2010th},
 \be
S^2=T^3=U^2=(ST)^3 = (SU)^2 = (TU)^2=(STU)^4 = 1 \ .
\ee
where $S$, $T$ and $U$ are the three generators.
If we drop $U$, this reduces to 
the presentation of $A_4$ \cite{Ma:2001dn}.
All the 24 group elements of $S_4$ may be obtained by multiplying the generators 
together, using the rules above.
Similarly, all the 12 group group elements of $A_4$ may be obtained by multiplying the generators 
$S$ and $T$ together, subject to the above rules.

The main interest of group theory from the point of view of physics, is that the group elements 
may be represented by matrices which respect the group multiplication laws. The smallest such matrices
which are not reducible to block diagonal form by a similarity transformation are called 
irreducible representations of the group. There are precise group theory rules for establishing the
irreducible representations of any group, but here we shall only state the results for
$S_4$ and $A_4$ in the $T$-diagonal basis, see \cite{Ishimori:2010au} for 
proofs, other 
examples and bases.

For $S_4$ there are two triplet matrix representations denoted ${\bf 3}$ and ${\bf 3'}$
which are independent and irreducible.
There are two singlet representations ${\bf 1}$ and
${\bf 1'}$. There also exists one irreducible doublet representation ${\bf 2}$.
For the $A_4$ subgroup there are three singlets
${\bf1},~{\bf 1^\prime}$ and ${\bf 1^{\prime \prime}}$ and one triplet ${\bf 3}$. 
The matrix representations in the diagonal $T$ basis are given in the
Table in Eq.\ref{table} \cite{King:2013eh} (where $\omega \equiv e^{i2\pi/3}$).
The Kronecker product rules for $S_4$ and $A_4$ are listed
in Appendix~\ref{app:CGs}.
\be
\begin{array}{c|ccc|c}\hline
S_4 & A_4 & S & T & U \\\hline
{\bf 1,1'} & {\bf 1} & 1&1 &\pm 1 \\[2mm]
{\bf 2} & \begin{pmatrix}{\bf 1''}\\{\bf 1'}\end{pmatrix} & 
\begin{pmatrix} 1 & 0 \\ 0&1 \end{pmatrix} 
&\begin{pmatrix} \omega & 0 \\ 0&\omega^2 \end{pmatrix}  
& \begin{pmatrix} 0& 1 \\ 1&0 \end{pmatrix} \\[4mm]
{\bf 3,3'} & {\bf 3}& 
\frac{1}{3}\begin{pmatrix} -1 & 2&2 \\ 2&-1&2 \\2&2&-1 \end{pmatrix} 
&\begin{pmatrix} 1&0&0\\ 0&\omega^2 &0 \\ 0&0&\omega \end{pmatrix}  
& \mp \begin{pmatrix} 1&0&0\\0&0& 1 \\ 0&1&0 \end{pmatrix} 
\\\hline
\end{array}
\label{table}
\ee

\subsection{Klein symmetry and direct models}

Suppose that the leptonic Lagrangian is invariant under the flavour symmetry associated with the group $G$.
Let us focus on the tranformations of the lepton doublet fields $L(x)$, where $x$ is spacetime and 
$L= (\nu_L, e_L )$ are the left-handed neutrino and charged lepton fields in a weak basis.
The lepton doublets transform under $G$ as,
\begin{equation}
L(x)\rightarrow \rho(g) L(x), 
\label{LG}
\end{equation}
where $\rho(g)$ is the group $G$ symmetry transformation
matrix associated with the group element $g$.
For example if $g=T$, and $L$ transforms as a triplet $\bf{3}$
of $S_4$, then $\rho(g)$ is the three dimensional matrix form of $T$
in Eq.\ref{table}, namely $\rho_{\mathbf{3}}(T)=\text{diag}(1, \omega^2, \omega)$.

The diagonal charged lepton mass matrix
$M_e$, appearing in the Lagrangian term $\overline{L}M_e e_R$,
may be combined into $M_eM_e^{\dagger}$ so that the right-handed transformations cancel,
and there is a phase symmetry,
\begin{equation}
T^{\dagger}(M_e M_e^{\dagger})T= M_e M_e^{\dagger}
\label{mesym}
\end{equation}
where for brevity we have written $T={\rm diag}(1, \omega^2 , \omega )$ 
which generates a subgroup $Z^T_3$ of $S_4$.

In a similar way, the Klein symmetry of the neutrino mass matrix, in this basis,
is given by,
\begin{equation}
m^{\nu}= S^Tm^{\nu} S, \ \ \ \ m^{\nu}= U^Tm^{\nu} U
\label{mnusym}
\end{equation}
where \cite{King:2009ap}
\begin{eqnarray}
S= U_{\rm PMNS}^*\ {\rm diag}(+1,-1,-1)\ U_{\rm PMNS}^T\\
U= U_{\rm PMNS}^*\ {\rm diag}(-1,+1,-1)\ U_{\rm PMNS}^T\\
SU= U_{\rm PMNS}^*\ {\rm diag}(-1,-1,+1)\ U_{\rm PMNS}^T
\end{eqnarray}
and 
\begin{equation}
{\cal K}=\{1, S, U, SU \}
\end{equation}
is called the Klein symmetry $Z^S_2\times Z^U_2$.
For the case that $U_{\rm PMNS}$ is equal to the tri-bimaximal mixing matrix 
$U_{\rm TB}$ in Eq.\ref{TB}, then $S$, $U$ and $T$ may be identified as the 
generators of $S_4$ in Eq.\ref{table}. In this way one may associate TB mixing with the 
discrete symmetry group $S_4$.
However, if the mixing matrix is something other 
than $U_{\rm TB}$ then $S$ and $U$ will differ from the generators in Eq.\ref{table}
and one must look for some other group. This exemplifies the so called ``direct'' approach
to model building whereby one postulates a discrete symmetry group $G$, whose generator
$T$ enforces the diagonal charged lepton mass matrix, while its generators $S$ and $U$
enforce a particular Klein symmetry associated with a particular PMNS matrix.
Different groups and generator embeddings will yield different predictions for the PMNS matrix.

From a dynamical point of view, the theory must organise itself so that the discrete symmetry 
group $G$ is broken by Higgs fields which know about flavour and are called flavons.
The flavons may be EW singlets or doublets.
There may be flavons $\phi^l$ whose VEVs preserve
$T$ (i.e. $T\langle  \phi^l \rangle = \langle \phi^l \rangle $)
and other $\phi^{\nu}$ whose VEVs preserve $S,U$
(i.e. $S\langle \phi^{\nu} \rangle = \langle \phi^{\nu} \rangle $ and 
$U \langle \phi^{\nu} \rangle = \langle \phi^{\nu} \rangle $). 
For example, consider the case of $S_4$ in the $T$ diagonal basis of Eq.\ref{table} \cite{King:2013eh},
where we emphasise that:
\be
U=\mp
\begin{pmatrix}
 1 & 0 & 0 \\
 0 & 0 & 1 \\
0 & 1 & 0
\end{pmatrix},\ \ \ \ 
SU=US=\mp \frac{1}{3}
\begin{pmatrix}
 -1 & 2 & 2 \\
 2 &  2 & -1 \\
2  & -1 & 2
\end{pmatrix}, \ \ \ \ { \rm for } \ \ {\bf{3}}, {\bf{3'}} \ \  { \rm respectively. }
\ee
In this basis one can check by explicit matrix multiplication 
(e.g. $T\langle  \phi_T \rangle = \langle \phi_T \rangle $, where $T$ is the matrix in Eq.\ref{table} and
$\langle \phi_T \rangle$ is the column vector given below)
that the symmetry preserving vacuum alignments are as follows \cite{King:2016yvg}:
$$
 \langle \phi_T \rangle \sim {\bf{3}} \sim  \begin{pmatrix}
 1 \\
 0\\
 0
\end{pmatrix},\   { \rm preserves  } \ T, \  { \rm breaks  } \ S,U, 
$$
$$
\langle \phi_T'  \rangle \sim {\bf{3'}} \sim  \begin{pmatrix}
 1 \\
 0\\
 0
\end{pmatrix},\  { \rm preserves  } \ T,U \  { \rm breaks  } \ S, 
$$
$$
\langle \phi_S  \rangle \sim {\bf{3}} \sim \begin{pmatrix}
 1 \\
 1\\
 1
\end{pmatrix},\  { \rm preserves  } \ S\  { \rm breaks  } \ T,U ,
$$
$$
\langle \phi_S'  \rangle \sim {\bf{3'}} \sim \begin{pmatrix}
 1 \\
 1\\
 1
\end{pmatrix},\  { \rm preserves  } \ S,U\  { \rm breaks  } \ T,
$$
$$
\langle \phi_{SU} \rangle  \sim {\bf{3}} \sim \begin{pmatrix}
 2 \\
 -1\\
 -1
\end{pmatrix},\  { \rm preserves  } \ SU\  { \rm breaks  } \ T,U ,
$$
and the two important $SU$ preserving alignments for ${\bf{3'}}$ flavons,
\be
\langle  \phi'_{\rm atm}  \rangle \sim {\bf{3'}} \sim \begin{pmatrix}
 0 \\
 1\\
 -1
\end{pmatrix},\  { \rm preserves  } \ SU\  { \rm breaks  } \ T,U ,
\label{phiatm}
\ee
\be
\langle  \phi'_{\rm sol}  \rangle \sim {\bf{3'}} \sim \begin{pmatrix}
 1 \\
 3\\
 -1
\end{pmatrix},\  { \rm preserves  } \ SU\  { \rm breaks  } \ T,U .
\label{phisol}
\ee

\begin{figure}[t]
\centering
\includegraphics[width=0.40\textwidth]{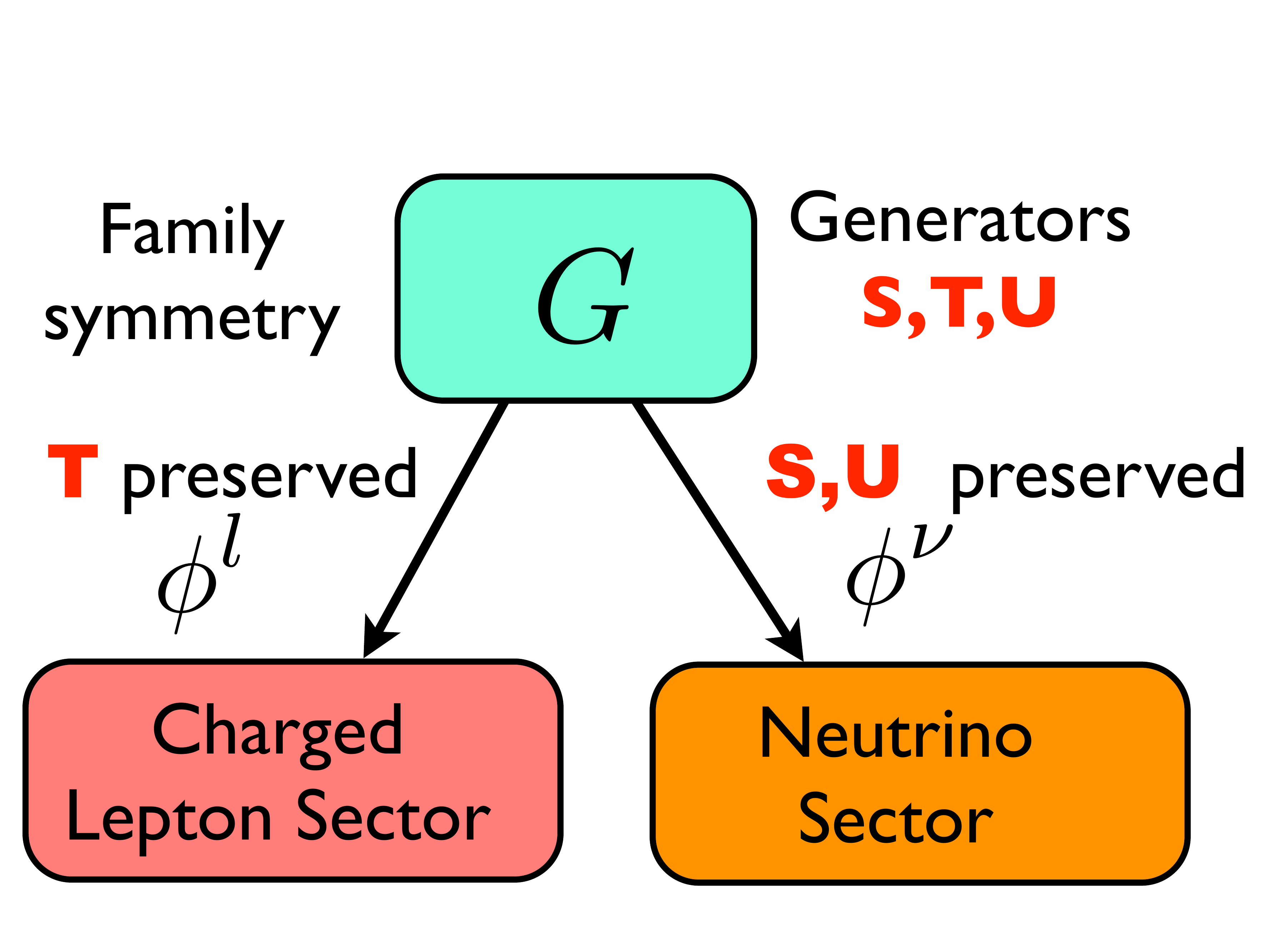}
\vspace*{4mm}
    \caption{This diagram illustrates the so called direct approach to models of lepton mixing. 
      } \label{direct}
\end{figure}

These flavons $\phi^l$ (identified with one or more of the $T$ preserving flavons) 
only carefully engineered to only appear in terms responsible for charged lepton masses.
The other flavons $\phi^{\nu}$ (identified with one or more of the $S,U$ preserving flavons) 
only couple to terms responsible for neutrino masses.

This is the so called ``direct approach'' illustrated in Fig.\ref{direct}.
For example $G=S_4$
    can lead to TB mixing
    if $T$ is preserved in the charged lepton sector, and $S,U$ are preserved in the neutrino sector, 
    which can be achieved dynamically by assuming that different symmetry preserving flavons are 
    confined to a particular sector. For example the charged lepton mass matrix $M_e$ may arise from 
    a non-renormalisable Lagrangian term $\frac{\phi^l}{\Lambda} L H_de^c$ 
    where $\Lambda$ is a heavy mass scale once the flavon $\phi^l$ and Higgs $H_d$ get VEVs.  
    Since only $\phi^l$ (not $\phi^{\nu}$) appears in the charged lepton sector, the mass matrix $M_e$ therefore respects the $T$ symmetry (see Eq.\ref{mesym}) preserved by the $\phi^l$ VEV. Similarly $m^{\nu}$ respects the $S,U$ symmetry
    (see Eq.\ref{mnusym})  preserved 
    by the $\phi^{\nu}$ VEV.

In such a ``direct approach''
the full Klein symmetry $Z^S_2\times Z^U_2$ of the neutrino mass matrix
arises as a subgroup of the initial family symmetry $G$.
Given the measurement of the reactor angle, the
only viable direct models are those based on $\Delta (6N^2)$
\cite{Holthausen:2012wt,King:2013vna,Fonseca:2014koa},
with quite large $N$ required. 
Such models generally predict $\text{TM}_2$ mixing and a \CP phase 
$\delta = 0,\pi$, both of which are disfavoured by current data.

\subsection{Semi-direct models}

\begin{figure}[t]
\centering
\includegraphics[width=0.50\textwidth]{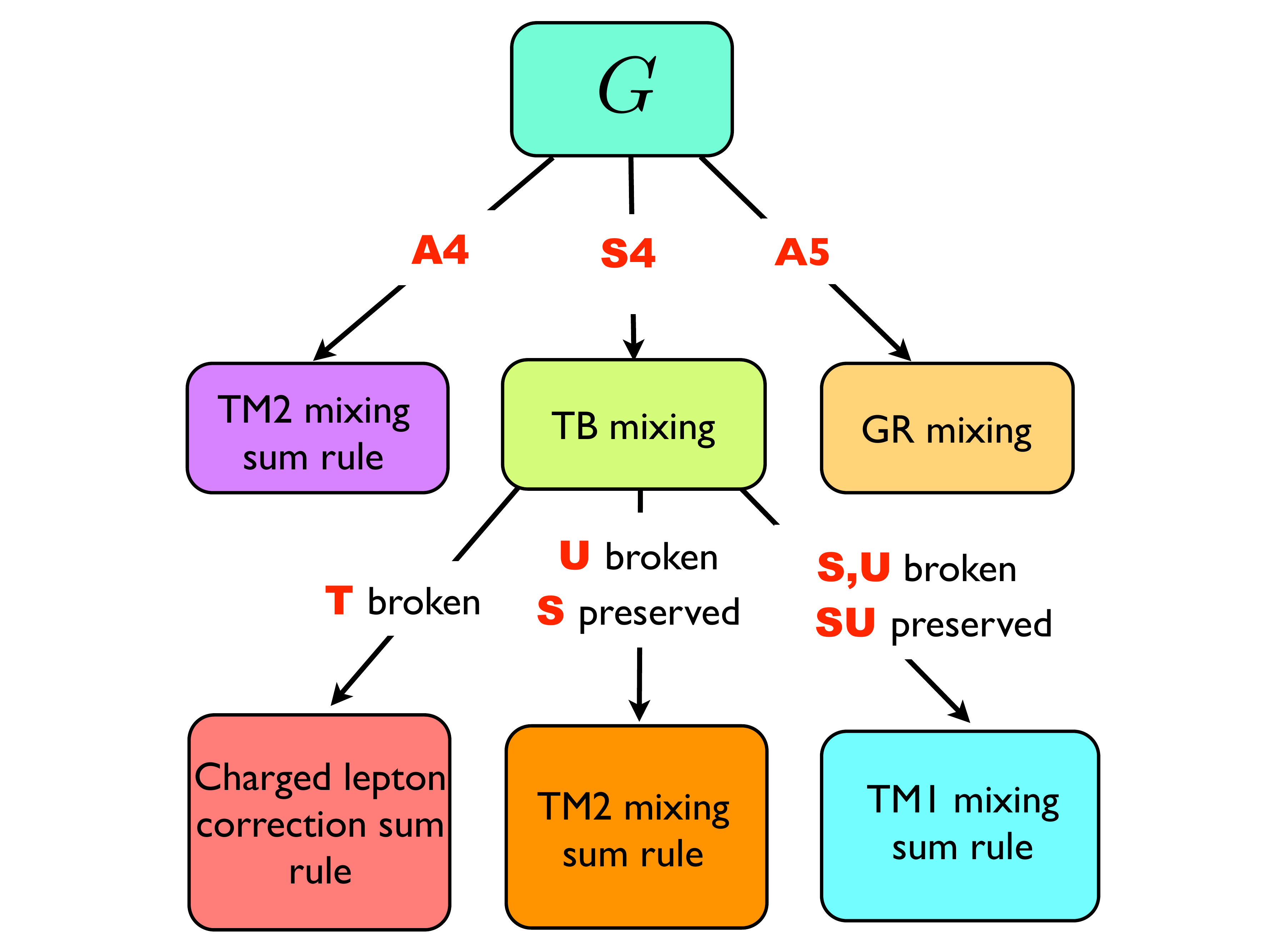}
    \caption{The semi-direct approach to models of lepton mixing. Since TB and GR mixing are excluded,
    some of the symmetry must be broken in either the charged lepton sector ($T$ breaking) or the neutrino sector
    ($U$ breaking). In the semi-direct approach some symmetry always remains as shown leading to mixing sum rule
    predictions. For the case of $S_4$, the figure indicates that $T$ breaking (with $S,U$ and hence TB preserved in the neutrino sector)
  leads to charged lepton correction sum rules.
    Similarly, $U$ breaking (with $T$ preserved in the charged lepton sector)
    can lead to $\text{TM}_1$ or $\text{TM}_2$ mixing and sum rules, depending on whether $SU$ or $S$ is preserved
    in the neutrino sector. The $A_4$ group does not contain $U$ and (with $T$ and $S$ preserved), leads to the $\text{TM}_2$ mixing and sum rule.
    Note that the $\text{TM}_2$ mixing sum rule is experimentally disfavoured.
         } 
    \label{semi-direct}
\vspace*{2mm}
\end{figure}

In the ``semi-direct'' approach, one may use 
smaller discrete family groups such as $S_4$ or $A_5$.
If applied in a ``direct'' way, such groups would 
lead to either TB or BM (for $S_4$) or 
GR mixing (for $A_5$), as in Fig.\ref{semi-direct}.
To obtain a non-zero reactor angle, 
one of the generators $T$ or $U$ must be broken.
Thus the semi-direct models do not enforce the full residual symmetry. 

Consider the following 
two interesting possibilities depicted in Fig.\ref{semi-direct}:
\begin{enumerate} 
 \item {The $Z_3^T$ symmetry of the charged lepton
 mass matrix is broken, but the full Klein symmetry $Z_2^S\times Z_2^U$ 
 in the neutrino sector is respected. This corresponds to having 
 charged lepton corrections, with 
 solar sum rules discussed in section \ref{solar}.}
\item{The $Z_2^U$ symmetry of the neutrino mass matrix is broken, but the  
$Z_3^T$ symmetry of the charged lepton
mass matrix is unbroken. In addition either $Z_2^S$ or $Z_2^{SU}$ (with $SU$ being the product of $S$ and $U$) is preserved.
This leads to either $\text{TM}_1$ mixing (if $Z_2^{SU}$ is preserved);
or $\text{TM}_2$ mixing (if $Z_2^S$ is preserved).
Then we have the atmospheric sum rules as discussed in section \ref{atmospheric}.}
\end{enumerate} 
In $A_4$ there is no $U$ generator to start with, but it is possible that 
$Z_2^S$ preserved. This could also arise of $S_4$ is broken to $A_4$ at higher order \cite{King:2011zj}.
In such cases, only half the Klein symmetry $Z_2^S$ is preserved, corresponding to 
the $S$ generator of $A_4$ or $S_4$, together with the $Z_3^T$ symmetry of the diagonal $T$ generator
enforcing the diagonality of the charged lepton mass matrix.
However, the $S$ generator implies $\text{TM}_2$ mixing and sum rules which are disfavoured 
due to the solar angle being smaller than its tri-bimaximal value.
Therefore below we shall focus on an example of the more successful $\text{TM}_1$ mixing with $SU$ preserved
 \cite{Luhn:2013vna}. We remark that, although this semi-direct approach was formalised in a general group theoretical analysis in  \cite{Hernandez:2012ra}, no other phenomenologically interesting examples were discovered,
 so the only case of interest remains $\text{TM}_1$.

\vspace{0.1in}

\underline{Example of a semi-direct model with $\text{TM}_1$ mixing: the Littlest Seesaw}

\vspace{0.1in}

Since the Littlest Seesaw model with 2RHN respects $\text{TM}_1$ mixing
(see Eq.\ref{CSD(n)a}), it is not too surprising that 
it can be realised as a semi-direct model, 
where $SU$ preserved in the neutrino sector and
$T$ in the charged lepton sector.
The novel feature of the model in \cite{King:2016yvg} is that it involves 
2RHNs, $N^c_{\rm sol}\sim {\bf 1}$, $N^c_{\rm atm}\sim {\bf 1}$
(unlike typical semi-direct models which involve 3RHNs in a triplet)
in addition to the lepton doublets which 
transform under $S_4$ as $L\sim {\bf 3}^{{\prime}}$, and the up-
and down-type Higgs fields $H_{u,d}\sim {\bf 1}$.
The neutrino Yukawa couplings of the model are of the form:
\be
\frac{\phi'_{\rm atm}}{\Lambda}LH_uN^c_{\rm atm}  
+ \frac{\phi'_{\rm sol}}{\Lambda}LH_uN^c_{\rm sol}  \ ,
\label{nucouplings}
\ee
where the non-renormalisable terms are suppressed by a dimensionful cut-off
$\Lambda$ and  the flavons $\phi'_{\rm atm}\sim {\bf 3}^{{\prime}}$ and
$\phi'_{\rm sol}\sim {\bf 3}^{{\prime}}$  have the $SU$ preserving vacuum
alignments in Eqs.\ref{phiatm}, \ref{phisol} \footnote{Vacuum alignment is fully discussed in \cite{King:2016yvg}.},
\be
\langle \phi'_{\rm atm} \rangle =\varphi'_{\rm atm}
\begin{pmatrix}
 0 \\
 1\\
 -1
\end{pmatrix},\ \ \ \ 
\langle \phi'_{\rm sol} \rangle =\varphi'_{\rm sol}
\begin{pmatrix}
 1 \\
 3\\
 -1
\end{pmatrix},
\label{CSD3}
\ee
i.e. $SU \langle \phi'_{\rm atm} \rangle = \langle \phi'_{\rm atm} \rangle $
and  $SU \langle \phi'_{\rm sol} \rangle = \langle \phi'_{\rm sol} \rangle $,
but break $T$ and $U$ separately, as shown in the previous subsection. The preserved $S_4$ subgroup $SU$ is instrumental
in enforcing TM$_1$ mixing.

The $S_4$ singlet contraction 
${\bf 3}^{{\prime}} \otimes {\bf 3}^{{\prime}} 
~\rightarrow ~{\bf   1}^{\phantom{\prime}}$  
implies $(L\phi')_{\bf 1}  = L_1\phi'_1 + L_2\phi'_3 +L_3\phi'_2 $  
(see Appendix~\ref{app:CGs}), 
which leads to the Dirac neutrino mass matrix~$m^D$
and RH neutrino mass matrix~$M_R$,
\begin{equation}
	m^D = \pmatr{0 & b \\ -a & -b \\ a & 3b } \equiv \pmatr{0 & b \\ a & b \\ a & 3b } ,
	\qquad
	M_{R}=
\left( \begin{array}{cc}
M_{\rm atm} & 0 \\
0 & M_{\rm sol}
\end{array}
\right),
			\label{mDn0}
\end{equation}
where the equivalence above follows after multiplying $L_2$ by a minus sign.
Using the mass matrices in Eq.\ref{mDn0}, the seesaw formula in Eq.\ref{seesaw} then implies\footnote{We
  follow the Majorana mass convention $- \frac{1}{2}\overline{\nu_L} m^{\nu}
  \nu_{L}^c $.}  the LSB low energy neutrino mass matrix in Eq.\ref{eq:matrix_LSB},
\begin{equation}
	m^\nu_{\rm LSB} = m_a 
	\left(
\begin{array}{ccc}
	0&0&0\\0&1&1\\0&1&1 
	\end{array}
\right)
	+ m_b e^{i\eta } 
	\left(
\begin{array}{ccc}
	1&1&3\\1&1&3\\3&3&9
	\end{array}
\right),
	\label{eq:mnu2p10}
\end{equation}
where without loss of generality, $m_a=|a|^2/M_{\rm atm}$, $m_b=|b|^2/M_{\rm
sol}$ may be taken to be real and positive while $\eta$ is a real phase parameter
which is not fixed by the semi-direct flavour symmetry $SU$.

In order to fix the phase $\eta$ to its desired value of $\eta = \pm 2\pi/3$ 
one can use the mechanism for spontaneous \CP violation first proposed in \cite{Antusch:2011sx}.
The idea is to impose a \CP symmetry in the original theory which is
spontaneously broken by complex flavon VEVs. 
In order to drive a flavon VEV to a complex
value whose phase factor is~$\omega^k$, one needs to introduce a further discrete symmetry like
$Z_3$, under which the flavons transform, leading to terms in the flavon potential like 
$(\phi^3/\Lambda - M^2)^2$. If \CP is respcted then all couplings and mass scales such as $\Lambda, M$ are real
and hence the potential is minimised for $\langle \phi \rangle= \omega^k|\Lambda M^2|^{1/3}$,
where $\omega = e^{i2\pi /3}$. When the flavon VEVs $\langle \phi'_{\rm atm}\rangle $ and 
$\langle \phi'_{\rm sol} \rangle$ 
are inserted into the seesaw formula this restricts the 
phase $e^{i\eta } $ to be one of the cube roots of unity, with 
the actual choice of $\eta = \pm 2\pi/3$ selected from a set of integer choices for $k$,
chosen randomly for different flavons.
Because the subject of spontaneous \CP violation

\subsection{Spontaneous \CP violation}

As we saw in the example in the previous subsection, models with 
discrete family symmetry may also possess \CP symmetry.
It is then possible 
spontaneously break the \CP symmetry along with the family symmetry.
In this subsection, we first recall a few basic facts about \CP symmetry,
and how it may be sponaneously broken, before going on to describe some recent approaches
to spontaneous \CP violation in models with discrete family symmetry.

Any Lagrangian may be written as follows: $\mathcal{L}=\mathcal{L_{CP}}+\mathcal{L}_{rem}$
where $\mathcal{L_{CP}}$ conserves \CP since it involves kinetic and gauge parts,
while $\mathcal{L}_{rem}$ includes 
the Yukawa couplings \cite{Bernabeu:1986fc}.
The remaining part $\mathcal{L}_{rem}$ may or may not respect one or more of the
general \CP transformations that leave $\mathcal{L_{CP}}$ invariant.
If it violates all of them then we are sure that $\mathcal{L}$ explicitly violates \CP .

For example, the quark Yukawa coupling Lagrangian in the SM explicitly violates \CP .
The same applies to the resulting quark mass matrices $M_u$ and $M_d$.
The signal of \CP violation in the quark sector of the SM is the non-vanishing of the 
rephasing invariant  \cite{Bernabeu:1986fc}, 
\be
I_1^q\equiv \det [M_u M_u^{\dag}, M_d M_d^{\dag}] =\frac{1}{3}\Tr \left( [M_u M_u^{\dag}, M_d M_d^{\dag}]^3 \right)
=6i \Delta^q J^q
\label{I1q}
\ee
where $\Delta^q $ is the product of the six quark mass squared differences, while $J^q$ is the Jarlskog invariant.
Explicitly,
\be
\Delta^q = (m_t^2-m_c^2) (m_t^2-m_u^2) (m_c^2-m_u^2) (m_b^2-m_s^2) (m_b^2-m_d^2) (m_s^2-m_d^2)
\label{Delta}
\ee
\be
J^q=\Im ( U_{us}U_{cb}U^*_{ub} U^*_{cs})= \frac{1}{8}\sin 2\theta^q_{12}\sin 2\theta^q_{13}\sin 2\theta^q_{23}\cos \theta^q_{13}\sin \delta^q
\label{J}
\ee
$I_1^q$ is also known as a \CP-odd invariant since its non-zero value is a signal of explicit \CP violation
in the theory. If it is zero then \CP is conserved, which may happen even if some of the Yukawa couplings 
in some basis are complex.

A similar \CP-odd invariant can be defined for the SM Lagrangian extended by Majorana neutrino masses as in 
Eq.\ref{lepton}. 
Due to the $SU(2)_L$ structure, the most general \CP transformation which leaves the leptonic gauge interactions invariant are (dropping spin and flavour indices),
\begin{equation}
L(x)\rightarrow X L^*(x_P), \ \ e_R(x)\rightarrow X' e_R^*(x_P),
\label{LCP}
\end{equation}
where $L= (\nu_L, e_L )$ are the left-handed neutrino and charged lepton fields in a weak basis,
and $x_P$ are the parity (3-space) inverted coordinates.
Typically $L$ will be a three dimensional column vector (corresponding to the three lepton families)
in a triplet representation of some flavour group $G$,
and $X$ will be a three dimensional matrix in flavour space.
In order for ${\cal L}^{\rm lepton}$ to be \CP invariant under Eq.(\ref{LCP}), the Lagrangian terms 
in Eq.\ref{lepton} go into their respective $H.c.$ terms and vice-versa leading to the
conditions on the mass matrices:
\begin{equation}
X^\dagger  m_{\nu} X^* = m_{\nu}^*, \ \ \ \ 
X^\dagger M_{e} X' = M_{e}^* \, ,
\label{mlCP}
\end{equation}
where we have written $m_{\nu} =m^{\nu_e}_{ij}$ and $M_{e} =v_dY^e_{ij}$.

The condition for \CP to be conserved is (analogous to the quark sector result)
in Eq.\ref{I1q} \cite{Bernabeu:1986fc}:
\begin{equation}
I_1^l \equiv\frac{1}{3} \Tr\left( \left[H_\nu , H_e \right]^3\right) = \frac{1}{3}\Tr\left( \left[H_\nu H_e -  H_e H_\nu \right]^3\right) = 
6i\Delta^l J^l =0\,,
\label{hhcube}
\end{equation}
where $H_\nu \equiv m_\nu m_\nu^\dagger$ and $H_e \equiv M_e M_e^\dagger$,
and $\Delta^l$ and $J^l$ are the analogues of the results for the quark sector
in Eqs.\ref{Delta} and \ref{J}, with $q\rightarrow l$ for the lepton mixing parameters, 
$u,c,t \rightarrow 1,2,3$ for the neutrino masses, and 
$d,s,b \rightarrow e, \mu , \tau$ for the charged lepton masses.
The condition $I_1^l =0$ is both a necessary and sufficient condition for Dirac \CP invariance.
If the mass matrices are chosen such that $I_1^l=0$ then Dirac type \CP is explicitly conserved 
while if $I_1^l\neq 0$ then Dirac type \CP is explicitly violated.
\footnote{
This is Dirac type \CP violation since it occurs both when neutrinos have both Dirac and Majorana masses. Apart from this, there may be 
two further 
necessary and sufficient conditions for low energy leptonic \CP invariance 
which only appear in the Majorana sector \cite{Branco:1986gr}.}

As shown in \cite{Branco:2015hea},
if a Lagrangian is specified, which is invariant under a family symmetry $G$
and some \CP transformation, then the 
consistency relations first introduced in \cite{Holthausen:2012dk,Feruglio:2012cw} are automatically satisfied,
namely,
\begin{equation}
X\rho(g)^*X^{\dagger}=\rho(g'),
\label{consistency}
\end{equation}
where $X$ is a \CP transformation matrix as in Eq.\ref{LCP} and $\rho(g)$ is the flavour transformation
matrix associated with a group element $g$ belonging to $G$ 
as in Eq.\ref{LG}, while $g'$ is another element of $G$.
The main point to emphasise is that the \CP tranformation matrix $X$ need not be the unit matrix, it can be any unitary matrix
that satisfies the consistency condition in Eq.\ref{consistency}.
If $X$ is the unit matrix then we refer to it as trivial \CP , while if $X$ is some other unitary matrix then we 
refer to it as non-trivial \CP , or sometimes, generalised \CP , although we emphasise that one \CP transformation
is as good as another, and both trivial and non-trivial \CP are equally valid and on the same footing,
indeed they are both basis dependent. Physical \CP violating observables 
only depend only on basis invariants such as $I_1^q$ and $I_1^l$,
which are independent on the matrix forms of $X$ which cancel by construction.

In the SM, the Yukawa matrices explicitly violate \CP therefore no transformation $X$ exists that 
leaves the theory \CP invariant.
However in theories beyond the SM, a new possibility arises, namely that the theory 
respects \CP at high energy, but \CP is spontaneously broken in the low energy effective theory.
Such theories are interesting since they allow for the possibility of being able to predict the
amount of \CP violation (e.g. the physical \CP violating phases in some basis).
We already saw an example of spontaneous \CP violation below Eq.\ref{eq:mnu2p10}.
In that example, we assumed that the high energy couplings in Eq.\ref{nucouplings} respected \CP symmetry,
which in that example implies that the Yukawa couplings are real. We then argued that the flavons
$\phi$ could develop VEVs with complex phases $\langle \phi \rangle= \omega^k|\Lambda M^2|^{1/3}$
which could break \CP spontaneously. In that example, we were implicitly assuming 
trivial \CP transformations where $X$ was identified with the unit matrix.

The question of {\em spontaneous} \CP violation amounts to whether the vacuum does or does not 
respect \CP symmetry. 
In order for the vacuum to be \CP invariant, the following relation has to be satisfied:
$ <0| \phi_i |0> = X_{ij} <0| \phi_j^* |0>$ \cite{Branco:1983tn}. 
The presence of $G$ usually allows for many choices for $X$. 
If any $X$ can be found then \CP is conserved by the vacuum.
If no choice of $X$ exists then the vacuum violates \CP .
In order to prove that no choice of $X$ exists one can construct
\CP-odd invariants.

In extensions of the Higgs sector of the SM, the \CP violation arising from the 
parameters of the scalar potential can be studied in a similar basis invariant way to 
the quark or lepton sector. For example, in the two Higgs Doublet Model (HDM)
(for a recent analysis see e.g.~\cite{Keus:2015hva})
a \CP odd invariant was identified in~\cite{Mendez:1991gp}. More generally, 
applying the invariant approach to scalar potentials has revealed relevant CPIs~\cite{Lavoura:1994fv, Botella:1994cs, Branco:2005em}, including for the 2HDM~\cite{Davidson:2005cw, Gunion:2005ja}. 
This analysis was recently extended to 
potentials involving three or six Higgs fields (which can be either electroweak doublets or singlets) which form irreducible triplets under a discrete symmetry \cite{Varzielas:2016zjc}.

\subsection{Residual \CP symmetry}

\begin{figure}[t]
\centering
\includegraphics[width=0.5\textwidth]{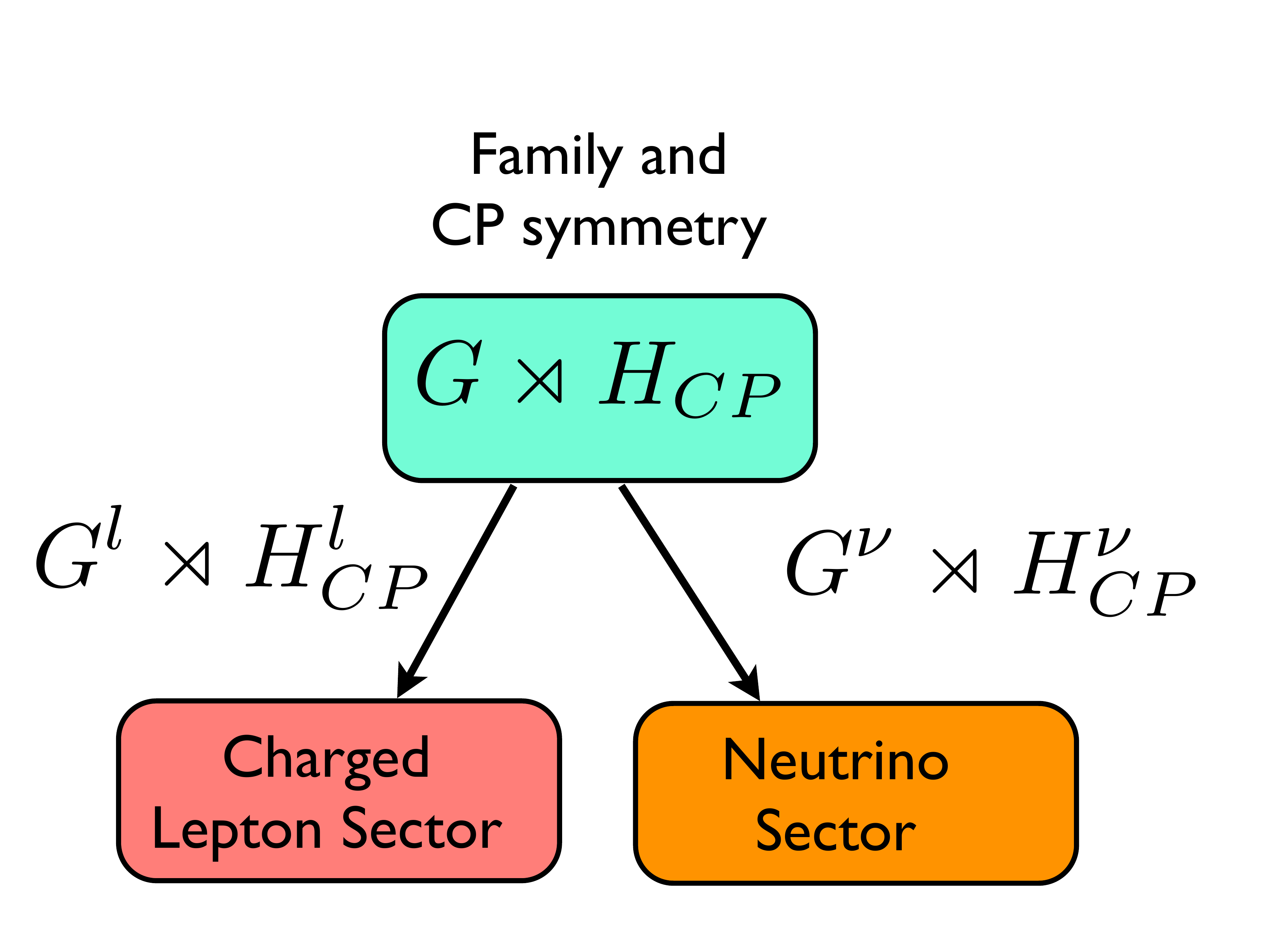}
\vspace*{4mm}
    \caption{The residual \CP symmetry approach to model building including both 
    discrete family (flavour) and \CP symmetry. The idea is that the original high energy theory conserves \CP but \CP is spontaneously broken in the low energy theory. Nevertheless one may define residual \CP 
    symmetries which are preserved in the charged lepton and neutrino sectors, which survive along with 
    preserved subgroups of the original family symmetry in each of these sectors. The semi-direct product 
    sign indicates that \CP does not always commute with flavour symmetry. 
    } \label{directCP}
\end{figure}

The residual \CP approach is based on models with discrete family symmetry, which are 
generalised to the case of a conserved \CP where $X$ may be non-trivial,
but must satisfy the consistency condition in Eq.\ref{consistency} (see e.g. \cite{Holthausen:2012dk} and references therein) which is spontaneously broken as shown in Fig.\ref{directCP}, i.e. preserving a different residual \CP
in the charged lepton and/or neutrino sectors. Of course the complete theory spontaneously violates \CP ,
but the preservation of different residual \CP symmetries (and flavour symmetries), 
in the two sectors provides predictive power, since it serves to constrain the charged lepton and neutrino
mass matrices separately. The residual flavour symmetry constraint on the mass matrices was given in Eqs.\ref{mesym} 
and \ref{mnusym}. The new residual \CP symmetry constraint on the mass matrices is as in Eq.\ref{mlCP}
but now different residual \CP symmetries are allowed for the LH charged leptons and LH neutrinos.
This is permitted since, below the electroweak symmetry breaking scale, they get their mass from
different flavons. 

In order to constrain \CP phases, one may suppose that the high energy theory respects
\CP , but it is spontaneously broken leaving some residual \CP symmetry
in the charged lepton and/or neutrino sectors but with \CP broken overall
~\cite{Feruglio:2012cw,Holthausen:2012dk}. This increases the predictivity of the theories,
since not only the mixing angles but also the \CP phases only depend on one single real 
parameter~\cite{Feruglio:2012cw}. The general \CP symmetry was originally discussed in the context of 
continuous gauge groups~\cite{Ecker:1981wv,Grimus:1995zi}.
It was subsequently applied to 
$\mu-\tau$ reflection symmetry~\cite{Harrison:2002kp,Grimus:2003yn,Farzan:2006vj},
where such theories predict a maximal Dirac \CP phase and maximal atmospheric mixing,
however non-maximality may arise from a simple extension~\cite{Chen:2015siy}.

As discussed above, it is nontrivial to give a consistent definition of general {\CP} 
transformations in the presence of discrete flavour symmetry, since namely the 
consistency condition in Eq.\ref{consistency} must be fulfilled~\cite{Holthausen:2012dk,Chen:2014tpa}. The relationship between neutrino mixing and \CP symmetry has been further refined in~\cite{Chen:2014wxa,Chen:2015nha,Everett:2015oka}, and a master formula to reconstruct the PMNS matrix from any given remnant \CP transformation has been derived~\cite{Chen:2014wxa,Chen:2015nha}. The phenomenological predictions and model building of combining discrete flavour symmetry with generalized CP have already been studied for a number of 
discrete groups in the literature, e.g. $A_4$~\cite{Ding:2013bpa}, $S_4$~\cite{Feruglio:2012cw,Ding:2013hpa,Feruglio:2013hia,Luhn:2013vna,Li:2013jya,Li:2014eia}, $A_5$~\cite{Li:2015jxa,DiIura:2015kfa,Ballett:2015wia,Turner:2015uta},
$\Delta(27)$~\cite{Branco:1983tn,Branco:2015gna},
$\Delta(48)$~\cite{Ding:2013nsa,Ding:2014hva}, $\Delta(96)$~\cite{Ding:2014ssa} and the infinite series of finite
groups $\Delta(3n^2)$~\cite{Hagedorn:2014wha,Ding:2015rwa}, $\Delta(6n^2)$~\cite{Hagedorn:2014wha,King:2014rwa,Ding:2014ora} and $D^{(1)}_{9n, 3n}$~\cite{Li:2016ppt}.
Recently leptogenesis has been considered in this approach
\cite{Chen:2016ptr,Hagedorn:2016lva}.
Below we give one illustrative example of a semi-direct analysis.

\vspace{0.1in}

\underline{Example of semi-direct models with $\text{TM}_1$ mixing and residual \CP}

\vspace{0.1in}

To illustrate the residual \CP approach, let us consider the semi-direct models based on $S_4$ with $\text{TM}_1$ mixing
\cite{Luhn:2013vna}, extended to include a residual \CP symmetry \cite{Li:2013jya}.
It turns out that the most general CP transformation consistent with $S_4$ flavor symmetry is of the same form as the flavor symmetry~\cite{Li:2013jya} (in the basis of Eq.\ref{table} \cite{King:2013eh}).  
Following \cite{Li:2013jya}, we shall consider the scenario that the $S_4$ and CP symmetry is broken down to the $Z^{T}_3$ subgroup in the charged lepton sector and $Z^{SU}\times CP$ in the neutrino sector. The residual flavor symmetry $Z^{SU}_2$ enforce that the lepton mixing matrix is the $\text{TM}_1$ pattern~\cite{Albright:2008rp}. The requirement that $Z^{T}_3$ is a symmetry of the charged lepton mass matrix entails that $M_{e}M^{\dagger}_{e}$ is invariant under the action of the element $T$,
\begin{equation}
\rho^{\dagger}_{\mathbf{3}}(T)M_{e}M^{\dagger}_{e} \rho_{\mathbf{3}}(T)=M_{e}M^{\dagger}_{e}\,.
\end{equation}
Since the representation matrix $\rho_{\mathbf{3}}(T)=\text{diag}(1, \omega^2, \omega)$ is diagonal, the charged lepton mass matrix $M_{e}M^{\dagger}_{e}$ has to be diagonal as well,
\begin{equation}
M_{e}M^{\dagger}_{e}=\text{diag}\left(m^2_{e},m^2_{\mu},m^2_{\tau}\right)\,,
\end{equation}
where $m_e$, $m_{\mu}$ and $m_{\tau}$ denote the electron, muon and tau masses, respectively.

In the neutrino sector the residual symmetry $Z_{2}^{SU}\times CP$ is preserved by the neutrino mass matrix. The residual CP transformation $X_{\nu}$ should be consistent with the remnant flavor symmetry $Z_{2}^{SU}$, and consequently the following  consistency equation (as in Eq.\ref{consistency}) has to be satisfied for $Z_{2}^{SU}$,
\begin{eqnarray}
\label{eq:consistent-two}
X_{\nu}\rho_{\mathbf{3}}^{*}(SU)X_{\nu}^{-1}=\rho_{\mathbf{3}}(SU)\,.
\end{eqnarray}
There are four consistent possible solutions for $X_{\nu}$, 
\begin{equation}
X_{\nu}=\rho_{\mathbf{3}}(1), \rho_{\mathbf{3}}(S), \rho_{\mathbf{3}}(U), \rho_{\mathbf{3}}(SU)\,.
\label{X}
\end{equation}
The light neutrino mass matrix $m_{\nu}$ is constrained by the residual family symmetry $Z^{SU}_2$ and residual CP symmetry $X_{\nu}$ as~\cite{Ding:2013hpa}:
\begin{subequations}
\begin{eqnarray}
\label{eq:remnant_flavor_nu} && \rho_{\mathbf{3}}^{T}(SU)m_{\nu}\rho_{\mathbf{3}}(SU)=m_{\nu}\,, \\
\label{eq:remnant_CP_nu} && X_{\nu}^Tm_{\nu}X_{\nu}=m^{*}_{\nu}\,,
\end{eqnarray}
\end{subequations}
where the second of these equations follows from Eq.\ref{mlCP}.
For $X_{\nu}= \rho_{\mathbf{3}}(S), \rho_{\mathbf{3}}(U)$,
the lepton mixing angles and CP phases are determined to be a special case of $\text{TM}_1$ mixing,
with maximal atmospheric mixing angle and maximal Dirac \CP violation $\delta_{CP}=\pm\frac{\pi}{2}$.
The Majorana phases are trivial with $\alpha_{21},\alpha_{31}=0,\pi$. 
The other two cases in Eq.\ref{X} predict zero \CP violation.

Finally we note that the Littlest Seesaw neutrino mass matrix in Eqs.\ref{eq:matrix_LSB}, \ref{eq:mnu2p10}
satisfies Eq.\ref{eq:remnant_flavor_nu} (after multiplying $L_2$ by a minus sign) but
can only satisfy Eq.\ref{eq:remnant_CP_nu} for $\eta = 0$, which is not acceptable,
therefore that model does not possess any remnant \CP symmetry in the neutrino sector.
Instead the LS prediction $\eta = \pm 2\pi/3$ arises from an extra $Z_3$ symmetry of the flavon potential,
as explained below Eq.\ref{eq:mnu2p10}.


\section{{\em Unification}: Grand Unified Theories of Flavour }
\label{ToF}

\begin{figure}[t]
\begin{center}
\includegraphics[width=0.93\textwidth]{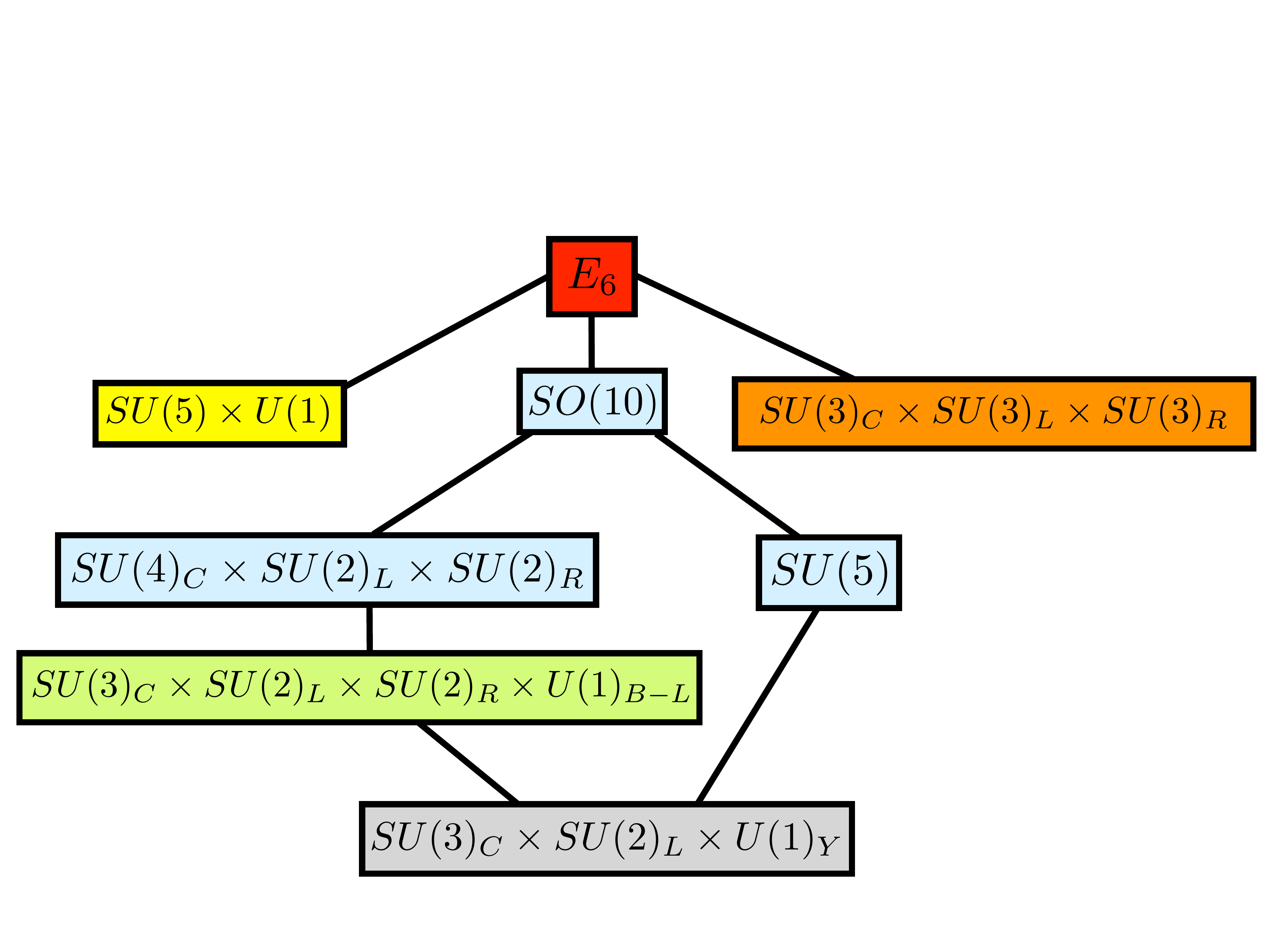}
\end{center}
\vspace*{-4mm}
    \caption{\label{GUTs}\small{Some possible candidate unified gauge groups which are subgroups of $E_6$.
    We shall focus on  $SU(5)$, $SO(10)$ and the Pati-Salam
gauge group $SU(4)_{C} \times SU(2)_L \times SU(2)_R$
    (in pale blue).}}
\vspace*{-2mm}
\end{figure}

We have argued that neutrino masses
and mixing angles are a part of the flavour puzzle, which includes charged leptons
and quarks. However lepton mixing angles are quite large, which seems to suggest 
discrete family symmetry. When the type I seesaw mechanism is also included, 
as a mechanism for small neutrino masses, then large scales may become involved,
possibly as large as the GUT scale.
In such a framework the origin of all quark and lepton masses and mixing could 
be related to some
GUT symmetry group $G_{{\rm GUT}}$, which unifies the
fermions within each family and therefore relates neutrino masses to charged quark and lepton masses.
Indeed, the inclusion of GUTs requires the problem of neutrino masses
and the problem of quark and lepton masses to be tackled simultaneously.
The choice of GUT group is quite large, but some possible candidate gauge groups are shown in Fig.~\ref{GUTs}.
In this section we shall focus mainly on $SU(5)$~\cite{Georgi:1974sy}, the Pati-Salam
gauge group $SU(4)_{C} \times SU(2)_L \times SU(2)_R$~\cite{Pati:1974yy}
and $SO(10)$~\cite{Fritzsch:1974nn} (shown in pale blue in Fig.~\ref{GUTs}).

\subsection{\label{sec:GUTs} SU(5)}

We first consider the gauge group $SU(5)$~\cite{Georgi:1974sy},
which is rank 4 and has 24 gauge bosons which transform as the ${\bf 24}$ adjoint representation.
A LH lepton and quark fermion family is neatly accommodated into the 
$SU(5)$ representations 
$F=\overline{\bf 5}$ and $T={\bf 10}$, where 
\begin{equation}
F= \left(
\begin{array}{c}d_r^c\\d_b^c\\d_g^c\\e^-\\-\nu_e \end{array}\right)_L  , \qquad
T= \left(
\begin{array}{ccccc} 0&u_g^c&-u_b^c&u_r&d_r\\
.&0&u_r^c&u_b&d_b\\ 
.&.&0&u_g&d_g\\ 
.&.&.&0&e^c\\
.&.&.&.&0
\end{array}\right)_L  \ ,
\end{equation}
where $r,b,g$ are quark colours and 
$c$ denotes \CP conjugated fermions. 

The $SU(5)$ gauge group may be broken to the SM 
by a Higgs multiplet in the ${\bf 24}$ representation developing a VEV,
\be
SU(5)\rightarrow SU(3)_C\times SU(2)_L\times U(1)_Y ,
\ee
with
\begin{equation}
\overline{\bf 5}=d^c(\overline{\bf 3},{\bf 1},1/3)\oplus  L({\bf 1},\overline{\bf 2},-1/2),
\end{equation}
\begin{equation}
{\bf 10} =u^c(\overline{\bf 3},{\bf 1},-2/3)\oplus  
Q({\bf 3},{\bf 2},1/6)\oplus e^c({\bf 1},{\bf 1},1),
\end{equation}
where $(Q,u^c,d^c,L,e^c)$ is a complete quark and lepton SM family.
This does not include the 
RH neutrinos, whose \CP conjugates are singlets of $SU(5)$, $\nu^c={\bf 1}$, and may be added separately.
Higgs doublets $H_u$ and $H_d$, 
which break EW symmetry in a two Higgs doublet model, may arise from
$SU(5)$ multiplets $H_{\bf 5}$ and $H_{\overline{\bf 5}}$,
providing the colour triplet components can be made heavy.
This is known as the doublet-triplet splitting problem.

The Yukawa terms for one family may be written as,
\begin{equation}
y_u H_{{\bf 5}i}T_{jk}T_{lm}\epsilon^{ijklm}+ y_{\nu}H_{{\bf 5}i}F^i\nu^c+
y_d H_{\overline{\bf 5}}^iT_{ij}F^j,
\end{equation}
where $\epsilon^{ijklm}$ is the totally antisymmetric tensor with
$i,j,j,k,l=1,\ldots, 5$. These give SM Yukawa terms,
\begin{equation}
y_u H_uQu^c+  y_{\nu}H_uL\nu^c+
y_d (H_dQd^c+H_de^cL).
\end{equation}
The Yukawa couplings for $d$ and $e$ are equal, at least 
at the GUT scale. Extending the argument to three families one finds 
that the Yukawa matrices are related,
\begin{equation}
Y_d=Y_e^T,
\end{equation}
which, though successful for the third family at the GUT scale, fails for the first and
second families. 

Georgi and Jarlskog (GJ)~\cite{Georgi:1979df} proposed that the (2,2) matrix entry 
of the Yukawa matrices may be given by,
\begin{equation}
(Y_{d})_{22} H_{\overline{\bf 45}}T_2F_2 ,
\end{equation}
involving a Higgs field $H_{\overline{\bf 45}}$, where 
$H_d$ is the light linear combination of the
electroweak doublets contained in $H_{\bf{\overline 5}}$~and~$H_{\bf{\overline {45}}}$.
This term reduces to 
\begin{equation}
(Y_{d})_{22}(H_dQ_2d_2^c-3H_de_2^cL_2),
\end{equation}
where the factor of $-3$ is a Clebsch-Gordan
coefficient.
With a zero Yukawa element (texture) in
the (1,1) position, this results in GJ relations, 
\begin{equation}
y_b = y_{\tau}, \quad  y_s = \frac{y_{\mu}}{3}, \quad y_d = 3y_e .
\end{equation}
These apply at the GUT scale.
After renormalisation group (RG) running effects are included, 
they approach consistency with the low energy masses.

The viability of the above GJ relations has been questioned in the light of 
precision determinations of quark 
masses such as $m_s$ from lattice gauge theory
(see, e.g.,~\cite{Ross:2007az}). In supersymmetric (SUSY) $SU(5)$, with low values 
of $\tan \beta = v_u/v_d$, the Yukawa relation for 
the third generation $y_b = y_{\tau}$ at the GUT scale remains viable.
However new $SU(5)$ relations like $ y_{\tau}/ y_b = -3/2$ and 
$y_\mu/y_s = 9/2$ ~\cite{Antusch:2009gu} are now phenomenologically preferred
to the GJ relations $ y_{\tau}/ y_b = 1$ and 
$y_\mu/y_s = 3$.

\subsection{Pati-Salam $SU(4)_{C} \times SU(2)_L \times SU(2)_R$}
\begin{figure}[t]
\centering
\includegraphics[width=0.40\textwidth]{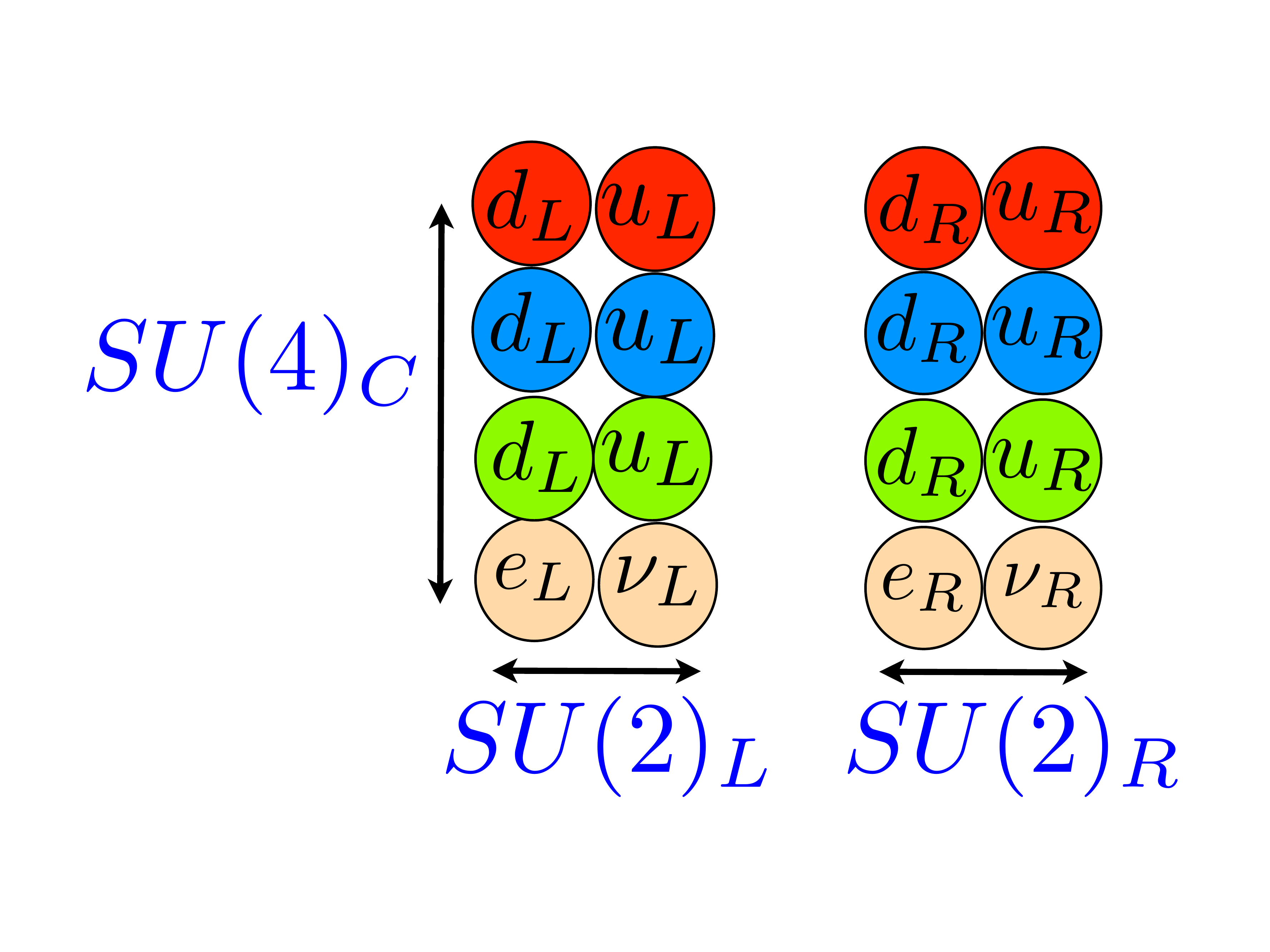}
\vspace*{-4mm}
    \caption{The Pati-Salam multiplets for one family of quarks and leptons where
    the leptons are the fourth colour and the assigment is left-right symmetric,
    so the $\nu_R$ is predicted.} \label{PS2}
\vspace*{-2mm}
\end{figure}

Historically, before $SU(5)$, Pati and Salam (PS) proposed the first type of unification of the SM, based on the 
gauge group~\cite{Pati:1974yy},
\be
SU(4)_{C} \times SU(2)_L \times SU(2)_R
\label{PS}
\ee 
where the leptons are the fourth colour and the assigment is left-right symmetric as shown in Fig.\ref{PS2}.

The LH
 quarks and leptons
transform under the PS gauge group as,
\begin{equation}
{\psi_i}(4,2,1)=
\left(\begin{array}{cccc}
u_r & u_b & u_g & \nu \\ d_r & d_b & d_g & e^-
\end{array} \right)_i
\label{psi}
\end{equation}
\begin{equation}
\psi^c_i(\bar{4},1,\bar{2})=
\left(\begin{array}{cccc}
u^c_r & u^c_b & u^c_g & \nu^c \\ d^c_r & d^c_b & d^c_g & e^c
\end{array} \right)_i
\label{psic}
\end{equation}
where $\psi^c_i$ are the \CP conjugated RH quarks and leptons (so that they become LH)
and 
$i=1\ldots 3$ is a family index.  Clearly the three RHNs (or rather strictly speaking their \CP conjugates $\nu^c_i$) are now predicted as part of the gauge multiplets. This is welcome since it means that neutrino masses, which arise
via the seesaw mechanism, will be related to quark and charged lepton masses as desired.

The Higgs
fields are contained in the following representations,
\begin{equation}
h(1,\bar{2},2)=
\left(\begin{array}{cc}
  {H_u}^+ & {H_d}^0 \\ {H_u}^0 & {H_d}^- \\
\end{array} \right) \label{h}
\end{equation}
where $H_d$ and $H_u$ are two low energy Higgs doublets.

The two heavy Higgs representations are
\begin{equation}
{H}(4,1,2)=
\left(\begin{array}{cccc}
u_H^R & u_H^B & u_H^G & \nu_H \\ d_H^R & d_H^B & d_H^G & e_H^-
\end{array} \right) \label{H}
\end{equation}
and
\begin{equation}
{\bar{H}}(\bar{4},1,\bar{2})=
\left(\begin{array}{cccc}
\bar{d}_H^R & \bar{d}_H^B & \bar{d}_H^G & e_H^+ \\
\bar{u}_H^R & \bar{u}_H^B & \bar{u}_H^G & \bar{\nu}_H
\end{array} \right). \label{barH}
\end{equation}

The Higgs fields are assumed to develop VEVs,
\begin{equation}
\langle \nu_H\rangle \sim M_{GUT}, \ \ \langle \bar{\nu}_H \rangle\sim M_{GUT}
\label{HVEV}
\end{equation}
leading to the symmetry breaking of the PS gauge group at $M_{GUT}$ down to that of the SM,
\begin{equation}
\mbox{SU(4)}\otimes \mbox{SU(2)}_L \otimes \mbox{SU(2)}_R
\longrightarrow
\mbox{SU(3)}_C \otimes \mbox{SU(2)}_L \otimes \mbox{U(1)}_Y
\label{422to321}
\end{equation}
in the usual notation.  Under the symmetry breaking in
Eq.\ref{422to321}, the Higgs field $H$ in Eq.\ref{h} splits into two
Higgs doublets $H_d$, $H_u$ whose neutral components subsequently
develop weak scale VEVs,
\begin{equation}
\langle H_d^0\rangle =v_d, \ \ \langle H_u^0\rangle =v_u \label{vevs}
\end{equation}
with $\tan \beta \equiv v_u/v_d$.

The Yukawa couplings for quarks and leptons are given by
combining the representations in Eqs.\ref{psi}, \ref{psic} and \ref{h} into a PS invariant,
\begin{equation}
y_{ij} h \psi_i \psi^c_j
\label{PSyuk}
\end{equation}
where $i,j=1,\ldots, 3$ are family indices.
Eq.\ref{PSyuk} reduces at low energies to the SM Yukawa couplings
\begin{equation}
y_{ij} (H_uQ_iu^c_j+  H_uL_i\nu^c_j+H_dQ_id^c_j+H_dL_ie^c_j).
\label{PSyuk2}
\end{equation}
Notice that the Yukawa couplings for quarks, charged leptons and neutrinos are equal
at the GUT scale, giving the prediction for Yukawa matrices,
\begin{equation}
Y_d=Y_u=Y_e=Y_{\nu},
\end{equation}
which fails badly at low energies for the first and
second families. As before, these relations may be fixed using Clebsch relations \cite{Antusch:2009gu}.

RH Majorana masses $M_R$ may be generated from the non-renormalisable operators,
\be
\frac{\lambda_{ij}}{\Lambda}\bar{H}\bar{H} \psi^c_i \psi^c_j
\rightarrow \frac{\lambda_{ij}}{\Lambda}\langle \bar{\nu}_H \rangle^2 \nu^c_i \nu^c_j
\equiv M_R^{ij}\nu^c_i \nu^c_j
\ee
where $\Lambda$ may be of order the Planck scale.

\subsection{SO(10)}

\begin{figure}[t]
\centering
\includegraphics[width=0.50\textwidth]{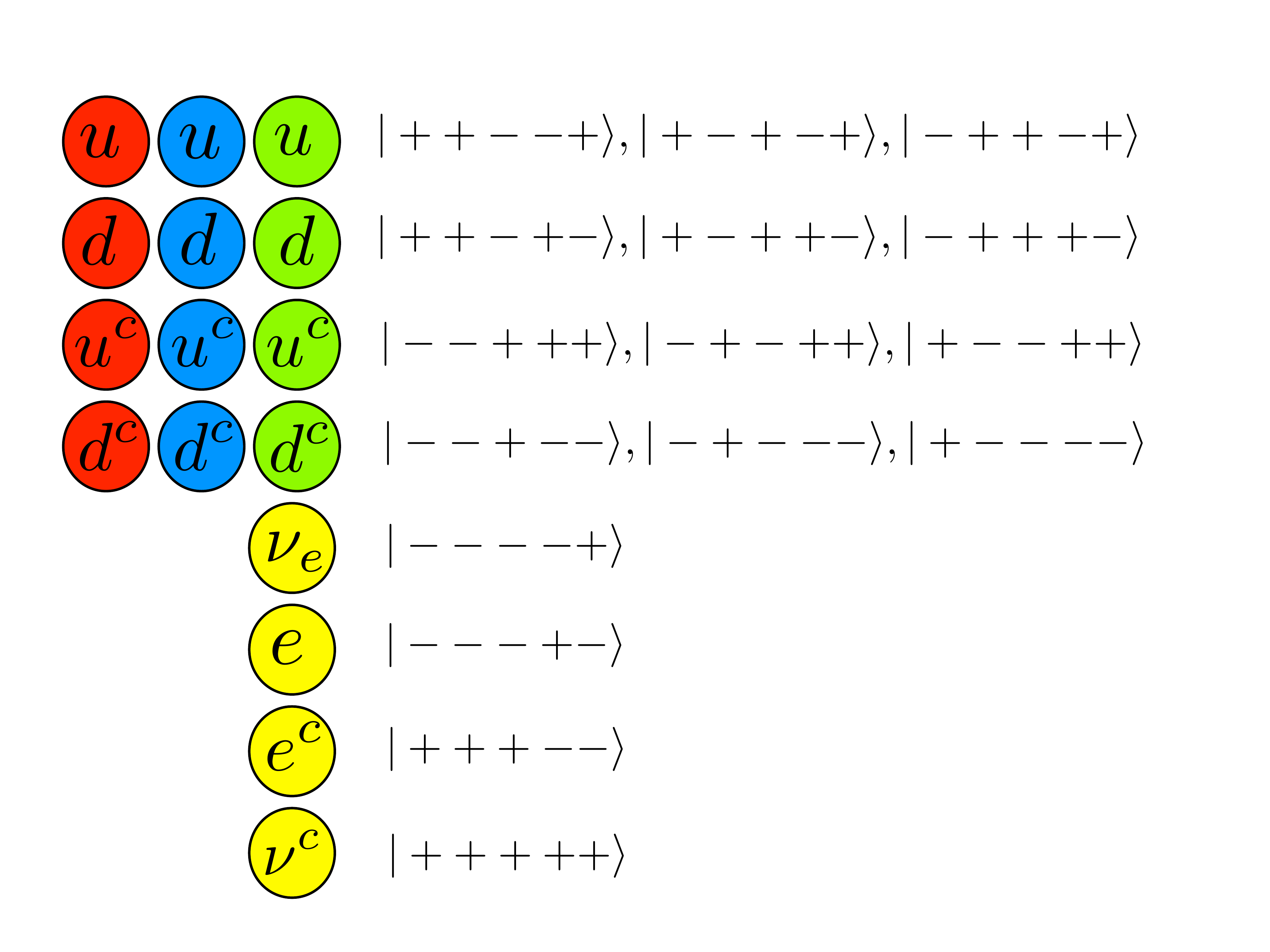}
\vspace*{-4mm}
    \caption{A complete family of LH quarks and leptons (where RH fermions are \CP conjugated)
    forms a single ${\bf 16}$ spinor representation of $SO(10)$, including the RHN (\CP conjugated as $\nu^c$). 
    The notation $|\pm \pm \pm \pm \pm \rangle$ labels the components of the spinor, in terms of a direct product of five
    Pauli matrices with eigenstates $| \pm \rangle$, respectively, with the constraint that there must be an even
    number of  $| - \rangle$ eigenstates. The embedding of the SM gauge group
    is such that the first three components of $|\pm \pm \pm \pm \pm \rangle$ is associated $SU(3)_C$, while the last two components
    are associated with the $SU(2)_L\times U(1)_Y$ gauge group.
    } \label{so10}
\vspace*{-2mm}
\end{figure}

We now consider $SO(10)$~\cite{Fritzsch:1974nn},
which is rank 5 and has 45 gauge bosons which transform as the ${\bf 45}$ adjoint representation.
A complete family of quarks and leptons neatly fits into a single 
${\bf 16}$ spinor representation of $SO(10)$, including the RHN (\CP conjugated as $\nu^c$),
as shown in Fig.\ref{so10}. 
The ${\bf 16}$ spinor representation of $SO(10)$ can be written as the direct product
of five Pauli matrices with eigenstates $|\pm \pm \pm \pm \pm \rangle$, with the constraint that there must be an even
number of $| - \rangle$ eigenstates, where each 
$| \pm \rangle$ is an eigenstate of a single $SU(2)$. The complex conjugate represenation ${\bf \overline{16}}$
corresponds to the states with an odd number of $| - \rangle$ eigenstates.

The theory of Lie groups is extensively covered in a number of textbooks, so here we only recall a few useful facts
which may help to understand the ${\bf 16}$ spinor representation of $SO(10)$.
Recall that $SO(3)$, which is locally isomorphic to $SU(2)$, has a ${\bf 2}$ spinor representation which can be written as 
a single set of Pauli matrices with eigenstates $| \pm \rangle \equiv | \pm {\frac{1}{2}} \rangle$.
The $SO(5)$ spinor representation ${\bf 4}$ can be written as the direct product of 
two Pauli matrices with eigenstates $| \pm \pm \rangle$. $SO(6)$, which is locally isomorphic to $SU(4)$,
has two complex spinor representations where the reducible ${\bf 4 \oplus \overline{4}}$ can be written as the direct product of three Pauli matrices with eigenstates $|\pm \pm \pm \rangle$,
where the ${\bf 4}$ corresponds to the states with an odd number of $| - \rangle$ eigenstates, while the 
${\bf \overline{4}}$ corresponds to the states with an even number of $| - \rangle$ eigenstates.
$SO(6)\sim SU(4)$ has an $SU(3)$ subgroup under which the ${\bf{4}}$ decomposes into a ${\bf 1 \oplus {3}}$
where the singlet is identified as the $| --- \rangle$ state and the triplet as the remaining $|++- \rangle, |+-+ \rangle,
|-++ \rangle$ states. Similarly, the ${\bf \overline{4}}$ decomposes into a ${\bf 1 \oplus \overline{3}}$
where the singlet is identified as the $| +++ \rangle$ state and the triplet as the remaining $|--+ \rangle, |-+- \rangle,
|+-- \rangle$ states. 

$SO(10)$ has a subgroup $SO(6)\times SO(4)$. The SM colour group $SU(3)$ corresponds
to precisely the subgroup of $SO(6)$ discussed in the preceeding paragraph,
where the first three components of $|\pm \pm \pm \pm \pm \rangle$ are associated with $SU(3)_C$
as in Fig.\ref{so10}. In fact the subgroup $SO(6)\times SO(4)$ is locally isomorphic to $SU(4)\times SU(2) \times SU(2)$
which is precisely the Pati-Salam gauge group, so one possible symmetry breaking direction is,
\be
SO(10) \rightarrow SU(4)_{C} \times SU(2)_L \times SU(2)_R
\ee
with 
\be
{\bf 16} \rightarrow ({\bf 4} , {\bf 2}, {\bf 1}) \oplus ({\bf \overline{4}},{\bf {1}},{\bf \overline{2}}).
\ee
Another possible symmetry breaking direction is,
\be
SO(10) \rightarrow SU(5)\times U(1)_X
\ee
with 
\be
{\bf 16} \rightarrow  {\bf \overline{5}}_{-3} \oplus {\bf 10}_{1}   \oplus {\bf {1}}_5
\ee
\be
{\bf 10} \rightarrow  {\bf 5}_{-2}   \oplus {\bf \overline{5}}_{2} .
\ee

The Kronecker product of two spinor representations gives:
\be
{\bf 16}\otimes {\bf 16}= {\bf 10}\oplus {\bf 126}\oplus {\bf 120}.
\ee
With quarks and leptons denoted as $\psi$ in the ${\bf 16}$ representation,
this allows Yukawa couplings if a Higgs $h$ in the ${\bf 10}$ representation of  $SO(10)$ is introduced, 
since ${\bf 10}\otimes {\bf 10}$ contains the singlet, namely,
\begin{equation}
y_{ij} h \psi_i \psi_j
\label{so10yuk}
\end{equation}
where $i,j=1,\ldots, 3$ are family indices.
Eq.\ref{so10yuk} reduces at low energies to the SM Yukawa couplings
\begin{equation}
y_{ij} (H_uQ_iu^c_j+  H_uL_i\nu^c_j+H_dQ_id^c_j+H_dL_ie^c_j),
\label{PSyuk2}
\end{equation}
where $y_{ij}$ is a symmetric matrix.
As in the PS model, 
the Yukawa couplings for quarks, charged leptons and neutrinos are equal
at the GUT scale, giving the prediction for Yukawa matrices,
\begin{equation}
Y_d=Y_u=Y_e=Y_{\nu},
\end{equation}
which may be fixed using Clebsch relations \cite{Antusch:2013rxa}.

RH Majorana masses $M_R$ may be generated from the non-renormalisable operators,
\be
\frac{\lambda_{ij}}{\Lambda}\bar{H}\bar{H} \psi_i \psi_j
\rightarrow \frac{\lambda_{ij}}{\Lambda}\langle \bar{\nu}_H \rangle^2 \nu^c_i \nu^c_j
\equiv M_R^{ij}\nu^c_i \nu^c_j
\ee
where $\Lambda$ may be of order the Planck scale,
and $\bar{H}$ are Higgs in the ${\bf \overline{16}}$ representation, whose RHN component gets a VEV,
breaking $SO(10)$ down to $SU(5)$ at the GUT scale.

\subsection{Flavoured GUTs }

\begin{figure}[t]
\centering
\includegraphics[width=0.60\textwidth]{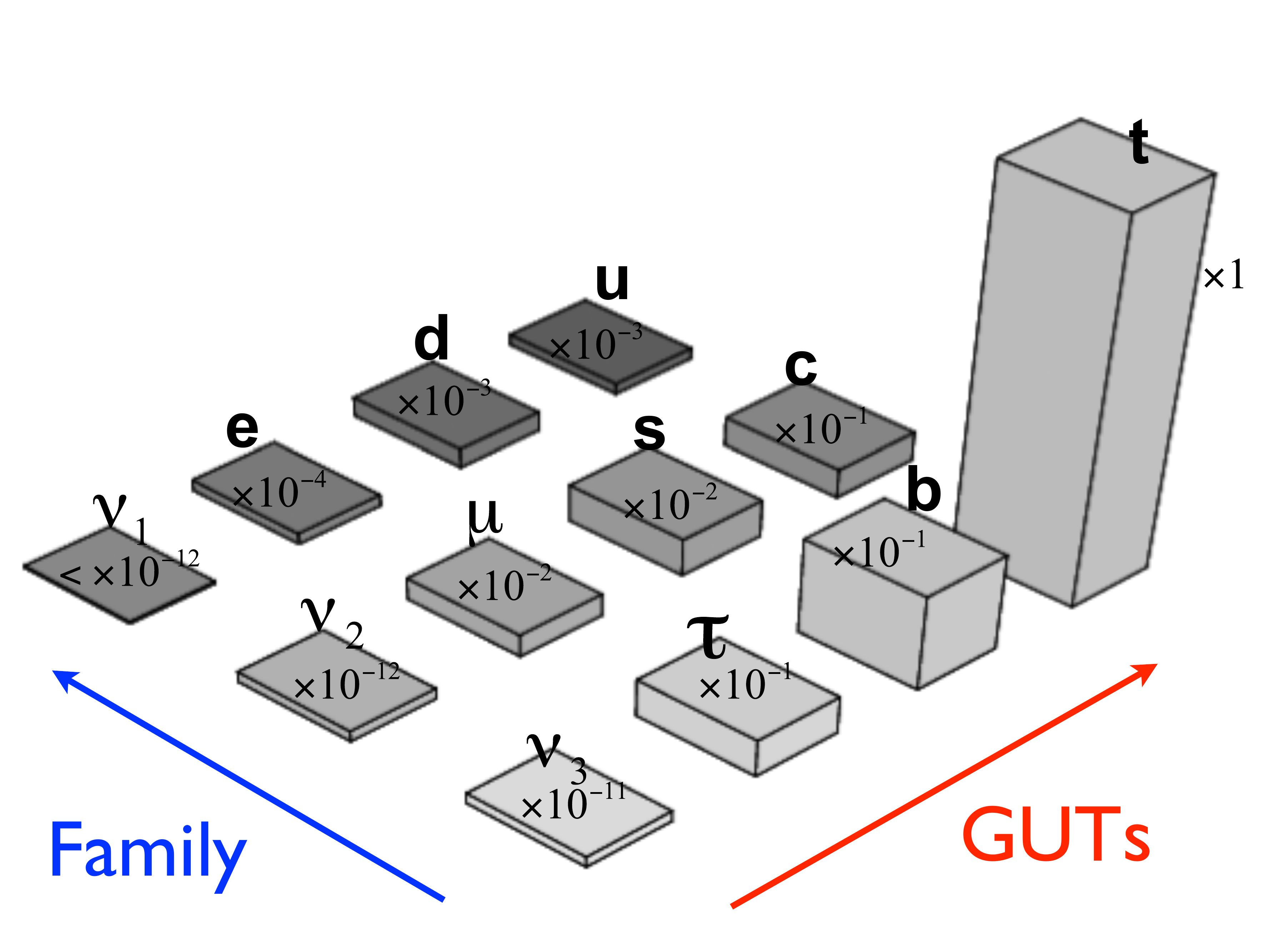}
\vspace*{0mm}
    \caption{Quark and lepton masses lego plot (true heights need to be scaled by the factors shown).
    The (scaled) heights of the towers representing the fermion masses,
show vast hierarchies which are completely mysterious in the SM. 
    GUTs and Family symmetries act in different directions as shown.
     } \label{masses}
\vspace*{0mm}
\end{figure}

The wider problem of the origin of the spectrum of quark and lepton masses suggests 
combining a Grand Unified Theory (GUT) as considered above \cite{Georgi:1974sy,Pati:1974yy,Fritzsch:1974nn}
with a Family Symmetry such as considered in the previous section, acting in different directions, 
as illustrated in Fig.\ref{masses}. 
Putting these two ideas together we are suggestively led to a framework of
new physics beyond the Standard Model based on commuting
GUT and family (FAM) symmetry groups,
\begin{equation}
G_{{\rm GUT}}\times G_{{\rm FAM}} .
\label{symmetry}
\end{equation}
Such Grand Unified Theories of Flavour 
(also known as Flavoured GUTs) 
would include the GUT predictions based on Clebsch relations 
\cite{Georgi:1979df,Ross:2007az,Antusch:2011qg,Antusch:2009gu,Zhang:2012mn}
as well as the prediction of neutrino mixing angles due to the discrete family symmetry,
as discussed in the previous section.
In principle this would allow connections to be made between 
smallest leptonic mixing angle, 
the reactor angle, and the largest quark mixing
angle, the Cabibbo angle, which are roughly
equal to each other up to a factor of $\sqrt{2}$ \cite{Zhang:2012mn},
as discussed in~\cite{Antusch:2011qg,Marzocca:2011dh}. 
Other relations such as the 
Gatto-Sartori-Tonin (GST) relation 
$\theta^q_{12} \approx\sqrt{m_d/m_s}$~\cite{Gatto:1968ss}
might also arise when combining GUTs with Family symmetry \cite{King:2001uz}.

\begin{table}[t]
\begin{center}
\begin{tabular}{|l|c|c|c|c||}
\hline
\hline
$\begin{array}{cc}
 & G_{{\rm GUT}} \\
G_{{\rm FAM}} &
\end{array}$
&
$\begin{array}{cc}
SU(2)_L\times U(1)_Y \\ ~
\end{array}$
&
$\begin{array}{cc}
SU(5) \\ ~
\end{array}$
&
$\begin{array}{cc}
{\rm PS} \\ ~
\end{array}$
&
$\begin{array}{cc}
SO(10) \\ ~
\end{array}$
\\\hline\hline

~$S_3$& \cite{survey-SM-S3}  & & & \cite{Anandakrishnan:2012tj} \\\hline

~$A_4$ &
\cite{survey-SM-A4,Brahmachari:2008fn,King:2011zj,Antusch:2011ic,King:2011ab,Hernandez:2012ra,Ma:2012ez,Altarelli:2012bn}
& 
\cite{survey-SU5-A4,Antusch:2013wn,Cooper:2012wf,Bjorkeroth:2015ora}
&\cite{King:2006np,King:2014iia,Belyaev:2016oxy}
& \\\hline

~$T'$ & \cite{Girardi:2013sza} &\cite{survey-SU5-Tprime} &&  \\\hline

~$S_4$ &
\cite{survey-SM-S4,Altarelli:2009gn,King:2011zj,Hernandez:2012ra,Altarelli:2012bn}
& \cite{Meloni:2011fx,Hagedorn:2012ut}&\cite{Toorop:2010yh} & \cite{survey-SO10-S4} \\\hline

~$A_5$ &\cite{Hernandez:2012ra,Cooper:2012bd}  & \cite{Gehrlein:2014wda} &  & \\\hline

~$T_7$ & \cite{survey-SM-T7,Luhn:2012bc}& &  & \\\hline

~$\Delta(27)$  & \cite{survey-SM-Delta27}& & &\cite{Bjorkeroth:2015uou}  \\\hline

~$\Delta(96)$   & \cite{Toorop:2011jn,Ding:2012xx}& \cite{King:2012in}& & \cite{Varzielas:2012ss}  \\\hline

~$D_N$ &\cite{survey-SM-DN} & & &  \\\hline

~$Q_N$ &  \cite{survey-SM-QN}& &  & \\\hline

~other  & \cite{survey-SM-other} & \cite{Antusch:2014poa} & \cite{Feldmann:2015zwa} &  \\
\hline
\hline
\end{tabular}
\end{center}
\caption{\label{table3}\small{Flavoured GUTs which include discrete family symmetry
    groups and the papers that use these symmetries to successfully describe
    the solar, atmospheric and reactor neutrino data.}}
\end{table}

There are many possible combinations of GUT and family symmetry groups, but not an infinite number.
The models may thus be classified according to the particular
GUT and family symmetry they assume as shown in Table~\ref{table3}. 
Unfortunately, even after specifying the GUT and family symmetry,
there remains a high degree of 
model dependence, depending on the details of the symmetry breaking
and vacuum alignment. In view of this, we shall restrict ourselves to just one example
from the Table~\ref{table3}, which is typical of the kind of approach taken for flavoured GUTs.

\vspace{0.1in}

\underline{Example of a flavoured GUT: $A_4\times SU(5)$}

\vspace{0.1in}

We now describe an example of a recent flavoured GUT from Table~\ref{table3}, namely an
$A_4 \times SU(5)$ SUSY GUT model~\cite{Bjorkeroth:2015ora}
with the following features:
\begin{itemize}
\item Renormalisable at GUT scale.
\item GUT breaking sector explicit, $ \mu $ term generated.
\item MSSM reproduced with R-parity from $\mathbb{Z}_4^R$. 
\item Doublet-triplet splitting via Missing Partner mechanism \cite{Masiero:1982fe}.
\item Proton decay suppressed.
\item Solves the strong \CP problem via Nelson-Barr mechanism \cite{Nelson:1983zb, Barr:1984qx}.
\item Up-type quark strong mass hierarchy explained.
\item Littlest Seesaw model arises with spontaneously broken \CP symmetry.
\end{itemize}

\begin{table}
\centering
\footnotesize
\begin{minipage}[b]{0.45\textwidth}
\centering
\begin{tabular}{| c | c c | c | c | c |}
\hline
\multirow{2}{*}{\rule{0pt}{4ex}Field}	& \multicolumn{5}{c |}{Representation} \\
\cline{2-6}
\rule{0pt}{3ex}			& $A_4$ & SU(5) & $\mathbb{Z}_9$ & $\mathbb{Z}_6$ & $\mathbb{Z}_4^R$ \\ [0.75ex]
\hline \hline
\rule{0pt}{3ex}%
$F$ 					& 3 & $\bar{5} $& 0 & 0 & 1 \\
$T_1$ 					& 1 & 10		& 5 & 0 & 1 \\
$T_2$ 					& 1 & 10		& 7 & 0 & 1 \\
$T_3$ 					& 1 & 10		& 0 & 0 & 1 \\
\rule{0pt}{3ex}%
$N_{\rm atm}^c$ 		& 1 & 1	 		& 7 & 3 & 1 \\
$N_{\rm sol}^c$ 		& 1 & 1 		& 8 & 3 & 1 \\
\rule{0pt}{3ex}%
$\Gamma$				& 1 & 1			& 0 & 3 & 1 \\[0.5ex]
\hline
\end{tabular}
\end{minipage}%
\qquad
\begin{minipage}[b]{0.45\textwidth}
\centering
\begin{tabular}{| c | c c | c | c | c |}
\hline
\multirow{2}{*}{\rule{0pt}{4ex}Field}	& \multicolumn{5}{c |}{Representation} \\
\cline{2-6}
\rule{0pt}{3ex}			& $A_4$ & SU(5) & $\mathbb{Z}_9$ & $\mathbb{Z}_6$ & $\mathbb{Z}_4^R$ \\ [0.75ex]
\hline \hline
\rule{0pt}{3ex}%
$H_5$					& 1 & 5			& 0 & 0 & 0 \\
$H_{\bar{5}}$			& 1 & $\bar{5}$	& 2 & 0 & 0 \\
$H_{45}$	 			& 1 & 45 		& 4 & 0 & 2 \\
$H_{\overline{45}}$ 	& 1 & $\overline{45}$ & 5 & 0 & 0 \\
\rule{0pt}{3ex}%
$\xi$ 					& 1 & 1			& 2 & 0 & 0 \\
$\theta_2$				& 1 & 1			& 1 & 4 & 0 \\
$\phi_{\rm atm}$		& 3 & 1 		& 3 & 1 & 0 \\
$\phi_{\rm sol}$		& 3 & 1 		& 2 & 1 & 0 \\[0.5ex]
\hline
\end{tabular}
\end{minipage}
\caption{Superfields containing SM fermions, the Higgses and relevant flavons.
The left table shows the matter fields which have odd $R$ charge and do not get VEVs.
The right table shows the Higgs fields with even $R$ charge, whose scalar components develop VEVs.
The $H_{45}$ with two units of $R$ charge breaks
$\mathbb{Z}_4^R$ down to $\mathbb{Z}_2^R$, which is identified as conventional R-parity.
}
\label{ta:SMF}
\end{table}

The model also requires the additional discrete symmetries 
$\mathbb{Z}_9\times \mathbb{Z}_6\times \mathbb{Z}_4^R$. 
The superfields relevant for quarks, leptons and Higgs,
including flavons, are shown in Table \ref{ta:SMF}.
SM quarks and leptons are contained in the superfields $F$ and $T_i$. The light MSSM Higgs doublet $H_u$ originates from a linear combination of $H_{5}$ and $H_{45}$, while $H_d$ 
arises from $H_{\overbar{5}}$ and $H_{\overbar{45}}$, in order to obtain acceptable 
relations between down-type quarks and charged leptons. 

Although renormalisable at the GUT scale, light 
fermion masses are suppressed when ``messenger fields''
are integrated out, resulting in effective non-renormalisable operators,
analogous to the way the seesaw mechanism works.
For example, 
the field $ \xi $, which gains a VEV $v_\xi \sim 0.06 M_{\mathrm{GUT}} $, results in a hierarchical fermion mass structure in the up-type quark sector through
effective operators like $ v_u T_i T_j (v_\xi/M)^{6-i-j}$, where $v_u$ is the VEV of $H_u$. 
The resulting symmetric Yukawa matrix for up-type quarks is
\begin{equation}
Y_{ij}^u  \sim \pmatr{\tilde{\xi}^4 & \tilde{\xi}^3 & \tilde{\xi}^2 \\ &\tilde{\xi}^2 & \tilde{\xi} \\& & 1}
\label{upYuk}
\end{equation}
where $ \tilde{\xi} = \braket{\xi}/M \sim 0.1 $ yielding a strong up-type mass hierarchy,
with quark mixing arising in large part from the up-sector.

The field $ \xi $ is in fact quite ubiquitous. As  well as explaining the structure of the up-type
quark mass matrix, 
it is also involved in the mass hierarchy for down-type quarks and charged leptons.
And it is responsible for the mass scales for the RH neutrinos.
Furthermore it yields a highly suppressed $ \mu $ term $ \sim (v_\xi/M)^8 M_{\mathrm{GUT}} $.

The down-type and charged lepton Yukawa matrices $ Y^d \sim Y^e $
are obtained from terms like $ F \phi T H $, leading to nearly diagonal matrices,
\begin{equation} 
	Y^d_{LR} \sim Y^e_{RL} \sim \pmatr{ \dfrac{\braket{\xi} v_{e}}{v_{\Lambda_{24}}^2} & \dfrac{\braket{\xi} v_{\mu}}{{v_{\Lambda_{24}}} {v_{H_{24}}}} & 0 \\[2ex] 0&\dfrac{{v_{H_{24}}} v_{\mu}}{M^2} & 0 \\[2ex] 0& 0& \dfrac{v_{\tau}}{M}}
\label{downYuk}
\end{equation}
where $v_{e, \mu , \tau}$ are flavon VEVs,
while $ v_{\Lambda_{24}} $ and $ v_{H_{24}} $ are VEVs of 
heavy Higgs $ \Lambda_{24} $ and $ H_{24} $.
Here we include the subscripts $LR$ to emphasise the role of the off-diagonal term to LH mixing from $Y^d$. This term introduces \CP violation into the CKM matrix via the phase of 
$\braket{\xi}$. Note that 
the off-diagonal term in $Y^e_{RL}$ gives mainly RH mixing,
with only a subleading negligible
contribution to LH charged lepton mixing $ \theta_{12}^e \sim m_e/m_\mu $.

The superpotential terms related to neutrino masses are,
\begin{equation}
	W_\nu = y_1 H_{5} F\frac{\phi_\mathrm{atm}}{\braket{\theta_2}} N_\mathrm{atm}^c 
	+ y_2 H_{5} F\frac{\phi_\mathrm{sol}}{\braket{\theta_2}} N_\mathrm{sol}^c 
	+ y_3 \frac{\xi^2}{M_\Gamma} N_\mathrm{atm}^c N_\mathrm{atm}^c 
	+ y_4 \xi N_\mathrm{sol}^c N_\mathrm{sol}^c,
\label{eq:neutrinomassWmodel}
\end{equation}
where $y_i$ are dimensionless and $\mathcal{O}(1)$. The first two terms on the RHS of 
Eq.\ref{eq:neutrinomassWmodel}
are analogous to Eq.\ref{nucouplings},
while the latter two terms generate diagonal RH neutrino masses. 
The model is formulated in the real basis of $A_4$ 
in Eq.~\ref{eq:ST}, where 
the vacuum alignment of the flavons may be shown to be: 
\begin{equation}
	\braket{\phi_{\mathrm{atm}}} = v_{\mathrm{atm}} \pmatr{0\\1\\1} , \qquad\qquad \braket{\phi_{\mathrm{sol}}} = v_{\mathrm{sol}} \pmatr{1\\3\\1}.
\end{equation}
This results in a low energy effective Majorana mass matrix of the LSA form
in Eq.~\ref{eq:matrix_LSA} namely,
\begin{equation}
	m^\nu = m_a 
	\left(
\begin{array}{ccc}
	0&0&0\\0&1&1\\0&1&1 
	\end{array}
\right)
	+ m_b e^{i\eta} 
	\left(
\begin{array}{ccc}
	1&3&1\\3&9&3\\1&3&1
	\end{array}
\right).
	\label{eq:mnu3}
\end{equation}
The Abelian flavour symmetry $\mathbb{Z}_9$ fixes the phase $ \eta $ to be one of the ninth roots of unity, through a variant of the mechanism used in \cite{Antusch:2011sx}, including the successful value 
$\eta = 2\pi/3$ in Table~\ref{tab:bfp2}.

\subsection{String theory approaches to flavoured GUTs}
\label{F}

Something is missing from the approaches considered so far: gravity. Any complete theory must make some accommodation for gravity,
at least conceptually. In our last subsection of this review we therefore turn to string theory, or in practice, superstring theory,
as a possible all encompassing framework
which could conceivably provide the origin of a Grand Unified Theory of Flavour - including gravity. Unfortunately, attempts to relate 
superstring theory to particle physics are inconclusive. Nevertheless, it is worth taking a peek at 
where superstring theory stands at present {\it vis \`a vis} flavoured GUTs.

\begin{figure}[t]
\centering
\includegraphics[width=0.40\textwidth]{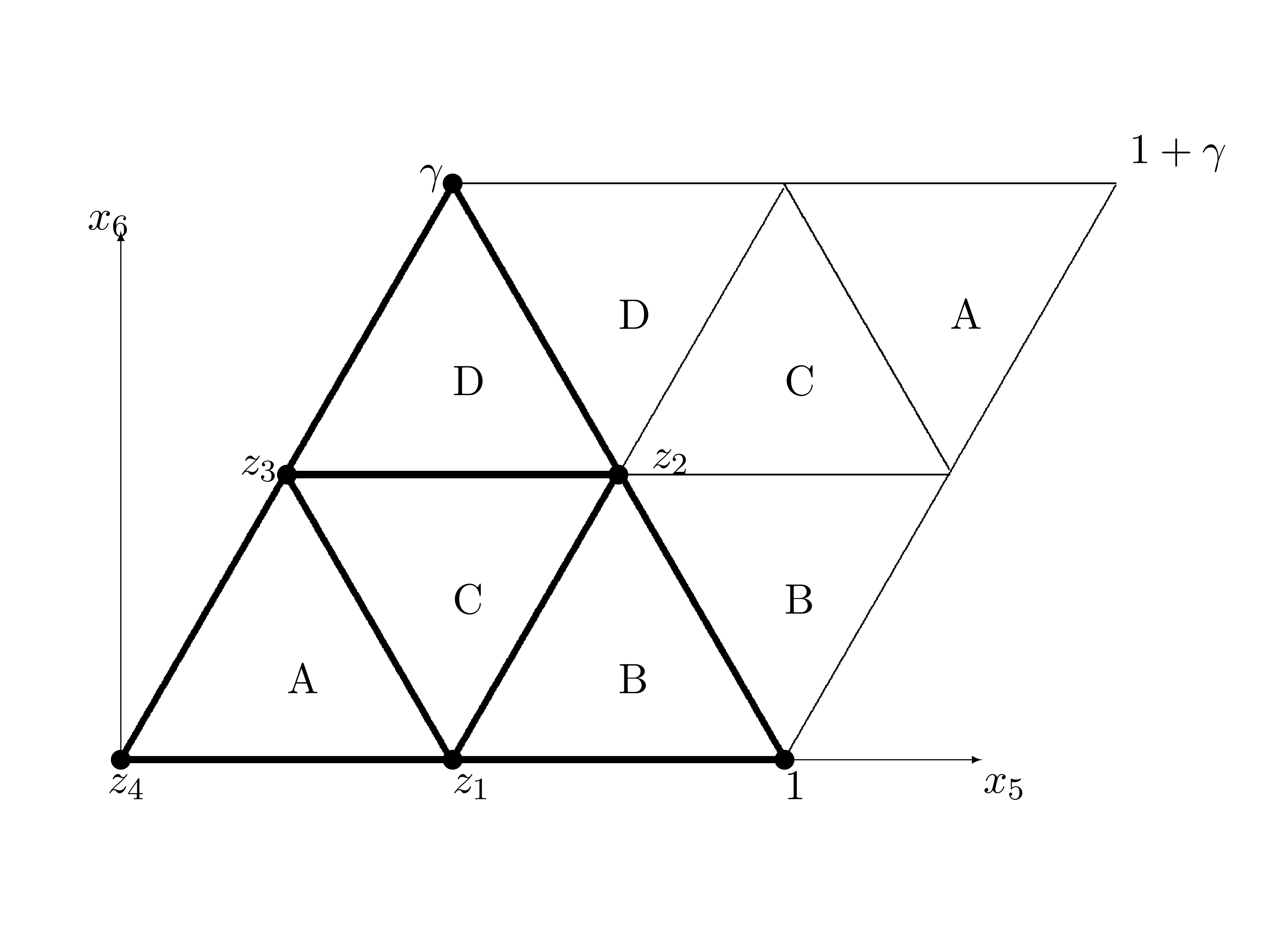}
\vspace*{-4mm}
    \caption{ In a 6-d theory the extra dimensions complexified as $z=x_5 +ix_6$ may be compactified into a torus $T^2$.
    The orbifold $T^2/Z_2$ is based on the twisted torus with a twist angle of $60^\circ$, with fixed points $z_i$.
     The $Z_2$ orbifolding then folds the rhombus into a tetrahedron (the fundamental domain in bold) giving rise to $A_4$ symmetry,
     with the regions $A,B,C,D$ identified respectively. 
    } \label{toby}
\vspace*{-2mm}
\end{figure}

Originally it was hoped that there would be a unique superstring theory based on heterotic string theory
with $E_8\times E_8$ (HE) or $SO(32)$ (HO) in $d=10$ dimensions, where the six extra dimensions are typically
compactified on an orbifold (for a review see e.g. \cite{Nilles:2008gq}).
It is possible to understand the origin of discrete family symmetry within the framework of HE theories with
orbifold compactification. Indeed there has been some interesting work on heterotic string theory in which
flavoured GUTs, i.e. GUTs together with discrete family symmetry, can arise from
orbifold compactification~\cite{Kobayashi:2006wq,Ko:2007dz}. 
For example, the origin of $A_4$ family symmetry can be understood by considering 
a $d=6$ theory compactified on a torus with the orbifolding $T^2/Z_2$
as shown in Fig.\ref{toby} which formed the basis of a model of leptons \cite{Altarelli:2006kg}.
The approach was subsequently extended to a SUSY GUT based on $SU(5)$ in $d=6$, where an $A_4$ family symmetry
was shown to emerge from orbifolding $T^2/(Z_2\times Z_2)$ \cite{Burrows:2009pi}. This approach 
was extended to $d=8$ \cite{Burrows:2010wz}, taking it one step closer
to full HE string theory with $d=10$.

\begin{figure}[t]
\centering
\includegraphics[width=0.40\textwidth]{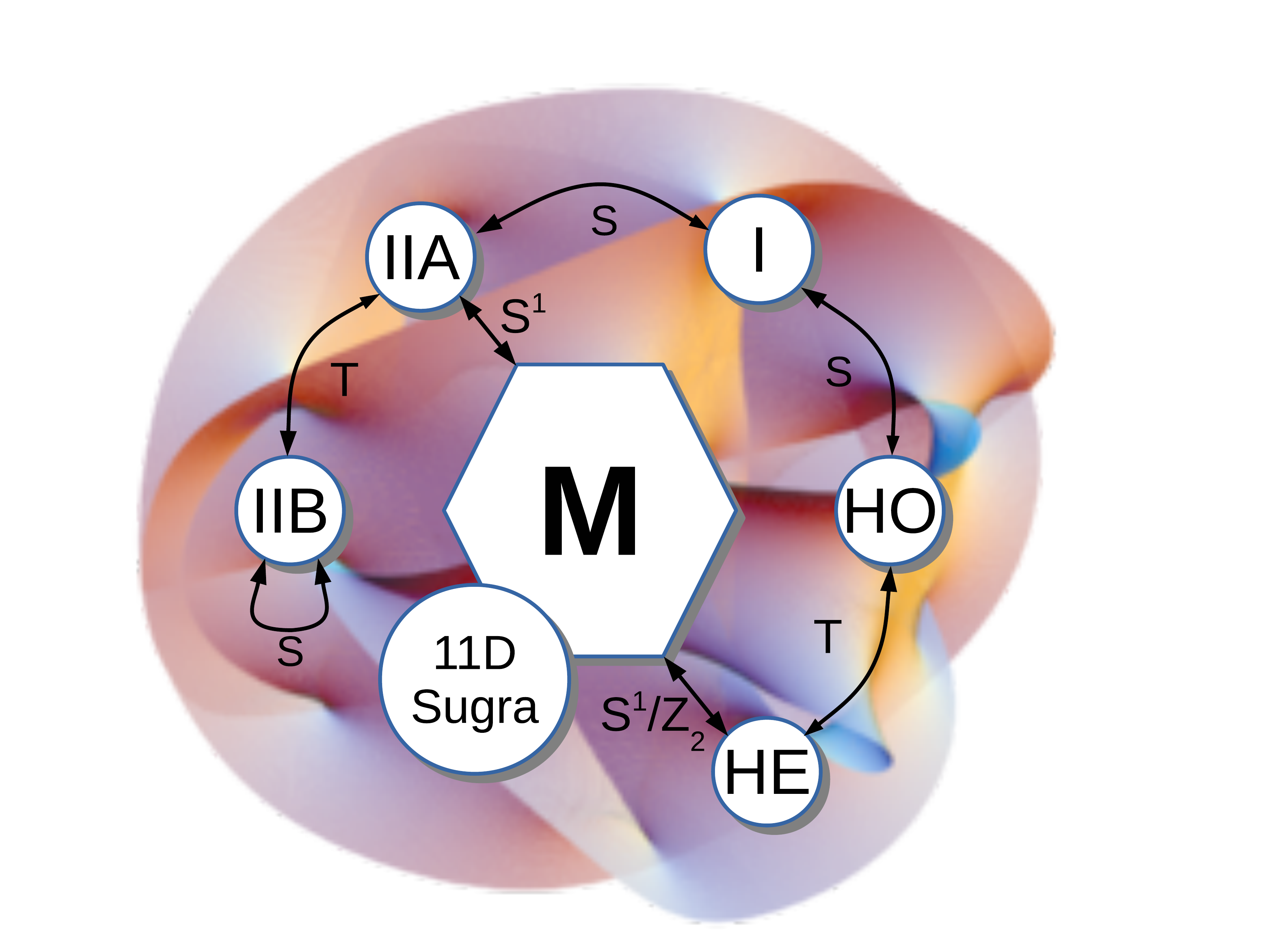}
\vspace*{-4mm}
    \caption{The duality web of string theories against the background of a Calabi-Yao manifold. } \label{string}
\vspace*{-2mm}
\end{figure}

Twenty years ago it was realised that strings also imply branes \cite{Polchinski:1996na},
which are solitonic sub-dimensional objects in $D$ spatial dimensions to which strings may attach themselves, and indeed must do so for 
consistency in certain string theories. Indeed it is possible that the SM gauge group is restricted to one or more of these branes.
Including such $D$-branes, there are other types of string theory denoted as type I, IIA, and IIB which are 
related by a complicated web of dualities, as depicted in Fig.\ref{string}, where $M$ theory is supposed to be the
Mother of all these string theories, whose low energy limit is 11-d supergravity. 
However a generic problem with $D$-brane models is how to achieve unification, for example based on $SU(5)$, and at the same time
a renormalisable top quark Yukawa coupling originating from $H_{\bar 5}F_{\bar 5}T_{10}$.
The issue is that this term is usually forbidden by $U(n)$ type symmetries arising from $D$-brane models.

One way round this, which has 
attracted considerable interest over the recent years, are the F theory models
based on $d=10$ type IIB string theory, but compactified on Calabi-Yao complex fourfold manifolds~\cite{Beasley:2008kw}.
This can be thought of as an elliptic fibration over the $d=10$ base manifold $B_3$, as shown in Fig.~\ref{F}.
Pinch points in the two-tori correspond to singularities in the base manifold where branes can intersect,
with gauge fields such as $SU(5)$ living on the branes and matter fields at the intersection between branes
(for a review see e.g.~\cite{Leontaris:2012mh}).
In Fig.~\ref{F}, the $SU(5)_{GUT}$ group lives on the $S$ brane, while
Yukawa couplings correspond to the intersection of matter curves.
Interestingly exceptional groups such as $E_6$ can be supported on the branes 
(not just $U(n)$) allowing Yukawa couplings to arise from the triple intersection 
of three fundamental multiplets $27^3$ \cite{King:2010mq}.

The $S'$ brane in Fig.~\ref{F} can also support an $SU(5)$ gauge group,
denoted as $SU(5)_{\perp}$, which is different from
the $SU(5)_{GUT}$ group lives on the $S$ brane. 
The full gauge group is then $SU(5)_{GUT}\times SU(5)_{\perp}$,
which is supposed to emerge from an $E_8$ point of enhancement \cite{Beasley:2008kw}.
however the gauge group is broken by fluxes which live on the branes,
analogous to magnetic fields in the extra dimensions. For example, 
$SU(5)_{GUT}$ may be broken to the SM gauge group by hypercharge flux,
where the mechanism naturally allows for doublet-triplet splitting.

The most common assumption is that $SU(5)_{\perp}$ is also broken to 
$U(1)_{\perp}^4$. The four $U(1)_{\perp}$ groups are usually identified 
by so called ``monodromy action'' down to a smaller symmetry $U(1)_{\perp}^n$,
where $n<4$. The surviving $U(1)_{\perp}^n$ group may be used as a family symmetry group,
which controls the number of copies of each chiral SM multiplet. It may be further broken
by additional singlet fields, which play the role of flavon fields, subject to the rules 
of F-theory, and such flavons may then appear in Yukawa operators from which 
the Yukawa matrices may be constructed \cite{King:2010mq}.

It was conjectured in \cite{Antoniadis:2013joa}, that instead of $SU(5)_{\perp}$ being broken to 
the Abelian subgroup $U(1)_{\perp}^4$, it might instead be broken to the discrete non-Abelian
subgroup $S_4$, or one of its discrete subgroups $A_4$, $D_4$, $Z_2\times Z_2$,
which might be identified as a family symmetry group. This possibility was studied in detail in 
~\cite{Karozas:2014aha}, where models were constructed along these lines. 
However this conjecture is far from being established, and it a matter of debate whether or not
such non-Abelian discrete family groups can emerge from F-theory.

\begin{figure}[t]
\centering
\includegraphics[width=0.60\textwidth]{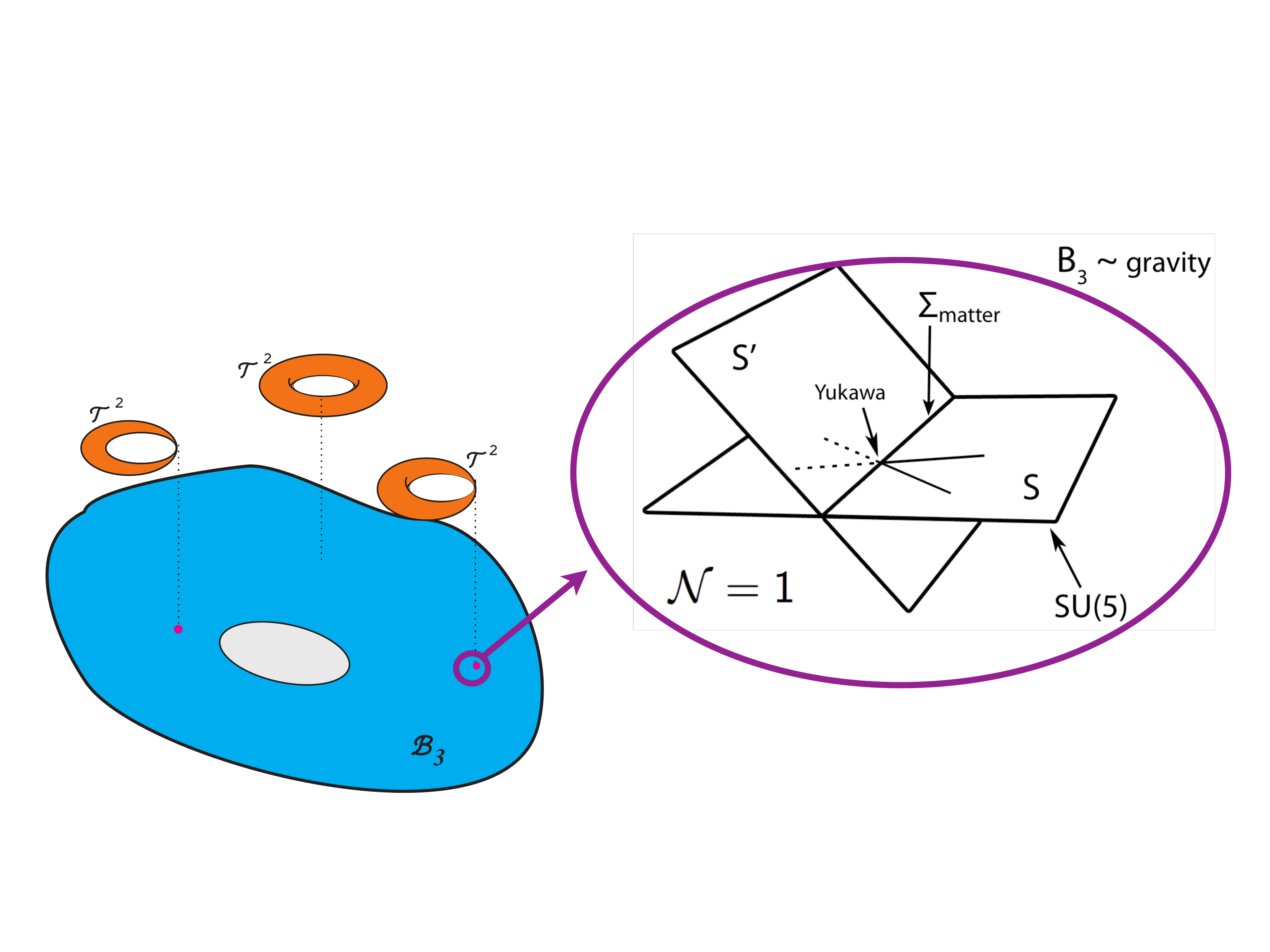}
\vspace*{-4mm}
    \caption{The F-theory construction
    based on $d=10$ type IIB string theory, but compactified on Calabi-Yao complex fourfold manifolds,
equivalent to an elliptic fibration over the compact $d=6$ (3 complex extra dimensions) base manifold $B_3$.
Pinch points in the two-tori correspond to singularities in the base manifold where branes 
which wrap $d=4$ (2 complex extra dimensions) can intersect,
with gauge fields of $SU(5)$ living on branes and matter fields along the compact $d=2$ (1 complex extra dimension) intersection curves between branes.
Yukawa couplings (which do not experience any extra dimensions) correspond to intersection of the matter curves.
} \label{F}
\vspace*{-2mm}
\end{figure}

It is worth to mention some recent developments in M theory compactified on
$G_2$ manifiold. The motivation for such an approach is that M theory is at the centre of the 
web of dualities in Fig.~\ref{string}, and is regarded by many as the most fundamental 
of all string theories. The phenomenological interest in $G_2$ compactification is
in formulating a consistent $SU(5)_{GUT}$, which is broken to the SM gauge group by Wilson line breaking which includes a natural mechanism for doublet-triplet splitting.
The phenomenological consequences of such an approach have been discussed in the 
review article in \cite{Acharya:2012tw}, which contains many original references.

The approach has been extended to $SO(10)_{GUT}$
\cite{Acharya:2015oea}. The Wilson line breaking mechanism preserves the rank of the gauge group,
so that it can break $SO(10)_{GUT}$ via $SU(5)_{GUT}\times U(1)_X$, 
down to the SM gauge group, but it can never break the $U(1)_X$ gauge group.
Furthermore, it was shown that the doublet-triplet splitting mechanism 
when applied to $SO(10)_{GUT}$ does not work in the same way as for $SU(5)_{GUT}$,
and results in extra vector-like states at roughly the TeV scale.
The spectrum of extra vector-like states have the quantum numbers of a complete extra
${\bf 16_X}\oplus {\bf \overline{16_X}}$ superfield representations of $SO(10)_{GUT}$,
although the GUT group is broken of course, and also the extra matter arises from
different high energy ${\bf 16}$ and ${\bf \overline{16}}$ states \cite{Acharya:2015oea}.

The importance of $SO(10)_{GUT}$ for this review
is of course that neutrino masses then become inevitable when it is broken to the SM gauge group.
However, neutrino masses can only arise once the $U(1)_X$ gauge group is broken,
and this can only occur at the field theory level, since Wilson lines cannot reduce rank
as mentioned above. The breaking of $U(1)_X$ can be acheived through the VEVs of the 
RH sneutrino components of the ${\bf 16_X}\oplus {\bf \overline{16_X}}$,
and neutrino masses then can arise via the operator 
$({\bf \overline{16_X}}\ {\bf \overline{16_X}}\ {\bf 16}\ {\bf 16})$.
However the origin of neutrino mass is more complicated than this, since R-parity breaking
is a generic consequence of the M theory approach, and the neutrino mass matrix
for a single physical neutrino mass turns out to be an eleven by eleven matrix!
We only remark here that a phenomenologically acceptable neutrino mass
can emerge from this framework with both the 
type I seesaw mechanism and R-parity violation 
contributing to neutrino mass \cite{Acharya:2015oea}.

\section{Conclusion}
\label{conclusion}

This concludes our review of Unified Models of Neutrinos, Flavour and \CP violation. 
We have come a long way, starting from neutrino experiments and ending up with string theory.
In the Introduction, we recalled the breathtaking advances in neutrino physics from 1998 onwards, then
we summarised what is known and what remains to be learned from neutrino experiments,
and why this means that we must go beyond the SM. After surveying the alternative mechanisms for the origin
of neutrino mass, we emphasised the biggest impact of neutrino physics, namely on the flavour problem,
then summarised the theoretical model building attempts to understand lepton mixing angles, which have had mixed success
so far, leaving the present state of neutrino model building in its present chaotic state. Moving forwards, we have identified
four pillars on which we advocate future models should be constructed, namely:
{\em predictivity},  {\em minimality}, {\em robustness} and {\em unification}.

We first gave an up to date discussion of the latest global fits on lepton mixing parameters
in which we saw that recent data from neutrino experiments gives intriguing hints on the pattern of neutrino masses, lepton mixing angles and the \CP violating phase. Present data (slightly) prefers a normal ordered (NO) neutrino mass pattern, with a
CP phase $\delta = -100^{\circ}\pm 50 ^\circ$, and (more significantly) non-maximal atmospheric mixing. Global fits for the NO case yield lepton mixing angle one sigma ranges:
$\theta_{23}\approx 41.4^\circ \pm 1.6^\circ$, $\theta_{12}\approx 33.2^\circ \pm 1.2^\circ$, $\theta_{13}\approx 8.45^\circ \pm 0.15^\circ$.
Cosmology and large scale structure further provide a limit on the sum of neutrino masses to be below about 
$0.23$~eV, favouring hierarchical neutrino masses over quasi-degenerate masses.

We then turned to the first pillar of any model: {\em predictivity}. Without this, there can be no discrimination between models
based on experiment, and therefore no lasting progress. We should not be embarrassed as theorists that our models are excluded
by experiment, since this represents progress; we should be much more concerned if our models do not make predictions
and so cannot be excluded! In this spirit, we reviewed 
simple patterns of lepton mixing such as bimaximal, golden ratio and tri-bimaximal,
which are not viable by themselves but may be combined with 
charged lepton corrections leading to solar mixing sum rules, or the structures may be partly preserved as in 
trimaximal lepton mixing leading to atmospheric mixing rules. Such sum rules are realistic targets for future experiments.
Indeed it seems that the $\text{TM}_2$ mixing sum rule is under severe tension, but the $\text{TM}_1$ sum rule survives.

The second pillar of any model, {\em minimality}, was then rigorously applied. Casting aside a wealth of viable models 
of neutrinos, some of which were reviewed in the Introduction, 
we have mainly focussed on the most minimal origin of neutrino mass based on 
the elegant type I seesaw mechanism, including the
one and two RH neutrino (RHN) models, the sequential dominance of three RH neutrinos,
constrained sequential dominance and the highly predictive littlest seesaw (LS) models, which includes
the $\text{TM}_1$ mixing sum rules amongst its predictions. We discussed the impact of future precision oscillation 
experiments on the LS models, which shows that the planned experiments are quite capable of excluding these models.
If they survive, then one must take such models seriously. If they are excluded then perhaps other models will emerge.
In this way, progress towards understanding the flavour puzzle can be made.

The third pillar on which any model should be based is that of {\em robustness}, meaning that any model
should not be {\it ad hoc}, but should have some theory behind it, or at least a symmetry.
After a brief review of finite group theory, we identified the Klein symmetry relevant for 
the Majorana neutrino mass matrix, and how this may be embedded into a non-Abelian family symmetry
spontaneously broken by flavons. We then described semi-direct models where only half the Klein symmetry
is preserved in the neutrino sector, and discussed the LS model as an example.
We then turned to spontaneous \CP violation, including invariants and the consistency condition,
before turning to the idea of residual \CP, which allows the \CP phases to be predicted.

Finally we turned to the fourth and final pillar which we advovate for models of flavour,
namely that of {\em unification}. Although seemingly rather esoteric, it has a solid motivation
in the history of physics going back to Maxwell's electromagnetism. It also has a practical motivation,
in that it necessarily brings in the quark sector into the same framework as the lepton sector,
so that any unified theory of leptons will also be a theory of quarks as well. This is important,
since any resolution to the flavour problem must include both quarks and leptons.
After an introductions to GUTs, we discussed models which combine family symmetry with GUTs,
the so called flavoured GUTs, limiting ourselves to a table of models in the literature, together with 
one example to illustrate the method. We finished off with some brief speculations 
about the possible string theory origin of such theories.

It is worth assessing where we stand in our quest towards a model of a unified model of neutrinos,
flavour and \CP violation, based on the four pillars of 
{\em predictivity},  {\em minimality}, {\em robustness} and {\em unification}.
At this moment in time, the Littlest Seesaw model has emerged as a possible
candidate which seems to satisfy all four requirements.
Indeed, all of the examples discusssed in this review involve the Littlest Seesaw as a common
thread which spans all four pillars. The reason for doing this is to show how 
any candidate theory should rest on these four principles. We could have chosen some other model
to demonstrate this, and it really does not matter which: we chose the Littlest Seesaw since it provides
a convenient example which highlights all four aspects of model building applied in a coherent way across
all of the desiderata. Let us therefore briefly give a critique
of the Littlest Seesaw model in all four categories.

The Littlest Seesaw is certainly {\em predictive}, 
with the neutrino masses and PMNS matrix fixed by two parameters, but on the other hand it is 
easy to rule it out by say the observation of an inverted ordering, or a definitive observation
of non-maximal atmospheric mixing in future experiments. The Littlest Seesaw is definitely {\em minimal},
involving just two RH neutrinos in the type I seesaw mechanism,
but we need to explain why in a particular basis 
the two right-handed neutrino mass matrix and the charged lepton mass matrix are diagonal,
and why in this basis the Dirac mass matrix has the CSD(3) form.
The Littlest Seesaw may be {\em robust}, in the sense that 
the required vacuum alignments for type B at least may 
arise from $S_4$ symmetry realised in a semi-direct way
with residual $Z_3^T$ in the charged lepton sector and $Z_2^{SU}$ in the neutrino sector,
but on the other hand the actual details of dynamical vacuum alignment (not discussed here)
are still quite complicated. The Littlest Seesaw can be incorporated into a {\em unified}
model based on $SU(5)$, but 
in practice we saw that such models are still rather complicated, involving 
rather large additional discrete symmetries as well as large numbers of flavon and messenger fields.
Fortunately the additional parameters which appear in the ultraviolet do not seem to be relevant 
for the low energy predictions of the model, but this does not alter the fact that these models are complicated.
Perhaps the ultraviolet completion of these models in the framework of string theory could
eventually lead to a simpler theory, at least in principle?

In conclusion, the discovery of neutrino mass and mixing continues to offer tantalising clues
that may help to unravel the mystery of fermion flavour, mass, mixing and \CP violation.
Although neutrino model building appears presently to be in disarray, the emerging experimental consensus
on some of the open questions in neutrino physics such as the ordering, scale and nature of neutrino mass
and the latest hints on the lepton mixing angles and \CP phase, will serve to shed 
light on the correct model building path. By constructing models based on the four pillars of
{\em predictivity},  {\em minimality}, {\em robustness} and {\em unification},
it may be possible for some young researcher reading this to 
eventually realise Feynman's dream of understanding flavour.

\vspace{0.1in}
The author acknowledges the STFC Consolidated Grant ST/L000296/1 and the European Union's Horizon 2020 Research and Innovation programme under Marie Sk\l{}odowska-Curie grant agreements 
Elusives ITN No.\ 674896 and InvisiblesPlus RISE No.\ 690575.
\vspace{0.1in}

\appendix

\section*{Appendix}


\section{\label{app:CGs}$\boldsymbol{S_4}$ and $\boldsymbol{A_4}$ group theory }
The Kronecker products of the groups are basis independent but the 
values of the Clebsch-Gordan coefficients
depend on the basis. 
We denote the Kronecker products and Clebsch-Gordan coefficients of $S_4$ 
in the basis of Eq.\ref{table} by the
following (where $n$ counts the number of primes which appear,
e.g. ${\bf 3}\otimes
{\bf 3}^\prime \rightarrow {\bf 3}^\prime$ has $n=2$ primes):
%
%
%
%
$$
\begin{array}{lll}
{\bf 1}^{(\prime)} \otimes {\bf 1}^{(\prime)} ~\rightarrow ~{\bf
  1}^{(\prime)} ~~
\left\{ \begin{array}{c} 
~\\n=\mathrm{even}\\~
\end{array}\right.
&%
\left.
\begin{array}{c} 
{\bf 1}^{\phantom{\prime}} \otimes {\bf 1}^{\phantom{\prime}} ~\rightarrow ~{\bf 1}^{\phantom{\prime}}\\
{\bf 1}^{{\prime}} \otimes {\bf 1}^{{\prime}} ~\rightarrow ~{\bf 1}^{\phantom{\prime}}\\
{\bf 1}^{\phantom{\prime}} \otimes {\bf 1}^{{\prime}} ~\rightarrow ~{\bf 1}^{{\prime}}
\end{array}
\right\}
&
\alpha \beta \ ,
\\[10mm]
{\bf 1}^{(\prime)} \otimes \;{\bf 2} \;~\rightarrow \;~{\bf 2}^{\phantom{(\prime)}}~~ \left\{
\begin{array}{c}
n=\mathrm{even} \\
n=\mathrm{odd}
\end{array} \right.
&%
\left.
\begin{array}{c}
{\bf 1}^{\phantom{\prime}} \otimes {\bf 2} ~\rightarrow ~{\bf 2} \\
{\bf 1}^{\prime} \otimes {\bf 2} ~\rightarrow ~{\bf 2}\\
\end{array}\;~
\right\}
&
 \alpha \begin{pmatrix} \beta_1 \\ (-1)^n \beta_2\end{pmatrix}  ,
\\[7mm]
{\bf 1}^{(\prime)} \otimes {\bf 3}^{(\prime)} ~\rightarrow ~{\bf 3}^{(\prime)}
~~ \left\{ \begin{array}{c}
~\\[3mm]n=\mathrm{even} \\[3mm]~
\end{array}\right.
&%
\left. 
\begin{array}{c}
{\bf 1}^{\phantom{\prime}} \otimes {\bf 3}^{\phantom{\prime}} ~\rightarrow ~{\bf 3}^{\phantom{\prime}}
\\
{\bf 1}^{{\prime}} \otimes {\bf 3}^{{\prime}} ~\rightarrow ~{\bf 3}^{\phantom{\prime}}
\\
{\bf 1}^{\phantom{\prime}} \otimes {\bf 3}^{{\prime}} ~\rightarrow ~{\bf 3}^{{\prime}}
\\
{\bf 1}^{{\prime}} \otimes {\bf 3}^{\phantom{\prime}} ~\rightarrow ~{\bf 3}^{{\prime}}
\end{array}
\right\}
&
 \alpha   \begin{pmatrix} \beta_1 \\  \beta_2\\\beta_3 \end{pmatrix}  ,
\\[12.2mm]
{\bf 2} \;\; \otimes \;\;{\bf 2} \;~\rightarrow \;~{\bf 1}^{(\prime)} ~~ \left\{\begin{array}{c}
n=\mathrm{even}\\
n=\mathrm{odd}
\end{array}\right.
&%
\left.
\begin{array}{c}
{\bf 2} \otimes {\bf 2} ~\rightarrow ~{\bf 1}^{\phantom{\prime}} \\
{\bf 2} \otimes {\bf 2} ~\rightarrow ~{\bf 1}^{{\prime}} 
\end{array}~\;
\right\}
&
 \alpha_1 \beta_2 + (-1)^n \alpha_2 \beta_1 \ , 
\\[7mm]
{\bf 2} \;\;\otimes \;\; {\bf 2} ~\;\rightarrow \;~{\bf 2}^{\phantom{(\prime)}} ~~ \left\{ \begin{array}{c}
~\\[-3mm] n=\mathrm{even}\\[-3mm]~
\end{array}\right.
&%
\left.
\begin{array}{c}
~\\[-3mm]
{\bf 2} \otimes {\bf 2} ~\rightarrow ~{\bf 2} \\[-3mm]~
\end{array}~~\,
\right\}
&
   \begin{pmatrix} \alpha_2 \beta_2 \\  \alpha_1\beta_1 \end{pmatrix} , 
%
%
\end{array}
$$
$$
\begin{array}{lll}
{\bf 2}\;\; \otimes \; {\bf 3}^{{(\prime)}} ~\rightarrow ~{\bf 3}^{{(\prime)}} ~~ \left\{\begin{array}{c}
~\\[-2mm] n=\mathrm{even}\\ \\[2mm]
n=\mathrm{odd}\\[-2mm]~
\end{array}\right.
&%
\left.
\begin{array}{c}
{\bf 2} \otimes {\bf 3}^{\phantom{\prime}} ~\rightarrow ~{\bf 3}^{\phantom{\prime}} \\
{\bf 2} \otimes {\bf 3}^{{\prime}} ~\rightarrow ~{\bf 3}^{{\prime}} \\[3mm]
{\bf 2} \otimes {\bf 3}^{\phantom{\prime}} ~\rightarrow ~{\bf 3}^{{\prime}} \\
{\bf 2} \otimes {\bf 3}^{{\prime}} ~\rightarrow ~{\bf 3}^{\phantom{\prime}} 
\end{array}\;
\right\}
&
 \alpha_1 \begin{pmatrix} \beta_2 \\  \beta_3\\\beta_1 \end{pmatrix} + (-1)^n
\alpha_2 \begin{pmatrix} \beta_3 \\  \beta_1\\\beta_2 \end{pmatrix}  ,
\\[13.5mm]
{\bf 3}^{(\prime)} \otimes {\bf 3}^{(\prime)} ~\rightarrow ~{\bf 1}^{(\prime)}
~~ \left\{ \begin{array}{c}
~\\n=\mathrm{even}\\~
\end{array}\right.
&%
\left.\begin{array}{c}
{\bf 3}^{\phantom{\prime}} \otimes {\bf 3}^{\phantom{\prime}} ~\rightarrow ~{\bf 1}^{\phantom{\prime}}
\\
{\bf 3}^{{\prime}} \otimes {\bf 3}^{{\prime}} ~\rightarrow ~{\bf 1}^{\phantom{\prime}}
\\
{\bf 3}^{\phantom{\prime}} \otimes {\bf 3}^{{\prime}} ~\rightarrow ~{\bf 1}^{{\prime}}
\end{array}\right\}
&
 \alpha_1 \beta_1 +\alpha_2\beta_3+\alpha_3\beta_2 \ ,
\\[9mm]
{\bf 3}^{(\prime)} \otimes {\bf 3}^{(\prime)} ~\rightarrow ~{\bf 2}^{\phantom{(\prime)}} ~~
\left\{ \begin{array}{c}
~\\[-3mm]
n=\mathrm{even}\\ \\[1mm]
n=\mathrm{odd}\\[-4.5mm]~
\end{array}\right.
&%
\left.\begin{array}{c}
{\bf 3}^{\phantom{\prime}} \otimes {\bf 3}^{\phantom{\prime}} ~\rightarrow ~{\bf 2} \\
{\bf 3}^{{\prime}} \otimes {\bf 3}^{{\prime}} ~\rightarrow ~{\bf 2} \\[3mm]
{\bf 3}^{\phantom{\prime}} \otimes {\bf 3}^{{\prime}} ~\rightarrow ~{\bf 2} \\
\end{array}\;
\right\}
&
\begin{pmatrix} \alpha_2 \beta_2 +\alpha_3 \beta_1+\alpha_1\beta_3\\ 
(-1)^n(\alpha_3 \beta_3+\alpha_1\beta_2+\alpha_2\beta_1) \end{pmatrix} ,
\\[10.5mm]
{\bf 3}^{(\prime)} \otimes {\bf 3}^{(\prime)} ~\rightarrow ~{\bf 3}^{(\prime)}
~~ \left\{\begin{array}{c}
~\\n=\mathrm{odd}\\~
\end{array}\right.
&%
\left.\begin{array}{c}
{\bf 3}^{\phantom{\prime}} \otimes {\bf 3}^{\phantom{\prime}} ~\rightarrow ~{\bf 3}^{{\prime}}
\\
{\bf 3}^{\phantom{\prime}} \otimes {\bf 3}^{{\prime}} ~\rightarrow ~{\bf 3}^{\phantom{\prime}}
\\
{\bf 3}^{{\prime}} \otimes {\bf 3}^{{\prime}} ~\rightarrow ~{\bf 3}^{{\prime}}
\end{array}\right\}
& 
\begin{pmatrix} 
2 \alpha_1 \beta_1-\alpha_2\beta_3-\alpha_3\beta_2 \\  
2 \alpha_3 \beta_3-\alpha_1\beta_2-\alpha_2\beta_1 \\  
2 \alpha_2 \beta_2-\alpha_3\beta_1-\alpha_1\beta_3 
 \end{pmatrix} ,
\\[9mm]
{\bf 3}^{(\prime)} \otimes {\bf 3}^{(\prime)} ~\rightarrow ~{\bf 3}^{(\prime)}~~\left\{\begin{array}{c}
~\\n=\mathrm{even}\\~
\end{array}\right.
&%
\left.\begin{array}{c}
{\bf 3}^{\phantom{\prime}} \otimes {\bf 3}^{\phantom{\prime}} ~\rightarrow ~{\bf 3}^{\phantom{\prime}}
\\
{\bf 3}^{{\prime}} \otimes {\bf 3}^{{\prime}} ~\rightarrow ~{\bf 3}^{\phantom{\prime}}
\\
{\bf 3}^{\phantom{\prime}} \otimes {\bf 3}^{{\prime}} ~\rightarrow ~{\bf 3}^{{\prime}}
\end{array}\right\}
&
\begin{pmatrix} 
\alpha_2\beta_3-\alpha_3\beta_2 \\  
\alpha_1\beta_2-\alpha_2\beta_1 \\  
\alpha_3\beta_1-\alpha_1\beta_3 
 \end{pmatrix}  .
\end{array}\\[3mm]
$$
%
%
%
%

The $A_4$ Clebsch-Gordan coefficients can be obtained from these
expressions by simply dropping all $S_4$ primes and identifying the components of
the $S_4$ doublet ${\bf 2}$ as the ${\bf 1''}$ and ${\bf 1'}$ representations
of $A_4$. We thus find the non-trivial $A_4$ products, explicitly,

%
%
%
%
$$
\begin{array}{lcl}
{\bf 1'} \otimes {\bf 1''} ~\rightarrow ~{\bf 1} 
&&
\alpha \beta \ ,\\[3mm]
{\bf 1'} \otimes {\bf 3} ~\rightarrow ~{\bf 3}  
&&
\alpha \begin{pmatrix} 
\beta_3 \\  
\beta_1 \\  
\beta_2 
 \end{pmatrix}  , \\[8mm]
{\bf 1''} \otimes {\bf 3} ~\rightarrow ~{\bf 3}  
&&
\alpha \begin{pmatrix} 
\beta_2 \\  
\beta_3 \\  
\beta_1 
 \end{pmatrix}  , \\[8mm]
{\bf 3} \otimes {\bf 3} ~\rightarrow ~{\bf 1} 
&&
\alpha_1\beta_1 +\alpha_2\beta_3+\alpha_3\beta_2 \ ,\\[2mm]
{\bf 3} \otimes {\bf 3} ~\rightarrow ~{\bf 1'} 
&&
\alpha_3\beta_3 +\alpha_1\beta_2+\alpha_2\beta_1 \ ,\\[2mm]
{\bf 3} \otimes {\bf 3} ~\rightarrow ~{\bf 1''} 
&&
\alpha_2\beta_2 +\alpha_3\beta_1+\alpha_1\beta_3 \ ,\\[3mm]
{\bf 3} \otimes {\bf 3} ~\rightarrow ~{\bf 3}  + {\bf 3} 
&&
\begin{pmatrix} 
2 \alpha_1 \beta_1-\alpha_2\beta_3-\alpha_3\beta_2 \\  
2 \alpha_3 \beta_3-\alpha_1\beta_2-\alpha_2\beta_1 \\  
2 \alpha_2 \beta_2-\alpha_3\beta_1-\alpha_1\beta_3 
 \end{pmatrix}
+
\begin{pmatrix} 
\alpha_2\beta_3-\alpha_3\beta_2 \\  
\alpha_1\beta_2-\alpha_2\beta_1 \\  
\alpha_3\beta_1-\alpha_1\beta_3 
 \end{pmatrix}   . 
\end{array}
$$
%

Although the table in Eq.\ref{table} shows the diagonal $T$ basis of $A_4$, it is sometimes convenient to 
work in diagonal $S$ basis in which all matrices are real in the triplet representation \cite{Ma:2001dn}, 
\begin{equation}\label{eq:ST}
S=\left(
\begin{array}{ccc}
1&0&0\\
0&-1&0\\
0&0&-1\\
\end{array}
\right), \ \ \ \ 
T=\left(
\begin{array}{ccc}
0&1&0\\
0&0&1\\
1&0&0\\
\end{array}
\right).
\end{equation}
From these generators one may obtain all 12 real $3\times 3 $ matrix group elements after multiplying these two matrices together in all possible ways \cite{Ma:2001dn}.
Note that although the basis in Eq.\ref{eq:ST} differs from Eq.\ref{table}, in both bases $T$ is traceless
since $1+\omega +\omega^2 = 0$ and is said to have zero character in all bases,
while $S$ has a character (or trace) of $-1$ in all bases.
In the basis of Eq.\ref{eq:ST} one has the following Clebsch rules for the multiplication of two triplets, 
${\bf 3}\otimes{\bf 3} = {\bf 1} \oplus {\bf 1'} \oplus {\bf 1''} \oplus {\bf 3_1} \oplus {\bf 3_2}$, with 
\begin{equation}\label{pr}
\begin{array}{lll}
(ab)_1&=&a_1b_1+a_2b_2+a_3b_3\,;\\
(ab)_{1'}&=&a_1b_1+\omega^2 a_2b_2+\omega a_3b_3\,;\\
(ab)_{1''}&=&a_1b_1+\omega a_2b_2+\omega^2 a_3b_3\,;\\
(ab)_{3_1}&=&(a_2b_3,a_3b_1,a_1b_2)\,;\\
(ab)_{3_2}&=&(a_3b_2,a_1b_3,a_2b_1)\,,
\end{array}
\end{equation}
where $a=(a_1,a_2,a_3)$ and $b=(b_1,b_2,b_3)$
are the two triplets and $\omega^3=1$. 
These differ from the Clebsch rules in the diagonal (but complex) $T$ basis given earlier,
showing that,
although the Kronecker product decomposition is valid in all bases, 
the Clebsch rules are basis dependent.

\end{document}